\documentclass[a4paper,fleqn,usenatbib]{mnras}
\usepackage[T1]{fontenc}
\usepackage{ae,aecompl}

\usepackage{graphicx}	
\usepackage{amsmath}	
\usepackage{amssymb}	
\usepackage{subfig}

\title[Fast and slow rotators in Illustris] 
{The Origin and Evolution of Fast and Slow Rotators in the Illustris Simulation}

\author[Z. Penoyre et al.]{
Zephyr Penoyre$^{1,2}$\thanks{zpenoyre@astro.columbia.edu},
Benjamin P. Moster$^{1,3,4}$,
Debora Sijacki$^{1}$,
Shy Genel$^{2,5}$
\\
$^1$ Institute of Astronomy and Kavli Institute for Cosmology, University of Cambridge, Madingley Road, Cambridge CB3 0HA, UK\\
$^2$ Department of Astronomy, Columbia University, NYC, NY 10027, USA\\
$^3$ Universit\"ats-Sternwarte, Ludwigs-Maximilians-Universit\"at M\"unchen, Scheinerstr. 1, 81679 M\"unchen, Germany\\
$^4$ Max-Planck-Institut f\"ur Astrophysik, Karl-Schwarzschild Stra\ss e 1, 85748 Garching, Germany \\
$^5$ Center for Computational Astrophysics, Flatiron Institute, 162 Fifth Avenue, New York, NY 10010, USA
}

\date{\today}

\pubyear{2017}

\begin{document}
\label{firstpage}
\pagerange{\pageref{firstpage}--\pageref{lastpage}}
\maketitle

\begin{abstract}
Using the Illustris simulation, we follow thousands of elliptical galaxies
back in time to identify how the dichotomy between fast and slow rotating
ellipticals (FRs and SRs) develops. Comparing to the
$\textrm{ATLAS}^\textrm{3D}$ survey, we show that Illustris reproduces similar
elliptical galaxy rotation properties, quantified by the degree of ordered
rotation, $\lambda_\textrm{R}$. There is a clear segregation between low-mass
($M_{\rm *} < 10^{11} M_{\rm \odot}$) ellipticals, which form a smooth
distribution of FRs, and high-mass galaxies ($M_{\rm *} > 10^{11.5} M_{\rm
  \odot}$), which are mostly SRs, in agreement with observations. We find that
SRs are very gas poor, metal rich and red in colour, while FRs are generally
more gas rich and still star forming. We suggest that ellipticals begin
naturally as FRs and, as they grow in mass, lose their spin and become
SRs. While at $z = 1$, the progenitors of SRs and FRs are nearly
indistinguishable, their merger and star formation histories differ
thereafter. We find that major mergers tend to disrupt galaxy spin, though in
rare cases can lead to a spin-up. No major difference is found between the
effects of gas-rich and gas-poor mergers and the amount of minor mergers seem
to have little correlation with galaxy spin. In between major mergers,
lower-mass ellipticals, which are mostly gas-rich, tend to recover their spin
by accreting gas and stars. For galaxies with $M_{\rm *}$ above $\sim 10^{11}
M_{\rm \odot}$, this trend reverses; galaxies only retain or steadily lose
their spin. More frequent mergers, accompanied by an inability to regain spin,
lead massive ellipticals to lose most of ordered rotation and
transition from FRs to SRs. 
\end{abstract}

\begin{keywords}
galaxies: elliptical and lenticular, cD -- galaxies: evolution -- galaxies: kinematics and dynamics -- galaxies: structure -- methods: numerical
\end{keywords}



\section{Introduction}
\label{sec:intro}

Observed elliptical galaxies may be classified into two groups with respect to their internal properties \citep{Davies:1983aa,Bender:1988aa,Bender:1990aa,Kormendy:1996aa,Kormendy:2012aa}. Low- and intermediate-mass elliptical galaxies typically have power-law surface brightness profiles \citep{Lauer:1995aa,Faber:1997aa}, show little or no radio and X-ray emission from hot gas \citep{Bender:1989aa}, tend to have discy isophotes \citep{Bender:1988ab,Bender:1989aa}, and exhibit significant rotation along the photometric major axis \citep{Emsellem:2007aa,Cappellari:2007aa}. On the other hand, massive ellipticals tend to have flat cores, hot gaseous haloes with strong radio and X-ray emission, and box-shaped isophotes. Kinematically they also differ strongly and experience slow rotation, exhibit kinematically decoupled components, and have a large amount of minor-axis rotation. The ATLAS$^\mathrm{3D}$ survey \citep{Cappellari:2011aa,Emsellem:2011aa,Krajnovic:2011aa}, which provides a comprehensive view of the properties of a volume-limited sample of local early-type galaxies, finds that only a minority of ellipticals falls into the second category. These have also been observed in the Virgo cluster \citep{Boselli:14}. A detailed summary of the properties of elliptical galaxies can be found in \citet{Kormendy:2009aa} and \citet{Kormendy:2016aa}.

The dichotomy of the physical properties between the two classes of elliptical galaxies led to the idea that they have different formation histories and form through different mechanisms. Early pre-$\Lambda$CDM studies \citep{Partridge:1967aa,Larson:1969aa} proposed that galaxy properties such as their isophotal shapes and rotation are determined by the angular momentum of infalling gas and the amount of turbulent viscosity. In this picture, slow rotators form in rapidly collapsing systems with efficient star formation and gas heating, while fast rotators form in more settled systems in which star formation and heating are inefficient. According to an alternative scenario, the `merger hypothesis' \citep{Toomre:1972aa,Toomre:1977aa}, elliptical galaxies form through the morphological transformation of disc galaxies in major mergers, and the properties of the remnants depend on the details of the mergers, such as the mass ratio of the galaxies, the amount of gas in the progenitor discs, and their orientation with respect to the orbital plane. This picture became very appealing at the advent of modern hierarchical cosmological models, in which mergers are an important component in the formation and evolution of dark matter haloes and galaxies \citep{White:1979aa,Fall:1980aa,Davis:1985aa}. Large samples of recent merger remnants are found in the local Universe \citep[e.g.][]{Schweizer:1982aa,Ellison:2013aa}, and a typical $M_*$ galaxy has experienced a major merger since $z\sim3$ \citep{Bridge:2007aa,Tasca:2014aa}.

Consequently, many studies have focused on numerical simulations of isolated, binary galaxy mergers and the properties of the merger remnants \citep{Gerhard:1981aa,Negroponte:1983aa,Barnes:1988aa,Hernquist:1993aa}. In recent years, this has lead to large libraries of merger simulations, considering both different initial conditions and different feedback schemes \citep[e.g.][]{Robertson:2006aa,Naab:2006aa,Cox:2006aa}. While collisionless simulations of major mergers between elliptical progenitors were shown to be in conflict with the observed properties of slow rotators \citep{White:1979aa,Bois:2010aa}, collisionless simulations of disc galaxies were more successful in reproducing elliptical-like properties. It was demonstrated that the mass ratio 
has a significant impact on the morphological and kinematic properties of the remnant: major merger simulations lead to slowly rotating, pressure supported, anisotropic remnants \citep{Negroponte:1983aa,Barnes:1988aa}, while more unequal mass ratios lead to more flattened faster-rotating ellipticals with more discy isophotes \citep{Barnes:1998aa,Naab:1999aa,Bendo:2000aa,Naab:2003aa}. However, remnants of gas-poor disc galaxies are in conflict with observations, as they do not reproduce the steep inner profiles. As according to Liouville's theorem, phase-space density is conserved during a collisionless process, the high central phase-space densities of elliptical galaxies cannot be produced from low phase-space density disc galaxies \citep{Carlberg:1986aa}. In dissipationless simulations this discrepancy could only be avoided by including large bulge components in the progenitors \citep{Hernquist:1993aa,Naab:2006aa}.

A different possibility to solve this problem is to take into account the gas
component in the progenitors. During a merger, the gas is torqued and loses
its angular momentum driving it to the central regions and increasing the
phase-space density \citep{Lake:1989aa}. The presence of gas thus leads to
more centrally concentrated remnants with rounder and less boxy centres
\citep{Barnes:1996aa,Bekki:1997aa,Hopkins:2008aa}. Using simulations of binary
major mergers, \citet{Cox:2006aa} confirmed that slowly rotating anisotropic
remnants can be formed when no gas is present. They also showed that if a cold
gaseous component is included in the progenitor discs, some merger orbits lead
to fast rotating remnants, which reproduce the observed distribution of
properties of ellipticals. However, the gas fractions that had to be used were
relatively high, and only a fraction of merger orbits produced fast rotators,
so that it remained unclear whether this scenario can lead to the large number
of observed fast-rotating ellipticals. Including a hot gaseous component in
the halo, \citet{Moster:2011aa} showed that fast rotators can be formed for
most orbits, even if the amount of cold gas in the progenitors is relatively
low. As gas subsequently cools from the hot halo and refuels the cold gas
disc, the initial amount of cold gas in the progenitors can be lower. In
massive haloes the hot gas is prevented from cooling by feedback processes,
such that more massive systems undergo gas-poor mergers and become slow
rotators, while low- to intermediate-mass systems undergo gas-rich mergers and
become fast rotators. On the other hand, some studies have argued that fast
rotators are created in minor mergers with varying mass ratio
\citep{Jesseit:2007aa,Jesseit:2009aa,Bois:2011aa}.

Although binary galaxy mergers have provided many insights into the formation
of fast and slow rotators, this approach has significant limitations. The
assembly histories of dark matter haloes and consequently galaxies are
considerably more complex than simple binary mergers or sequences
thereof. While early growth is dominated by the accretion of gas and
subsequent star formation, major and numerous minor mergers with a large range
of mass ratios play a significant role in their mass growth
\citep{DeLucia:2007aa,Moster:2013aa}. Moreover, \citet{Moster:2014aa} showed
that multiple mergers, where a second satellite galaxy enters the main halo
before the first satellite has merged with the central galaxy, are more common
than sequences of isolated binary mergers. It is therefore not possible to
simply string the results from binary merger simulations together, but events
with three or more galaxies involved have to be considered, which can only be
achieved with full cosmological simulations. Using high-resolution
cosmological zoom-in simulations,  such as \citet{Naab:2014aa} (hereafter N+14) and \citet{Choi:17} the formation histories of slow and fast rotators can be studied in a cosmological context. N+14 found that fast rotators are formed when the galaxies have
late assembly histories, and that both gas-rich major and minor merger
scenarios are common. However, they also identified fast-rotating merger
remnants that have formed in gas-poor major mergers of fast-rotating
progenitors. Slow rotators can also be formed in various scenarios: either by
major mergers (both gas-rich and gas-poor), or by gas-poor minor
mergers. While cosmological zoom-in simulations provide a much more detailed
view into the different formation mechanisms of slow and fast rotators, they
share a significant limitation with binary merger simulations. As the initial
conditions (i.e. the systems to be re-simulated) are chosen rather
arbitrarily, they do not form a representative sample of local early-type
galaxies. This can only be achieved with a volume-limited sample.

In the last years, the field of hydrodynamical simulations of galaxy formation
has made large advancements. Modern hydrodynamical codes produce very good
results in standard hydrodynamical tests such as fluid instabilities,
turbulence, and shocks
\citep{Teyssier:2002aa,Springel:2010aa, Sijacki:2012, Hu:2014aa,Bryan:2014aa,Hopkins:2015aa}. Moreover, state-of-the-art feedback methods are able to reduce the baryon conversion efficiency significantly, such that the stellar masses of the simulated galaxies are in good agreement with empirical constraints \citep{Moster:2010aa,Behroozi:2010aa,Guo:2010aa}. Several studies have now used these powerful codes to run hydrodynamical simulations of cosmological volumes producing galaxy populations that are in good agreement with many observational constraints, such as the stellar mass function, star formation rates, and galaxy sizes. Amongst the most detailed simulations are the Magneticum simulation \citep{Hirschmann:2014aa,Remus:2017aa}, the Illustris simulation \citep{Vogelsberger:2014aa,Genel:2014aa}, the Horizon-AGN simulation \citep{Dubois:2014aa,Welker:2014aa}, the MassiveBlack-II simulation \citep{Tenneti:2014aa,Khandai2015aa}, and the Eagle simulation \citep{Schaye:2015aa,Crain:2015aa}. As these simulations trace the formation of galaxies in a cosmological context for a representative sample of galaxies they are ideally suited to study morphological and kinematical properties in a statistical manner.

The aim of this paper is twofold: firstly we analyse the morphological and kinematical properties of galaxies in the Illustris simulation and compare them to the ATLAS$^\mathrm{3D}$ observations to judge if the simulated galaxies are a good representation of observed slow and fast rotators. Secondly we trace the simulated galaxies through cosmic time and investigate which formation channels are the most important ones in the formation of slow- and fast-rotating ellipticals. 

The paper is organized as follows. In Section~2 we provide a brief summary of the Illustris simulation. We also explain how we analyse the simulated galaxies and the mergers. In Section~3 we present our results for the slow and fast rotators at $z=0$ and compare them to the observed sample, including their kinematic properties, the central profiles, the isophotal shapes, and the X-Ray luminosities. We further present the dependence of the kinematic properties on various galaxy properties, such as their stellar mass, gas fraction, star formation rate, metallicity, size, and colour. In Section~4, we investigate the merger histories of slow and fast rotators, and identify which channels are the most important ones in their formation. Finally, in Section~5 we summarise and discuss our results and compare them to previous studies. Throughout this paper, we assume a 9-year Wilkinson Microwave Anisotropy Probe \citep[WMAP9;][]{Hinshaw:2013aa} $\Lambda$CDM cosmology with $\Omega_\mathrm{m}=0.2726$, $\Omega_\mathrm{\Lambda}=0.7274$, $\Omega_\mathrm{b}=0.0456$, $h=0.704$, $n=0.963$ and $\sigma_\mathrm{8}=0.809$, and we employ a \citet{Chabrier:2003aa} initial mass function (IMF).

\section{Methods}

\subsection{The Illustris Simulation}
Starting with a box of $106.5$~Mpc (comoving) on a side, the Illustris simulation tracks
dark matter (DM) and baryonic matter in a standard $\Lambda$CDM cosmology consistent with the Wilkinson Microwave Anisotropy Probe 9-year data release \citep{Hinshaw:2013aa}. The simulation is performed using the moving mesh code AREPO \citep{Springel:2010aa} solving hydrodynamics in a quasi-Lagrangian way, taking advantage of the Voronoi tessellation. For this work we will focus solely on the largest and best resolved of a suite
of simulations, Illustris-1, from now on just referred to as Illustris,
following $1820^3$ DM particles and approximately as many gas elements, with
masses and gravitational softenings of $m_{\rm DM} = 6.26 \times 10^6 \, {\rm
  M_{\odot}}$, $m_{\rm gas} = 1.26 \times 10^6 \, {\rm M_\odot}$, $\epsilon_{\rm DM} = 1.42 \,{\rm kpc}$ and $\epsilon_{\rm gas} = 0.71 \,{\rm kpc}$, respectively. For further details of the simulations see \citet{Vogelsberger:2014aa} and \citet{Genel:2014aa}.

In addition to gravity and hydrodynamics in
an expanding universe with a uniform ionising background, a suite of sub-grid
models are used that are crucial to galaxy formation and evolution. Radiative
heating and cooling processes including both primordial and metal-line cooling
are incorporated and high density gas can cool to form star particles. These
have their own associated 
IMF, return mass and metals in accord with the stellar evolutionary tracks 
and produce corresponding feedback through supernovae, leading to
galactic-scale winds. A simple recipe for the formation of massive seed black
holes is adopted, motivated by the direct collapse scenario. Black holes then can grow through accretion and mergers, and AGN feedback
 including quasar, radio and radiative modes are modelled \citep{Sijacki:2015aa}. The free parameters
 of these sub-grid models are tuned to reproduce the $z = 0$ stellar mass
 function, cosmic star formation rate (SFR) evolution and the mass-metallicity
 relation. \cite{2013MNRAS.436.3031V} presents full details of the methods
 employed and how the free parameters are set.  

A total of 136 snapshots are taken at particular redshifts, spaced for the
latter part of the simulation since $z = 3$ by $\Delta a \approx 0.01$, where
$a$ is the cosmological scale factor. Together with the data stored in the
snapshots, the on-the-fly Friends-of-Friends (FOF) and SUBFIND algorithms
\citep{2001NewA....6...79S,2009MNRAS.399..497D} provide catalogues of
virialized dark matter halos and their bound subhalos, including a number of
their properties. For further details see the public release paper
\citep{2015A&C....13...12N}.

At $z = 0$ there are 4,366,546 gravitationally bound subhalos. 309,166 of these
subhalos contain star particles, and henceforth we call these galaxies. We
define, in Section~\ref{sec:spirals}, the limit of a well resolvable galaxy to
be one with over 20,000 star particles and at $z = 0$ we find 4,591 qualifying
galaxies in the Illustris simulation, including satellite galaxies as well as
centrals. Taking advantage of the merger trees constructed following \citet{2015MNRAS.449...49R} we can trace back in time the identity and properties of the progenitor galaxies, and by combining this information with subhalo and particle catalogues analyse the nature and effect of galaxy mergers and the evolution of galactic properties over cosmic time.


\begin{figure*}
\begin{center}
\includegraphics[width=0.9\textwidth]{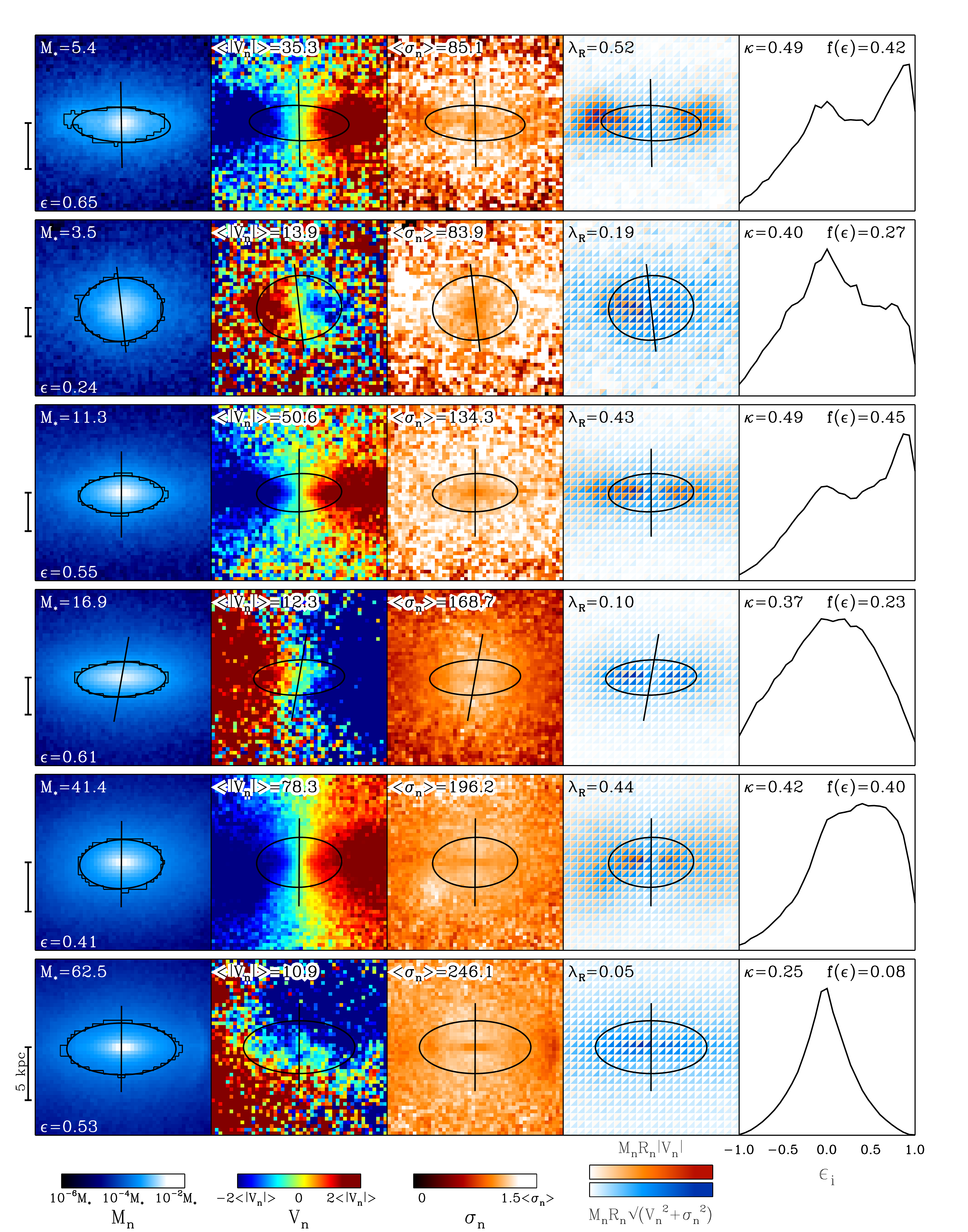}
\caption{Projected properties of 6 illustrative elliptical galaxies at $z =
  0$, showing in 
  descending order, a low mass FR and SR, an intermediate mass FR and SR and
  finally a high mass FR and SR. From left to right plots show projected
  density, line-of-sight velocity, line-of-sight velocity dispersion,
  contribution of pixels to ordered vs. disordered components of $\lambda_\mathrm{R}$, and finally the distribution of the
  circularity parameter (as detailed in Appendix~\ref{ap:Spirals}). Each galaxy is projected
  along the medium axis of the inertia tensor, and plotted with the minor axis
  in the vertical direction. In the first column the contour of pixels
  containing half the galaxy's mass is shown, to which the ellipse is then
  fitted. The black line on each galaxy shows the projected direction of the
  total angular momentum. Scale bars are all $5\, {\rm kpc}$ long, $M_*$ is
  the total 
  stellar mass in units of $10^{10} \, {\rm M_\odot}$, $\langle v \rangle$ and
  $\langle \sigma \rangle$
  are mass-weighted line-of-sight velocity and velocity dispersion,
  respectively (in ${\rm km \, s}^{-1}$). $\lambda_{\rm R}$ is the degree of
  ordered rotation and $\epsilon$ is the projected ellipticity. $\kappa$ is
  the fraction of kinetic energy devoted to circular orbits in the plane and
  $f(\epsilon)$ is the fraction of stars with a disk-like circularity
  parameter. The mass and spin histories of these six galaxies are shown and
  discussed in Figure~\ref{fig:FullHist_j}.} 
\label{fig:GalPlots}
\end{center}
\end{figure*}

\subsection{Modelling Galaxies via Stellar Kinematics}
As we will be comparing to observational data all our kinematic analysis is based almost solely on the motions of star particles, ignoring black holes, dark matter and gas. We also work directly with the mass of each star particle, which means that where we compare to observational measures that examine a galaxy's light profile we make the implicit assumption of a constant mass-to-light ratio.

To find the centre of the galaxy we take the position of the most bound star particle. We determine the galaxy's velocity via the centre of mass velocity of the most bound 50\% of particles; this avoids substructure at large radii with high velocities biasing the result. We also exclude stars outside of $50\,{\rm kpc}$
(comoving) of the centre, which aids in the exclusion of satellites at large
radii, particularly a problem for the highest mass galaxies. In the rest frame centred on the galaxy, each star particle denoted with index
$i$ has a position $\mathbf{r}_i$, velocity $\mathbf{v}_i$ and mass $m_i$. We
hence calculate the specific angular momentum of each star particle $\mathbf{j}_i$ and
to find the plane of the galaxy we use the total angular momentum
of the most bound 50\% of particles, $\hat{\mathbf{J}}$.

\subsection{Classifying Spirals and Ellipticals}
\label{sec:spirals}

We limit our sample only to elliptical galaxies, and hence must cut any spirals from our sample. This is easy enough to do via visual examination when having a small
sample, but it is unfeasible for such a large sample as we have from
Illustris. Instead we use the method outlined in \citet{Sales21062012}
(hereafter S+12) to compute the $\kappa$ parameter, which measures the
fraction of kinetic energy invested in ordered rotation, i.e.

\begin{equation}
\begin{centering}
\label{kappa}
\kappa = \dfrac{K_{\rm rot}}{K} \,,\,\,\,\,\, K_{\rm rot}= \sum \dfrac{1}{2}  m_i \left( \dfrac{j_{z,i}}{R_i} \right)^2\,,
\end{centering}
\end{equation} where $K_{\rm rot}$ is the total kinetic energy invested in ordered rotation, dependent on the particle's mass, cylindrical radius $R_i=(r_i^2-(\mathbf{r}_i\cdot\hat{\mathbf{J}})^2)^{1/2}$ and specific angular momentum perpendicular to the disk $j_{z,i}=\mathbf{j}_i\cdot\hat{\mathbf{J}}$. $K$ is the total kinetic energy of the galaxy, and both measures of kinetic energy are calculated over all star particles.

For a perfectly rotating, completely disk dominated galaxy $\kappa$ will tend
to $1$, whereas for a bulge dominated or elliptical galaxy $\kappa$ will tend
to $1/3$. We define the cut-off between elliptical galaxies and spiral
galaxies to be $\kappa = 0.5$, which gives a realistic ratio of galaxies at
high masses. These cuts are in good agreement with observations \citep{Conselice21122006} for all large galaxies. However for galaxies with $M_* \lessapprox 10^{10.5} M_\odot$ we start to see unexpectedly high fractions of ellipticals. We present these results, as well as a discussion of other possible methods to seperate the populations, in Appendix~\ref{ap:Spirals}.

Similar results were seen in Illustris by
\citet{2015arXiv150207747S} and \citet{Gomez:2014aa}, where low mass galaxies
are bulge dominated by kinematic measures, but have SFRs and disc properties
characteristic of spiral galaxies. The conclusion which we draw from this is
that low mass galaxies in Illustris, 
with less than $\sim 10^4$ star particles, cannot be satisfactorily examined
on a kinematic basis. This could be due to deficiencies in the feedback
mechanisms which are needed to create an extended disk in simulation, and/or
the poor resolution of these low mass galaxies. We thus draw a line between
galaxies which can reasonably be resolved and kinematically classified and
those which cannot. From now onwards we will restrict our analysis to galaxies
with over $20,000$ star particles, roughly corresponding to a stellar mass of
$10^{10.5} \, {\rm M_\odot}$.

\subsection{Examining Galaxies in Projection}
\label{sec:projection}

To compare with observational data we must project our 3-dimensional galaxy
data onto a two dimensional image seen from a particular line-of-sight. We can
choose an arbitrary, fixed, line-of-sight throughout and thus our results will
be statistically comparable to observable data, or we can rotate each galaxy
such that we observe it edge-on allowing us to take intrinsic values of shape
and motion. We will do both in the course of this analysis to examine the
effects of projection on our data set. 

When using an arbitrary line-of-sight we choose to look directly along the
z-axis, such that line-of-sight velocity $V_i=v_{z,i}$ and two-dimensional
projected position $\mathbf{R}_i = r_{x,i} \hat{\mathbf{i}} + r_{y,i}
\hat{\mathbf{j}}$. To gauge the intrinsic properties of galaxies we use the inertia tensor
\citep[see e.g.][]{2013MNRAS.431..477J},
\begin{equation}
I_{\mu,\nu} = \dfrac{\sum m_i r_{\mu,i} r_{\nu,i}}{\sum m_i}\,,
\end{equation}
(where the sum is performed over the 50\% most bound star particles) to project the galaxy so as to extremise its
ellipticities. From the eigenvectors of $\mathbf{I}$, $\mathbf{e}_1$,
$\mathbf{e}_2$ and $\mathbf{e}_3$ (corresponding to the long, medium
and short axes of the approximated ellipsoid, respectively) we find the projected line-of-sight velocities $V_i=\mathbf{v}_i \cdot \mathbf{e}_2$ and two-dimensional projected positions $\mathbf{R}_i = (\mathbf{e}_1 \cdot \mathbf{r}_i) \hat{\mathbf{i}} + (\mathbf{e}_3 \cdot \mathbf{r}_i) \hat{\mathbf{j}}$.

We find the projected circular stellar half-mass radius, $r_{\rm h}$ then make a
grid of $48$ by $48$ square pixels, scaled initially to make an image $4
\,r_{\rm h}$ wide, centred on the galaxy. The grid can be dynamically resized to ensure that all relevant data is included at the highest possible resolution, though the factor of $4$ is chosen to ensure this is rarely necessary.

Using the projected positions of each star particle, $\mathbf{R}_i$ we create
20 pseudo particles, each with $\frac{1}{20}$th the original mass, and
randomly distribute them around the original projected position with
displacements drawn from a two-dimensional Gaussian with a standard deviation
of $0.3 \,{\rm kpc}$, from which we determine which, if any, pixel they reside
in. Each pixel, which we label $n$ of $N$ ($=\,48 \times 48 \,=\, 2304$), now
contains $I_{\rm n}$ star pseudo-particles, which in turn we index using
$i_{\rm n}$,
corresponding to the subset of indices of star particles $i$ contained in
pixel $n$. We then find the total mass of stars projected in each pixel
$M_{\rm n}$,
their centre of mass line-of-sight velocity $V_{\rm n}$ and the mass-weighted
dispersion in line-of-sight velocity, $\sigma_{\rm n}$, computed relative to
$V_{\rm n}$. To further analyse the galaxy we then examine a contour of pixels which
contains half the galaxy's mass. It is from this subset of pixels that we
analyse the rotation properties of the galaxy, and from the outline of this contour that we measure the galaxy's shape.

In finding this contour we use a novel approach. Starting from a pixel at the
centre of our image we sequentially add the next heaviest pixel adjacent to
any of the pixels already included. We continue to add pixels in this manner,
filling any internal gaps when they arise, until the total mass of the pixels
contained has exceeded half the total mass of the galaxy. This method creates
a single connected contour that well represents the shape of the galaxy's iso-density contour. We
then fit an ellipse to the pixels which lie on the edge of this contour to
find the semi-major and semi-minor axes, $a$ and $b$, respectively and thus
the ellipticity of the galaxy, $\epsilon$.

We experimented with an alternative expression for ellipticity, as presented in \citet{2007MNRAS.379..401E}, which takes account of the internal structure of the galaxy as well as the external contour 
\begin{equation}
    \label{ellip2}
    \epsilon' = 1 - \sqrt{\dfrac{\langle x^2 \rangle}{\langle y^2 \rangle}}\,,
\end{equation}
where $x$ and $y$ correspond to the components of the projected position
vector $\mathbf{R}_i$ and the angular brackets denote the mass-weighted mean
of the pixels included in the contour. This was previously used as a
flux-weighted mean but with our assumption of constant mass to light ratio the
two are equivalent. 

We found both measures to be in very good agreement, suggesting that the shape of the external contour well represents the internal structure and we have chosen to classify the galaxies in our sample using $\epsilon$ rather than $\epsilon'$ as this quantity is more readily available to observers.

Finally we use our projection to find $\lambda_\mathrm{R}$, the degree of ordered rotation. We have
\begin{equation}
\label{lambda}
    \lambda_\mathrm{R} = \dfrac{\langle R\vert V\vert \rangle}{\langle R \sqrt{V^2 + \sigma^2}\rangle}\,,
\end{equation}
where $\langle \, \rangle$ refer to a mass-weighted mean.

For perfectly ordered rotation we expect the velocities of particles to be much greater than the local velocity dispersion and hence $\lambda_\mathrm{R}$ will tend to $1$. In practice values higher than $\sim 0.8$ are very rare in elliptical galaxies, perhaps as their past mergers have flung some stars into non-uniform orbits. For the most perturbed galaxies we expect there to be no clear direction of motion for each pixel and hence the projected velocity dispersion should hugely outweigh any ordered rotation, giving values of $\lambda_\mathrm{R}$ tending to zero.

We applied each of the above analyses to every galaxy with over $20,000$ star
particles at each snapshot from $z=4$ to $z=0$, such that we could track the
type, shape and degree of rotation of each galaxy back up to $12$ billion
years ago. Figure~\ref{fig:GalPlots} shows an example of our methods for 6 illustrative
elliptical galaxies at $z = 0$. There are three pairs of fast and slow
rotators (cutoff criteria are explained in Section~\ref{sec:Properties}) at
low mass (top, $M_*/M_{\rm \odot} < 10^{11}$), intermediate mass (middle,
$10^{11} < M_*/M_{\rm \odot} < 10^{11.5}$) and high mass (bottom, $10^{11.5} <
M_*/M_{\rm \odot}$), as indicated in the first column. All are relatively
elongated and of similar half-mass 
radii. As shown on the panels in the second and third columns, the fast
rotators show a clear velocity signature, with rotation well 
aligned from small to large radii, whilst slow rotators are much more
disordered with generally lower speeds and comparable or higher velocity
dispersions. In the fourth column we show a visualisation of the relative
contribution of different parts of the galaxy to the numerator and denominator
of equation (\ref{lambda}) (normalised by the maximum pixel value of either
for the galaxy). These show that there is similar structure in the measure of
disorder of all galaxies (the denominator) but that fast rotators have a much
larger degree of ordered rotation (numerator) corresponding to disk-like
rotation and that this continues out to large radii. Finally, in the fifth
column we show the disk--bulge comparison and can see that FRs have
significant disk-like components (though not dominant) although towards higher
mass the disk and the bulge become difficult to separate.

\begin{figure}
\begin{center}
\includegraphics[width=0.95\columnwidth]{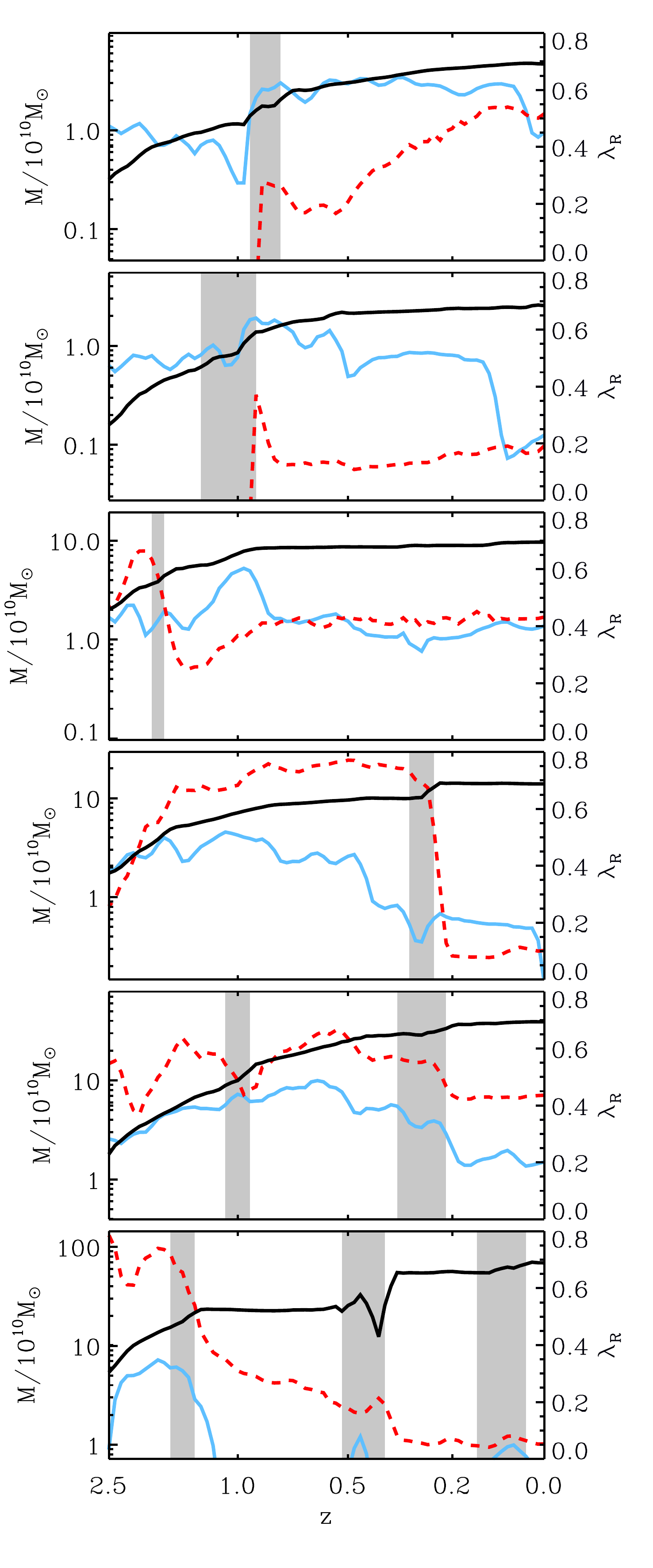}
\caption{Evolution of the six galaxies shown in Figure~\ref{fig:GalPlots}
  since $z = 2.5$. The solid black line shows the stellar mass, $M_*$, and
  solid blue line the gas mass, $M_{\rm g}$, the scale for both is shown on
  the left. The dashed red line shows the degree of ordered rotation,
  $\lambda_\mathrm{R}$, the scale for which is shown on the right. The grey
  regions are the period in which the galaxy is undergoing a major merger,
  starting at the point of maximum stellar mass of the incoming body (roughly
  the start of the encounter). From top to bottom low mass FR and SR, an
  intermediate mass FR and SR and a high mass FR and SR are shown.}
\label{fig:FullHist_j}
\end{center}
\end{figure}

\begin{figure*}
\begin{center}
\includegraphics[width=0.95\textwidth]{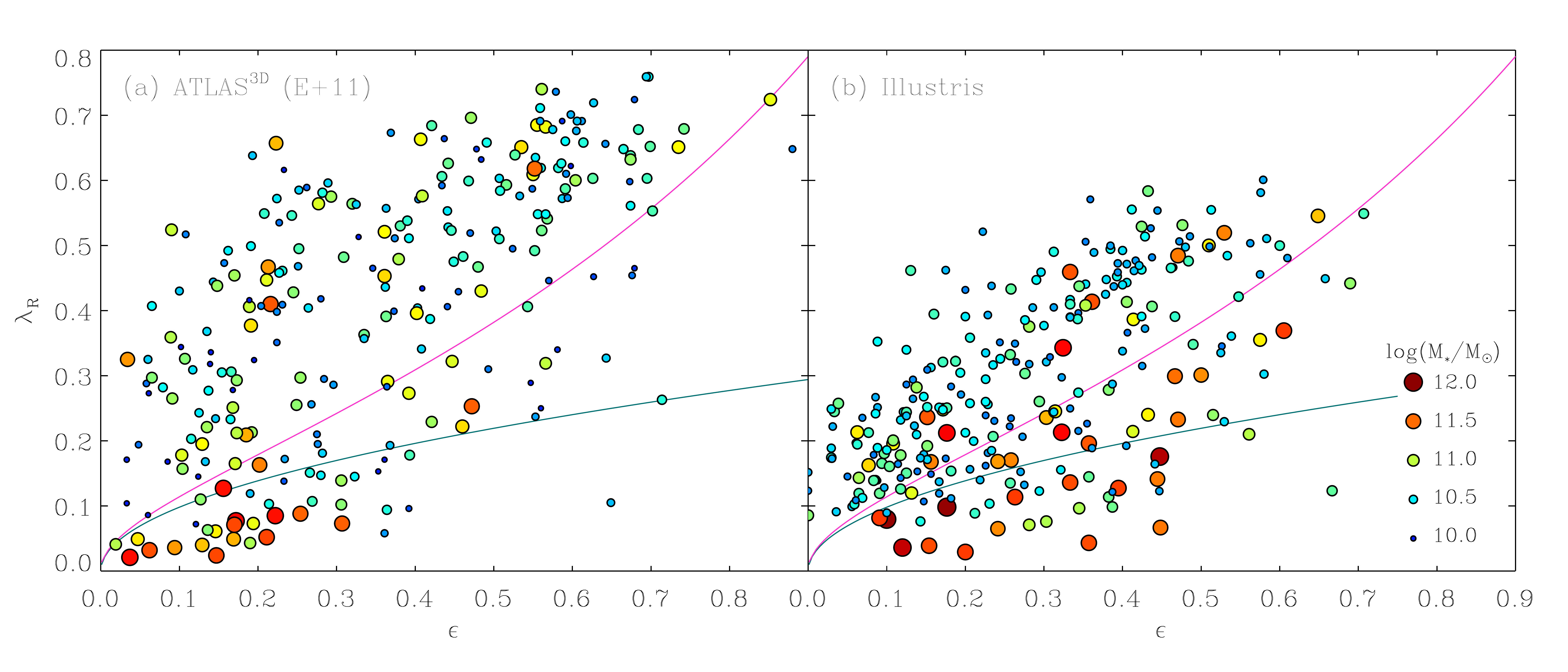}
\caption{Distribution of ETGs in the $\epsilon$ - $\lambda_{\rm r}$ plane, (a)
  reproduced from the $\text{ATLAS}^{\text{3D}}$ survey (E+11) compared to (b)
  those created in the Illustris simulation, sampled based on the same
  selection criteria (see text for more details). The magenta line shows the line on which edge-on elliptical galaxies with the same degree of rotational support would lie, as described in text and re-derived in Appendix~\ref{ap:Magenta}. The green line follows equation (\ref{cutoff}), defining the cutoff below which galaxies are classified as SRs. Each galaxy is represented by a circle whose colour and size scale with the stellar mass.}
\label{fig:LambdaEms}
\end{center}
\end{figure*}

\subsection{Following the merger histories of massive galaxies}

We take advantage of the galaxy merger trees to look at the accretion
histories of the galaxies in our sample, examining how mergers influence
galaxies with different intrinsic properties at $z = 0$. The merger trees are
generated using the SubLink algorithm, as detailed in
\citet{2015MNRAS.449...49R}. We combine the calculated values of
$\lambda_\mathrm{R}$, $\epsilon$, $\kappa$ and $j$ with the data stored in the
SubLink trees for stellar mass $M_*$, and gas mass $M_\mathrm{g}$, within two
stellar half-mass radii. We follow these properties back in time along the main progenitor branch of each galaxy at $z = 0$.

Following the prescription of \citet{2015MNRAS.449...49R} for each subhalo
that merges with the main progenitor at a particular redshift we trace that
subhalo back to the point at which its stellar mass is maximum, approximately
the point at which the galaxies begin tidally interacting, and we record its
properties. We classify major mergers as
those for which the stellar mass of the merging subhalo, $M_\mathrm{*,i}$
satisfies $M_\mathrm{*,i} > \frac{1}{4} M_*$ and all other mergers are
classified as minor mergers. Thus we calculate the total stellar mass accreted
at each snapshot through major mergers, $\Delta M_\mathrm{*,maj}$, and through
minor mergers, $\Delta M_\mathrm{*,min}$, and similarly for the gas mass
accreted through major, $\Delta M_\mathrm{g,maj}$, and minor, $\Delta
M_\mathrm{g,min}$, mergers.  
 
We find a rough estimate for the change of stellar mass due to in-situ star
formation at some snapshot $n$ as $\Delta M_\mathrm{*,situ}(n) = M_{*}(n) -
M_{*}(n-1) - \Delta M_\mathrm{*,maj}(n) - \Delta M_\mathrm{*,min}(n)$,
i.e. any stellar mass change not accounted for by mergers is assumed to stem
from in-situ star formation. However, as stellar mass is transferred between galaxies throughout a merger, and more galaxies may join the interaction during the infall period, this simple treatment is unlikely to perfectly capture the mass changes(e.g. see \citet{Rodriguez-Gomez:2016aa} for more details). Also, this formulation assumes no mass loss through stripping or stellar evolution and hence in practice may underestimate the in-situ star formation for galaxies with violent collisions which fling out a significant mass fraction (and negative values of $\Delta M_\mathrm{*,situ}(n)$ are thus possible).

Figure~\ref{fig:FullHist_j} shows the evolution since $z=2.5$ of the 6
galaxies from Figure~\ref{fig:GalPlots}. There are several interesting
features to note. Both of the lower mass galaxies only have
$\lambda_\mathrm{R}$ defined for the latter part of this period, because
before that point they are not categorised as well resolved ellipticals. All
galaxies grow in mass by about an order of magnitude in this period (solid
black curves), whilst
their gas content stays roughly constant or reduces (solid blue
curves). Major mergers (indicated with gray bands) are associated with 
large changes in their 
rotation (dashed red curves) and also their angular momentum (not shown here
but well correlated with $\lambda_\mathrm{R}$), but substantial changes can
also occur over periods without major mergers. Section~\ref{sec:History}
details analysis of the histories of the whole population of thousands of
well-resolved ellipticals in Illustris.

\begin{figure*}
\begin{center}
\includegraphics[width=0.95\textwidth]{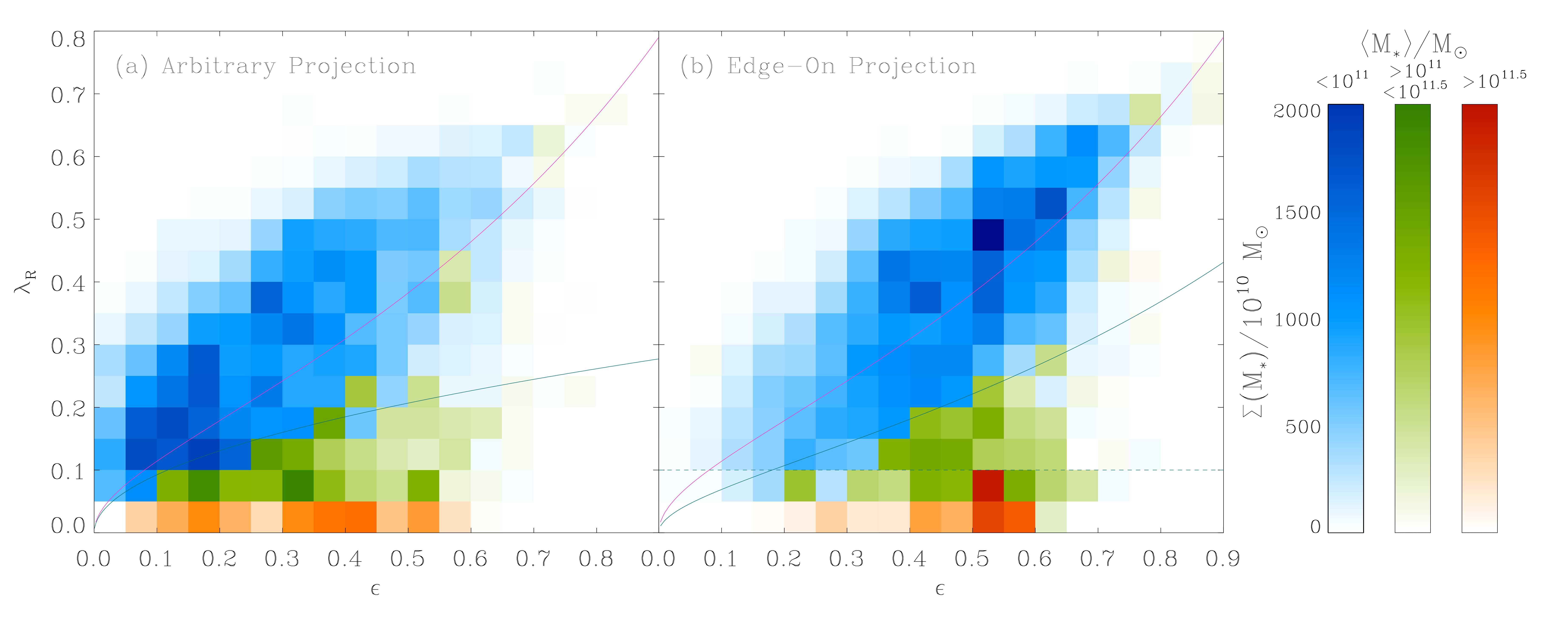}
\caption{Left panel (a): Properties of galaxies when viewed from an
  arbitrary projection angle, shown for every elliptical galaxy between
  $z = 0.1$ and $z = 0$. Right panel (b): Similar figure for edge-on
  projections of each galaxy between $z = 0.1$ and $z = 0$. The colour of pixels is
  defined by the average mass of the galaxies contained and its colour
  saturation by the total mass of the galaxies contained in that pixel
  (normalised compared to highest total mass pixel). The solid green lines
  refer to the FR-SR cutoff for both projections. The dotted green line shows
  the previously suggested cutoff of a fixed value of $\lambda_{\rm R}$ (see
  text for more details).}
\label{fig:Lambda_z0}
\end{center}
\end{figure*}

\section{Populations of Fast and Slow Rotators}
\label{sec:Properties}
\subsection{Comparing to the $\text{ATLAS}^{\text{3D}}$ survey}
\label{sec:ATLAS3D}
The first question we seek to answer is whether the Illustris simulation
produces a representative sample of fast and slow rotators. The
$\textrm{ATLAS}^{\textrm{3D}}$ survey 
provides an excellent set of observational data to compare to, \citet{Emsellem:2011aa} (hereafter E+11), examining the
distribution of ellipticity and $\lambda_\textrm{R}$ for 260 early type
galaxies (ETGs) in a volume- and magnitude-limited sample. We reproduce their
results in the left-hand panel of Figure~\ref{fig:LambdaEms}.

E+11 selected ETGs based on whether they could clearly discern spiral arm
structures in the galaxy. Taking into account the variation due to projection
effects in observed galaxy properties, E+11 suggested making a divide between
FRs and SRs based on their $\epsilon$ and $\lambda_{\rm R}$ properties using
the curve
\begin{equation}
\label{cutoff}
    \lambda_{R,{\rm cutoff}} = 0.31 \sqrt{\epsilon}\,,
\end{equation}
above which galaxies are classified as FRs and below as SRs. Of their sample
$224$ galaxies, or $86\pm2\%$, were classified as FRs and the remaining $36$,
$14\pm2\%$, as SRs, using (\ref{cutoff}) to divide the population.

They found FRs to be roughly oblate, with the ellipticity of a galaxy
correlating linearly with its anisotropy, $\beta$. In other words, the FRs are roughly
rotationally supported with stars on mostly circular orbits. Thus the motions
of stars and the projected shape of the galaxy has a simple, linear
correlation. SRs are instead galaxies with low degrees of rotational support,
with many stars on radial or generally irregular orbits. Thus we can identify
galaxies whose shape is well represented by the assumption of rotational
support, those with ellipticity ($\epsilon$) proportional to their degree of
velocity anisotropy ($\delta$). \citet{Cappellari:2007aa} and E+11 find an
approximate expression for the variation of $\lambda_\mathrm{R}$ with
$\epsilon$ for galaxies with $\delta \propto \epsilon$. We detail this
expression in Appendix~\ref{ap:Magenta}, and follow their lead in showing the
relationship for $\delta = 0.7 \epsilon$ as a magenta line in the left-hand
panel of Figure~\ref{fig:LambdaEms}.

The E+11 sample of galaxies is volume- and magnitude-limited to those galaxies
within $42\,{\rm Mpc}$, brighter than $M_{\rm K_s} = -21.5$~mag, which are
clearly visible from the William Herschel telescope (declination
$|\delta-29^{\circ}|<35^{\circ}$) and which are far from the Galactic plane to
avoid contamination (galactic 
latitude $|b|>15^{\circ}$). Using the same criteria (arbitrarily setting the
equatorial 
direction aligned with the positive z axis, and the galactic plane relative to
that) we took a sample of galaxies from Illustris, shown in
the right-hand panel of Figure~\ref{fig:LambdaEms}, finding 270 galaxies. Of
this sample $38$, or $14\%$ are SRs and $86\%$ FRs. 

The number of galaxies and fraction of FRs and SRs is in very good agreement
between the $\text{ATLAS}^{\text{3D}}$ survey and the Illustris
simulation. We also find that the majority of FRs are lower mass galaxies
whilst the highest mass galaxies are mostly SRs, as found in observations as
well. However, there are fewer high $\lambda_\mathrm{R}$ galaxies than the
$\text{ATLAS}^{\text{3D}}$. This may be a selection effect, as in the $\text{ATLAS}^{\text{3D}}$ survey ETGs are selected by excluding any galaxies with visible spiral arms. Thus it is possible some fast spinning disk-like object, with no clearly visible spiral arms, are included. Our elliptical galaxies are selected by comparing the fractional energy in the disk and the bulge, thus excluding the fastest rotating and most disk-like structures. It could also suggest, as we discuss in Section~\ref{sec:spirals}, there may be limitations to the simulation at producing extended rotationally supported low mass objects.

Simulated high mass SRs show higher
ellipticities than we might expect. We believe this may be due to the specific
sub-grid physics choices of the Illustris simulation, particularly due to the role of gas in major mergers. It has been shown that the presence of a high gas fraction can lead to a more rounded slow rotating galaxy, as in \citet{Jesseit:2007aa}, J+09 and N+14, and that the success in reproducing observed kinematics from major mergers is strongly dependent on the gas component for SRs, though FRs are not as affected \citep{Bois:2010aa}.

This tendency to slower spinning, more rounded observed galaxies shows the limitation of the simulation's ability to capture the full detail of
these galaxies, but the very good overall qualitative and quantitative
agreement leads us to conclude that the Illustris galaxies comprise a
sufficiently representative sample of fast and slow rotating galaxies from
which to base the analysis that will follow.

\begin{figure*}
\begin{center}
\includegraphics[width=\textwidth]{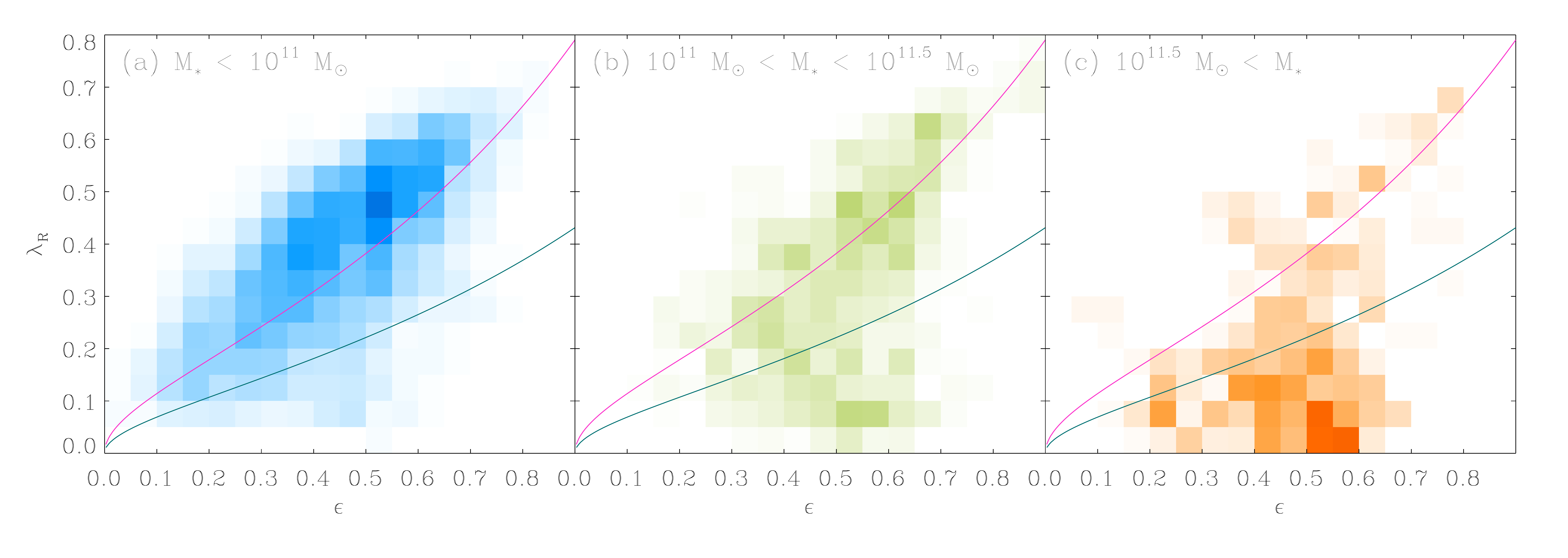}
\vspace{-0.5cm}
\caption{Intrinsic $\epsilon$ - $\lambda_{\rm R}$ plots as in previous
  figures, now 
  separated by galaxy mass. From left to right low mass ($<10^{11}\,
  \mathrm{M_\odot}$), intermediate mass (between $10^{11}\, \mathrm{M_\odot}$
  and $10^{11.5}\, \mathrm{M_\odot}$) and high mass ($>
  10^{11.5} \,\mathrm{M_\odot}$) galaxies are plotted. Note the clear
  distinction between the regions occupied by low and high mass galaxies,
  which could explain the observed distinct populations of fast and slow
  rotating galaxies.}  
\label{fig:Lambda_z0_Mass}
\end{center}
\end{figure*}

\subsection{Simulated FR and SR populations for a larger sample}

By including the whole population of well resolved elliptical galaxies and
considering multiple snapshots, we create a large sample of galaxies from
which distributions of galaxy properties can be derived. Starting from our
arbitrary line-of-sight we analyse $\lambda_{\rm R}$ and
$\epsilon$ for 15,774 elliptical galaxies with over 20,000 star particles
each. Of these $86.8\%$ are classified as FRs and $13.2\%$ as
SRs. 

For this large population sample between $z = 0.1$ and $z = 0$, the left-hand panel
of Figure~\ref{fig:Lambda_z0}, shows similar
characteristics to the subsets shown in Figure~\ref{fig:LambdaEms}, with a
clear segregation based on the mass of a galaxy. More
massive galaxies tend to have low values of $\lambda_\mathrm{R}$ and are
classified as SRs, while 
the lower mass galaxies form the bulk of FRs, the majority of which
reside above and to the left of the magenta line, showing significant
rotational support. Note that there is a particularly high 
number of FRs concentrated around $(\epsilon,\lambda_\mathrm{R}) \approx
(0.1,0.15)$. Intermediate mass galaxies with $10^{11}\,{\rm M_{\rm \odot}} \le
M_{*} \le 10^{11.5} \,{\rm M_{\rm \odot}}$ occupy the region in between and are
classified as a mixture of FRs and SRs.

We also present results for galaxies projected edge-on, allowing us to gauge
their intrinsic properties, as shown in the right-hand panel of
Figure~\ref{fig:Lambda_z0}. Of 15,774 galaxies, $95.7\%$ are FRs and $4.3\%$
are SRs, based on the previously suggested cutoff for the intrinsic spin of
galaxies of $\lambda_{\rm R} = 0.1$ (J+09). We suggest an improvement to this simplistic cutoff between FRs and SRs. We use a linear scaling of the line defining galaxies with a constant degree of rotational support, with a constant of proportionality chosen to recover similar fractions of SRs and FRs as found for an arbitrary projection angle. We thus use the approximate relation shown as the magenta line in Figure~\ref{fig:LambdaEms} (the form of which is reproduced in Appendix~\ref{ap:Magenta}), i.e. 
\begin{equation}
\label{newcutoff}
\lambda_{\rm R,cutoff} = \alpha \lambda_\mathrm{R}\,\,\mathrm{for}\,\,\delta = 0.7 \epsilon 
\end{equation}
for some constant $\alpha$. For $\alpha=0.65$ we find $86.6\%$ FRs and $13.4\%$ SRs. We also find good agreement between whether any
individual galaxy is classified as an SR in an arbitrary projection and an
edge on projection, with over $90\%$ overlap for more massive galaxies
($\sim 75\%$ for low mass). Comparing the fraction of FRs predicted by this cutoff
with those found for galaxies projected along an arbitrary line-of-sight we
see very good agreement. It should be noted that the
chosen value of $\alpha=0.65$ is left purposefully imprecise as the cutoff
line is not strongly physically motivated and thus is at best a useful rough approximation. Fine-tuning of $\alpha$ can give an even tighter correlation but also a misleading impression of the precision of the cutoff criteria.

Examining the two panels in Figure~\ref{fig:Lambda_z0} we see that many
galaxies have shifted towards higher $\epsilon$ and $\lambda_{\rm R}$, in
agreement 
with the expectation that these values are maximal when viewed edge-on. We
see, as before, a clear distinction between the majority of FRs being lower mass
galaxies and the majority of SRs being intermediate and higher mass galaxies,
with the most massive galaxies dominating at the lowest values of
$\lambda_{\rm R}$.

\begin{figure*}
\begin{center}
\includegraphics[width=0.8\textwidth]{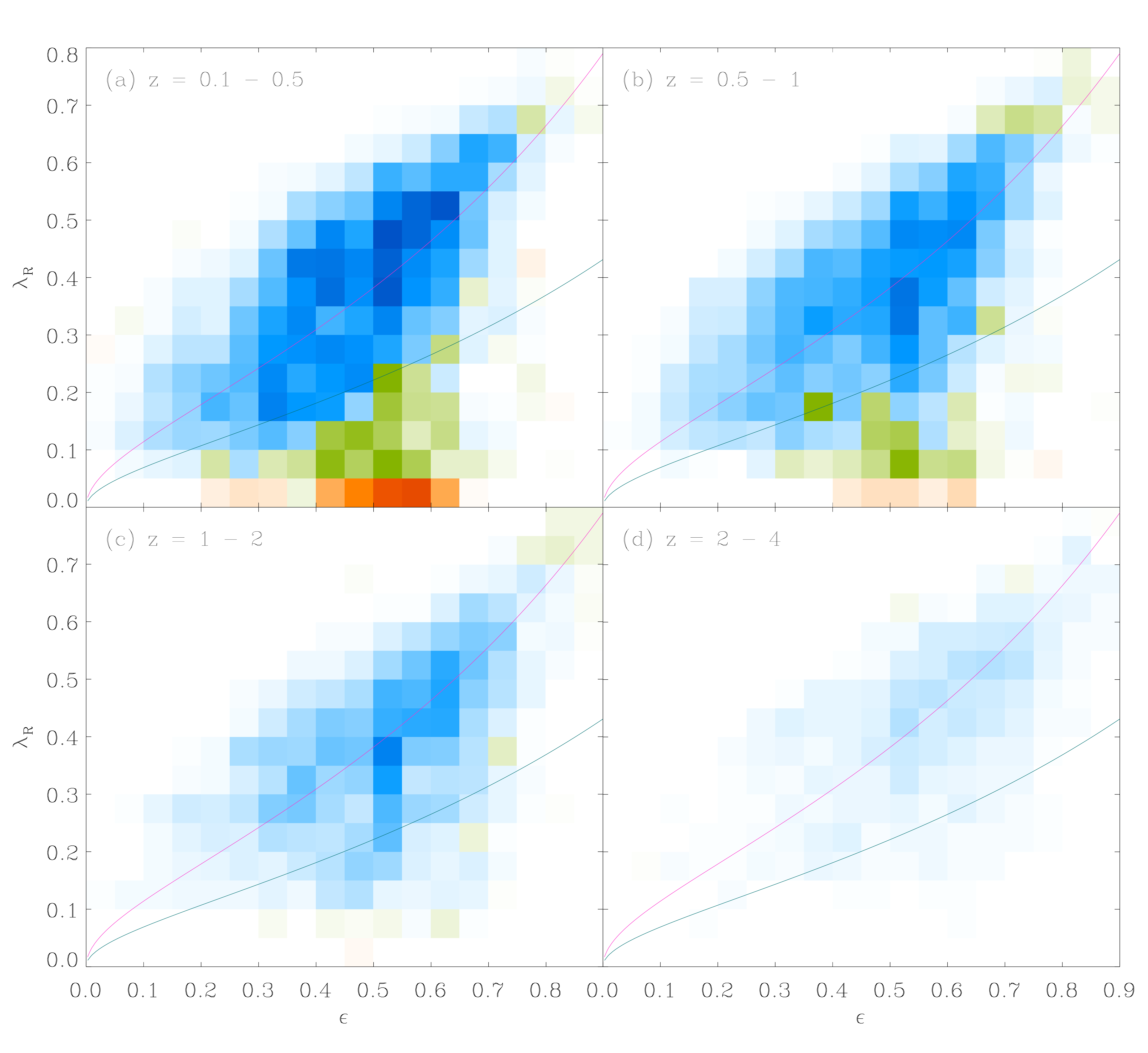}
\caption{Similar to Figure~\ref{fig:Lambda_z0} showing properties of
  elliptical galaxies, in edge-on projection, over intervals of redshift going
  back in time to $z=4$. While FRs are present at all epochs, SRs exist
  only for $z < 1$ in agreement with the establishment of the galaxy red
  sequence in Illustris at about this epoch.}
\label{fig:Lambda_z}
\end{center}
\end{figure*}

\subsection{Kinematic properties of galaxies as a function of their mass}
\label{subsec:MassBracket}

Our divisions between low, intermediate and high mass galaxies
($M_*/{\rm M_\odot} < 10^{11}, 10^{11} \leq M_*/{\rm M_\odot} <
10^{11.5} $ and $10^{11.5} < M_*/{\rm M_\odot}$, respectively) whilst set
quite arbitrarily, actually serve to provide strongly contrasting trends in
their kinematics at $z=0$. In Figure~\ref{fig:Lambda_z0_Mass} we present the
edge-on distributions of populations of low, intermediate and high mass galaxies
independently, which together make up all the resolvable galaxies in the
interval from $z = 0.1$ to $z = 0$.

Starting with the low mass bracket, we now see a very clear locus of points
around which these galaxies are clustered along a line roughly parallel with
equation (\ref{vsigma}) passing through the point $(\epsilon,\lambda_{\rm R})
\approx (0.5,0.5)$. Though some low mass galaxies cross the FR-SR boundary
these do not seem to be a separate population, just a natural spread in the
distribution. This may suggest that low mass galaxies classified as SRs are
not in fact part of a separate population from low mass FRs; they are the tail
end of the FR distribution.

Next examining the highest mass range we see a clear tendency to low
$\lambda_\mathrm{R}$ values, with the majority being classified as SRs. There
is a much more scatter, due to the lower numbers of higher mass galaxies, but
the galaxies seem to be roughly clustered around a point at
$(\epsilon,\lambda_{\rm R}) \approx (0.55,0.05)$. It is worth noting that
there also exist some high mass galaxies with high $\lambda_{\rm R}$ values,
which could either be the high mass tail of the lower mass galaxy
distribution, or could be galaxies spun up by a fortunate major merger (whilst
we expect the majority of major mergers to cause spin down, see
Section~\ref{subsec:Mergers}). 

The boundaries for the intermediate mass
bracket have been picked rather arbitrarily, but seem to well separate two
regions with much more pronounced properties, and intermediate mass galaxies
themselves seem to broadly fall into one of these two groupings. 

We also examined the distribution of galaxy morphologies for the central 
galaxies (corresponding to the primary dark matter halos) compared to
satellite galaxies and find that both seem to follow the same distributions in
$(\epsilon,\,\lambda_{\rm R})$ space, a trend that has since been observed in the MaNGA survey, Greene et al. (in prep.). This suggests that environmental
effects, on the scale of individual dark matter halos, do not have a large
impact on spin properties.

The sharp distinction between the low and high mass galaxies, as seen in
Figure~\ref{fig:Lambda_z0_Mass} and their correspondingly discrete
distributions of $\lambda_{\rm R}$ is striking, and could feasibly explain the
observed distinct populations of fast and slow rotating galaxies. 

\subsection{Cosmic evolution of SR and FR fractions}

In Figure~\ref{fig:Lambda_z}, we show a series of $\epsilon$, $\lambda_\mathrm{R}$ 
plots for redshift intervals:
$0.1 \le z \le 0.5$, $0.5 \le z \le 1$, $1 \le z \le 2$, and $2 \le z \le 4$. We observe a decreasing population of high mass galaxies with increasing $z$, in accord with
the hierarchical galaxy assembly, and of SRs as well, highlighting their
dependence on past merger histories. Of those galaxies that are classified as
SRs, particularly at higher redshifts, many appear to be part of a natural
spread of FRs, as we have previously suggested in
Section~\ref{subsec:MassBracket} and excluding these we see negligible SR
fractions until roughly $z = 1$, in agreement with the establishment of the
galaxy red sequence at about this epoch in the Illustris simulation. At all
redshifts FRs still seem to be centred at roughly the same ($\epsilon$,
$\lambda_\mathrm{R}$).

\subsection{Alternative methods for classifying fast and slow rotators}

Whilst the $\lambda_{\rm R}$ parameter is perhaps the best measure of the
kinematic properties of an elliptical galaxy there are a variety of other
properties that separate the population of fast and slow rotating ellipticals
\citep[see e.g.][]{Kormendy:2009aa,Kormendy:2016aa}. They relate to
many different properties of these galaxies, from their density profiles to
their luminosities, and we summarise here tests of some of these criteria on
the massive ellipticals in Illustris.

\subsubsection{Cusps vs. Cores}
Slow rotators generally exhibit flattened density profiles at their centres
(cores) whilst most fast rotators have rapidly rising density at small radii
(cusps). These may be the result of binary super-massive black holes at the
centres of galaxies scouring the local stellar population and leading to an
evacuated core \citep{2006ApJ...648..976M}. This would not be realisable by
the Illustris simulation, where black hole merging is not followed at small
scales due to the spatial resolution limitations.

Due to the gravitational softening length of stellar particles being of the
order of $1 \, {\rm kpc}$ it is impossible to judge whether low mass galaxies
have cuspy density profiles as the scale of a cuspy or cored profile is of the
same order as the softening length. For the most massive galaxies, whose cores
have characteristic lengths larger than the softening length and so are
resolvable, we see cored density profiles. These galaxies are almost all SRs
so this is well in line with observations. For galaxies with $M_* > 10^{12}\,
{\rm M_\odot}$ we see core radii mostly between $3$ and $5\, {\rm kpc}$.

\subsubsection{Boxy vs. Disky Isophotes}
Fast rotating ellipticals, with a significant fraction of stars still on disk-like orbits, should appear slightly disky when viewed edge on, with a central bulge and elongated fringes. Conversely slow rotators are more rounded and less disky, and can even have boxy isophotes, with bulges off axis \citep{2007ApJ...658..710N,2010gfe..book.....M}.

If we fit an ellipse with radius $R_{\rm ell}(\phi)$ to the isophote with radius
$R_{\rm iso}(\phi)$ and express the residuals as a Fourier series
\begin{equation}
    \label{eqn:Fourier}
    \Delta(\phi) = R_{\rm iso}(\phi) - R_{\rm ell}(\phi)
    = \Sigma^{\infty}_{n=1} a_n \cos(n\phi) + b_n \sin(n\phi)\,,
\end{equation}
the sign of the $a_4$ parameter tells us if the elliptical is boxy ($a_4 < 0$) or disky ($a_4 > 0$).

We perform this analysis for every central galaxy in Illustris at $z = 0$,
fitting ellipticals to isodensity contours and fitting Fourier series to the
residuals. In Appendix~\ref{ap:Contour} we discuss the methods used in more
detail, and present a visual example of the success and failures of capturing
the contours of a galaxy with this technique.

\begin{figure}
\begin{center}
\includegraphics[width=1.0\columnwidth]{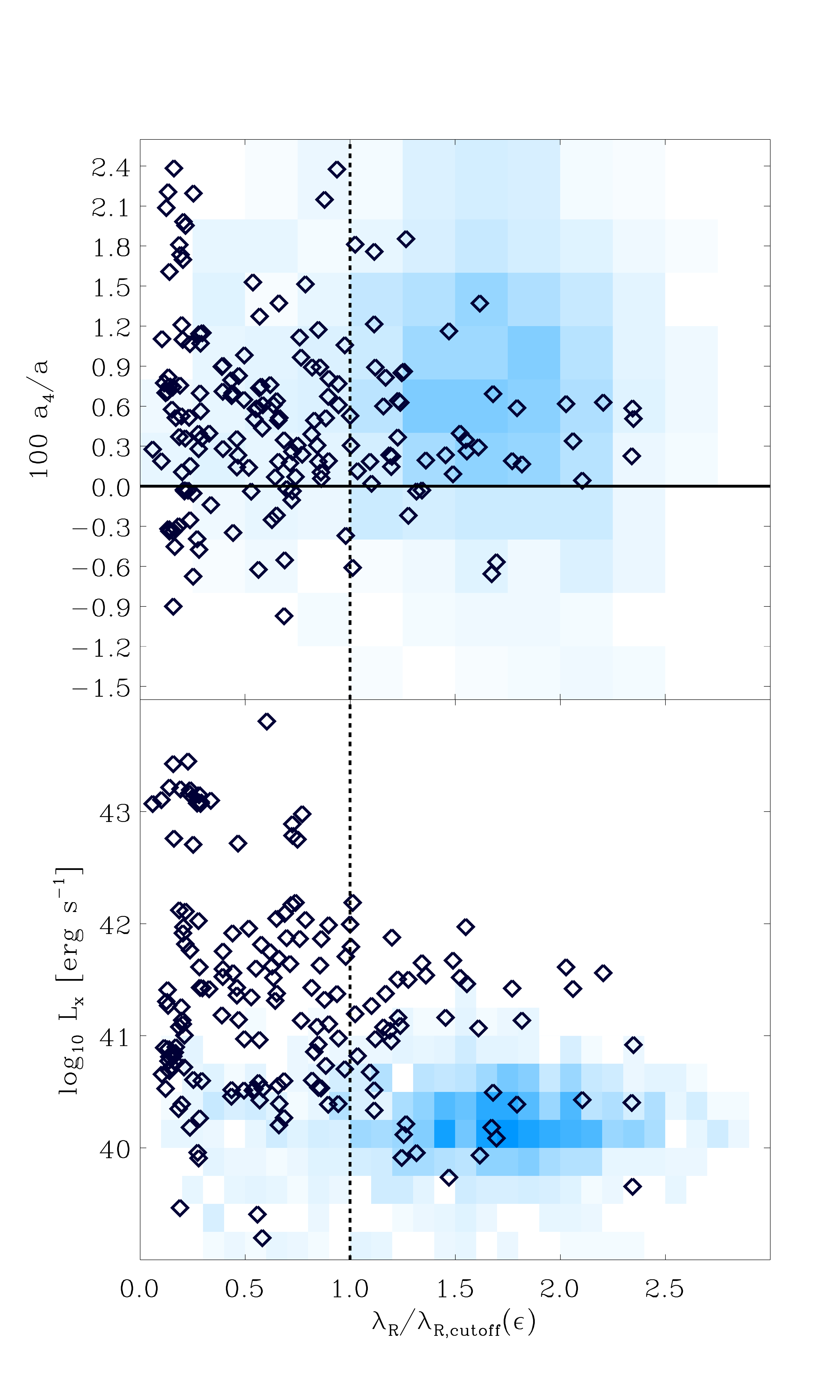}
\vspace{-0.5cm}
\caption{Correlation of $\lambda_\mathrm{R}$ with shape and X-ray luminosity
  of central elliptical galaxies at $z = 0$. Diamond symbols are for
  individual galaxies with $M_*>10^{11.5} M_\odot$, while 2D histograms show
  galaxies with $M_*<10^{11.5} M_\odot$. Horizontal axis shows spin normalised
  by the corresponding cutoff value for each galaxy's ellipticity $\epsilon$,
  such that those below a value of 1 are SRs and above are FRs. The vertical
  axis on the top panel shows the $a_4$ parameter normalized by the semi-major
  axis, $a$, of the fitted ellipse. Solid horizontal line shows the division
  between disky ellipticals (above) and boxy ellipticals (below). The bottom
  panel shows the X-ray luminosity, calculated via equation \ref{eqn:Lum}.} 
\label{fig:Lambda_ContourXray}
\end{center}
\end{figure}

The upper panel of Figure~\ref{fig:Lambda_ContourXray} shows the distribution
of boxy and disky FRs and SRs, plotted as pixels for galaxies with $M_* <
10^{11.5} \, {\rm M_\odot}$ and as individual points for more massive
galaxies. The population of galaxies is similar to the observed
$\text{ATLAS}^{\text{3D}}$ sample (Figure~13 of E+11) and with simulated
merger remnants including winds and gaseous halos \citep{Moster:2011aa}. 

The
majority of low mass galaxies are disky FRs, as expected, but with significant
variation encompassing some disky SRs and boxy FRs. However, examining the
fitting of contours to individual galaxies shows that for these smaller, less
well-resolved galaxies the contour fit can be erroneously dominated by small
morphological features. Hence much of this spread of properties may be
numerical rather than representative of galaxy properties, as detailed in
Appendix~\ref{ap:Contour}.

Looking at the high mass galaxies we see almost no boxy FRs and a high number
of disky FRs, as would be expected. Amongst SRs though there is a large
spread, with disky SRs being the most common morphology. Assuming that
galaxies evolve from fast rotation to slow via mergers we would expect to see
very few boxy FRs, as boxy shapes could only occur through disrupting a disky
ellipsoid. Some merger geometries may still lead to irregular, or even disky
SRs, and thus the wide variety of SRs morphologies is not necessarily a cause
for concern. That said specific choices of the Illustris sub-grid physics,
especially the very energetic AGN feedback in the radio mode, leading to fewer
wet mergers at lower redshift, probably also play a part in this skew towards
disky SRs.  

\subsubsection{X-Ray Luminosity}

SRs are often observed to have much brighter
X-ray emission with respect to the FRs due to
significant amounts of gas at $\sim 10^7 \,{\rm K}$ or above
\citep{2005MNRAS.364..169P,2006MNRAS.367..627E}. Maintaining gas at these high
temperatures may require a large energy input to the gas, which could be
delivered by AGN feedback, also explaining the shut-off of star
formation. Infalling gas from mergers can also be a source of heating through
shocks, though this is likely a secondary effect \citep{2006MNRAS.368....2D}. 

We adopt a simple estimate of the X-ray luminosity, $L_{\rm X}$, from a Bremsstrahlung approximation
\begin{equation}
L_{\rm X} = 1.2 \times 10^{-24} \dfrac{1}{m_{\rm p}^2} \sum^{N_{\rm
    gas}}_{i=1} m_{\rm g,i} \, \rho_{\rm g,i} \, \mu_{\rm i}^{-2} \, T_{\rm
  g,i}^\frac{1}{2} \,\, {\rm (erg \,s^{-1})}\,,
\label{eqn:Lum}
\end{equation}
where $m_{\rm p}$ the mass of a proton, $m_{\rm g,i}$ is the mass of the
$i^{th}$ gas cell, $\rho_{\rm g,i}$ and $T_{\rm g,i}$ are gas density and
temperature, respectively, and $\mu_{\rm i}$ is the mean molecular weight. We include all gas cells within a radius whose mean density is 200 times the
critical density of the Universe and exclude dense, star forming regions
sitting on the effective equation of state \citep{Springel2003} which would
unphysically skew the galaxy to higher luminosities.

The distribution of X-ray luminosities as a function of spin is shown in the
bottom panel of Figure~\ref{fig:Lambda_ContourXray}. As before, lower mass
($M_* < 10^{11.5} \, {\rm M_\odot}$) galaxies (plotted here as pixels) show no
clear 
trends, and most have similar luminosities of $L_{\rm X} \approx 10^{40}
\,{\rm erg\, s}^{-1}$ regardless of spin, though it should be noted there are
few SRs in this sample. Higher mass galaxies (plotted individually as symbols)
are generally more luminous, especially some SRs which are 2 to 3 orders of
magnitude brighter. There is again much variation and only a slight trend, for
slower spinning high mass galaxies to be X-ray bright. Note a clear divide between the bulk of galaxies and galaxies with $L_\mathrm{x} > 10^{42}$. We suggest these particularly bright galaxies are undergoing energetic AGN feedback, heating the gas much more effectively. 

There are a variety of other characteristics distinguishing fast and slow
rotators, such as the steepness of the S\'ersic index, the level of
anisotropy/triaxiality and the presence of a strong radio source which we
will not examine in this work, while the separation between FRs and SRs in
terms of stellar ages and hence metallicities is discussed in the next
section. 

\subsection{Properties of Elliptical Galaxies}

\begin{figure*}
\begin{center}
\includegraphics[width=0.85\textwidth]{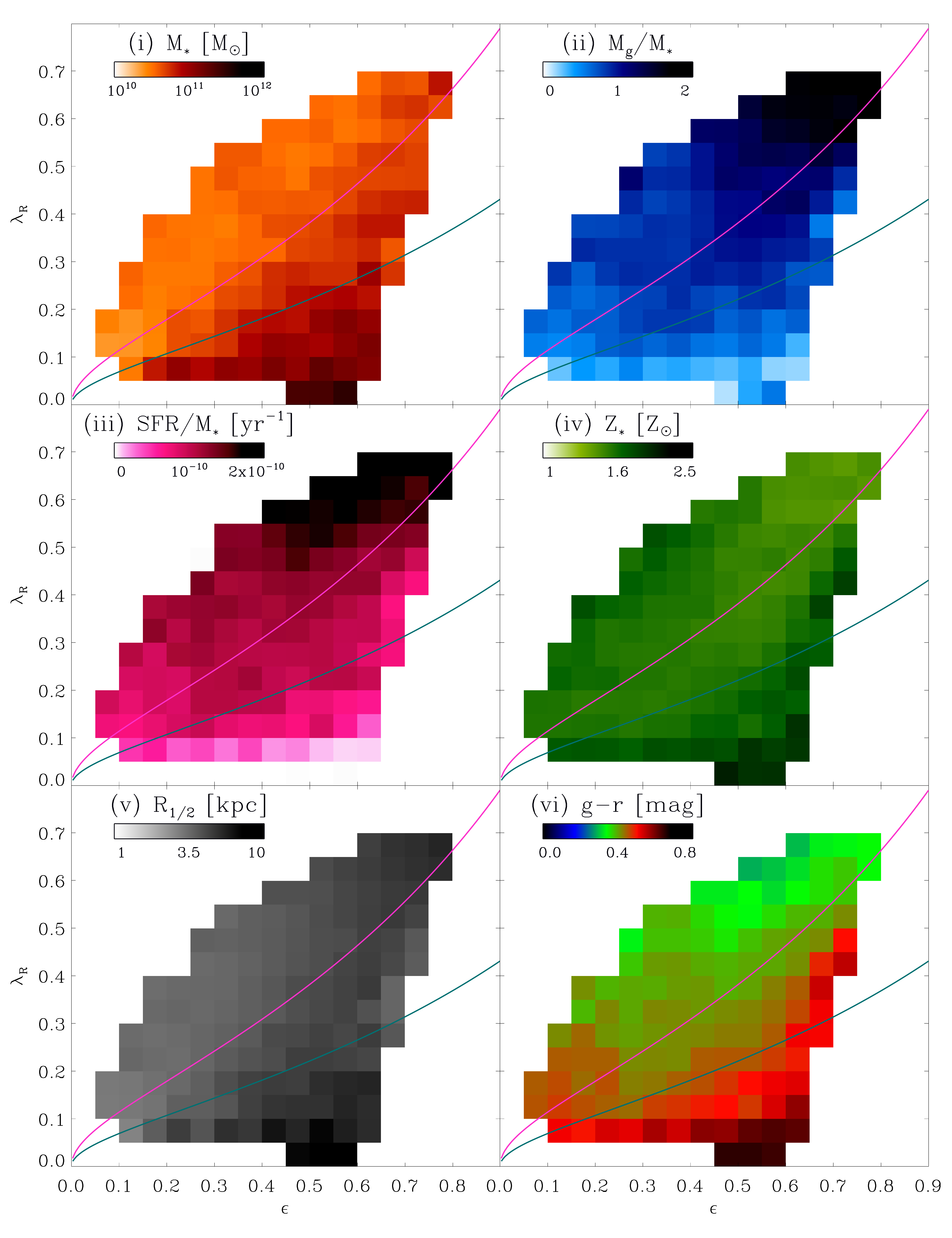}
\caption{Distribution of various properites of galaxies in the
  ($\lambda_\mathrm{R},\epsilon$) space. Top panels show stellar mass (left)
  and gas mass normalized to the stellar mass (right). Middle panels show
  specific star formation rate (left) and stellar metallicity (right). Bottom
  panels show stellar half-mass radius (left) and g-r colours (right). Results
  are averaged over snapshots ranging from $z = 0.2$ to $z = 0$ and any pixel
  which does not contain at least 30 galaxies over this period is
  excluded. Note that SRs are characterized by several distinct properties:
  large 
  stellar masses, sizes and metallicities, low gas content and specific star
  formation rates, and red colours.}
\label{fig:Lambda_Extra}
\end{center}
\end{figure*}

Here we discuss the distribution of various galaxy properties with relevance
to their degree of spin and ellipticity, as shown in
Figure~\ref{fig:Lambda_Extra}. 
\begin{itemize}
\item[i)] Stellar mass (top left): as previously stated, there is a very strong correlation between a galaxy's spin and its stellar mass, and we see here that SRs (below the green line) are significantly more massive. We also see a gradient with mass in the FR population (particularly above the magenta line), with rounder, slower spinning FRs being slightly less massive.
\item[ii)] Gas fraction (top right): the fastest rotating galaxies are by far
  the most gas rich, and the fraction of gas decreases with spin down to SRs
  which are almost devoid of gas. There is also a trend among FRs of decreasing gas fraction for rounder slower spinning galaxies.
\item[iii)] Specific star formation rates (middle left): SRs are forming
  almost no stars whilst the fastest spinning FRs are still undergoing
  significant star formation. Rounder, slower spinning FRs are forming stars
  at a lower rate than more elongated FRs in agreement with their lower gas
  fraction.  
\item[iv)] Stellar metallicity (middle right): The fastest rotating galaxies
  are the least enriched, while SRs have highly enriched stars. These results are mirrored in gas metallicity (not shown).
\item[v)] Stellar half-mass radius (bottom left): The distribution of galaxy sizes is very
  similar in form to the stellar masses (as shown in panel (i)). The fastest
  spinning FRs are larger than rounder, slower spinning FRs, and very massive
  SRs are by far the largest galaxies in the sample. 
\item[vi)] Colour (bottom right): SRs are the reddest galaxies in the sample, followed by
  round FRs and finally elongated FRs, in agreement with gas fraction and
  metallicity distributions.
\end{itemize}

In all of these panels we see strong gradients tracing from the center of the
FR distribution, at $(\epsilon,\lambda_{\rm R}) \approx (0.5,0.5)$ to the slowest spinning galaxies in the sample at $\sim (0.5,0.0)$. In many we also see a gradient along the distribution of FRs, spanning from bottom left to top right of the plots.

Almost all gradients of the former kind can be explained by the gradient in
mass. More massive galaxies are likely to undergo more mergers and more AGN
feedback, leading to gas being expelled. Gas poor galaxies then cannot sustain
star formation, thus have less short-lived high mass blue stars and will be
redder in colour. The galaxy metallicity is closely related to the age of a galaxy, showing that more massive galaxies must have formed earlier and undergone more bursts of star formation in their lifetimes, possibly fuelled by star-burst events during galaxy mergers. 

The gradients of the latter kind, across the distribution of FRs, may be driven
by efficient galactic wind feedback or by environmental differences. More
efficient feedback would expel gas, cutting off star formation and stunting
the growth of stellar mass. Alternatively galaxies in a less gas rich
environment, with only a small supply of inflowing gas, would have less star
formation and thus be less massive and smaller. Both mechanisms could explain
the transition from spheroid to disk-like morphologies if we assume the
angular momentum of a galaxy comes from that of its inflowing
material. Cutting off the supply of gas, either due to a gas poor environment
or by feedback preventing inflowing material reaching the galaxy, could lead
to rounder, slower spinning objects.

\begin{figure*}
\includegraphics[width=\columnwidth]{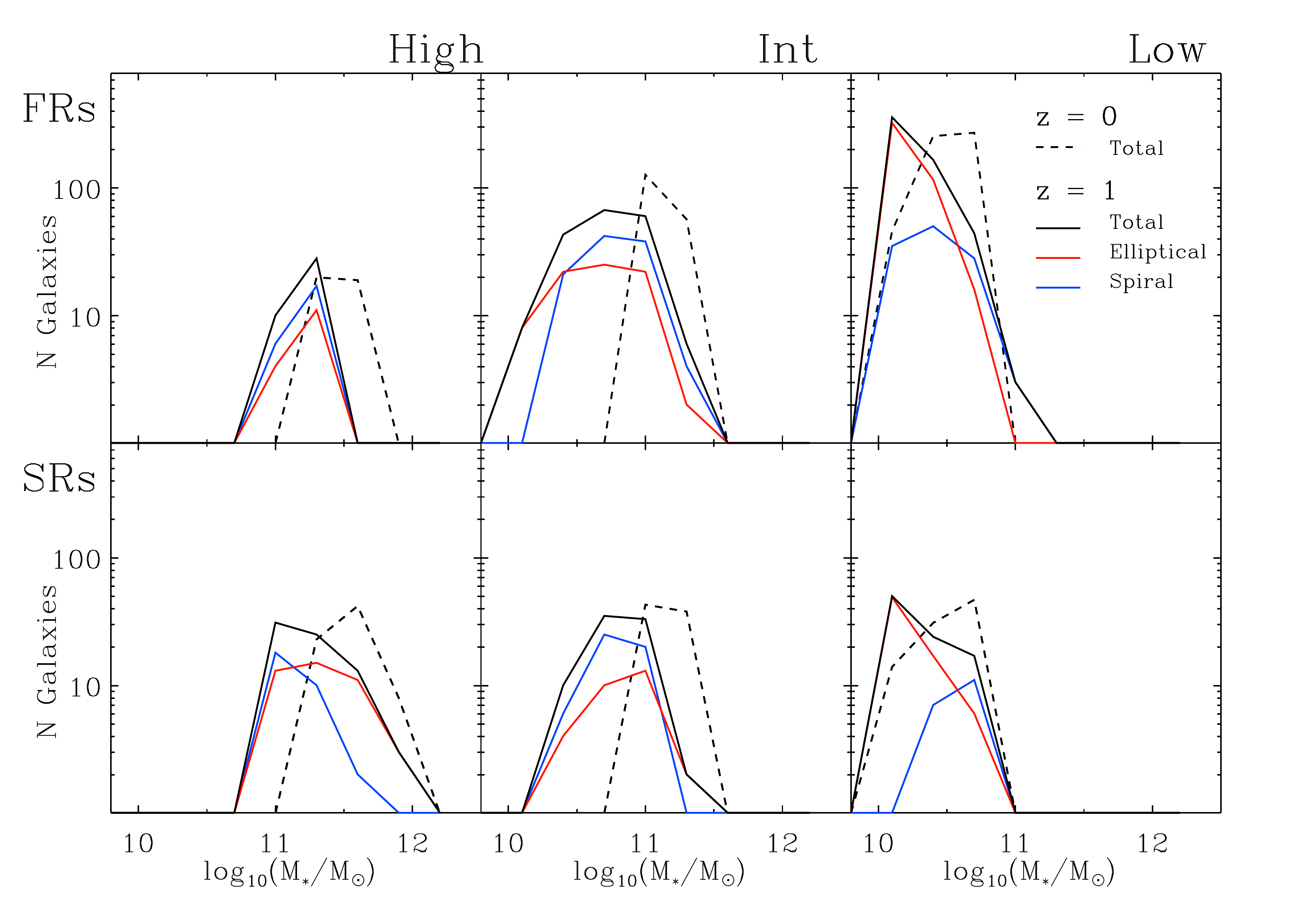}
\hspace{5mm}
\includegraphics[width=\columnwidth]{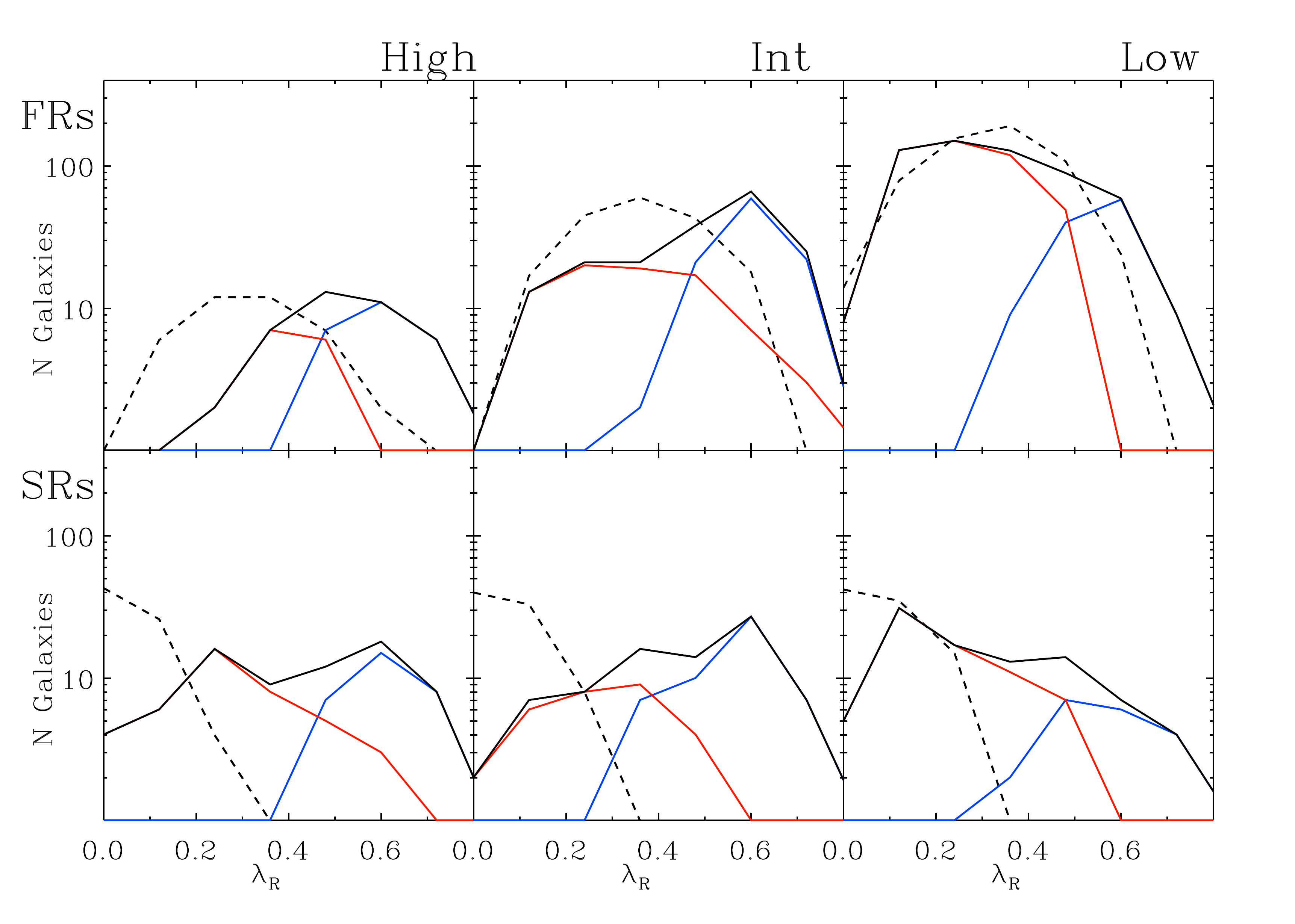}
\caption{Distribution of main progenitor galaxies at $z = 1$, separated by
  their properties at $z = 0$ (FRs, SRs and low, intermediate and high mass
  brackets) in terms of the galaxy's stellar mass (left-hand panel) and
  $\lambda_\mathrm{R}$ parameter (right-hand panel) at $z = 0$. The population
  of galaxies is separated into spirals and ellipticals at $z = 1$. Only
  galaxies which are well resolvable (over 20,000 star particles) at $z = 1$,
  and which are classified as ellipticals at $z = 0$, are included.}
\label{fig:ProgType_z1}
\end{figure*}

\begin{figure*}
\includegraphics[width=\columnwidth]{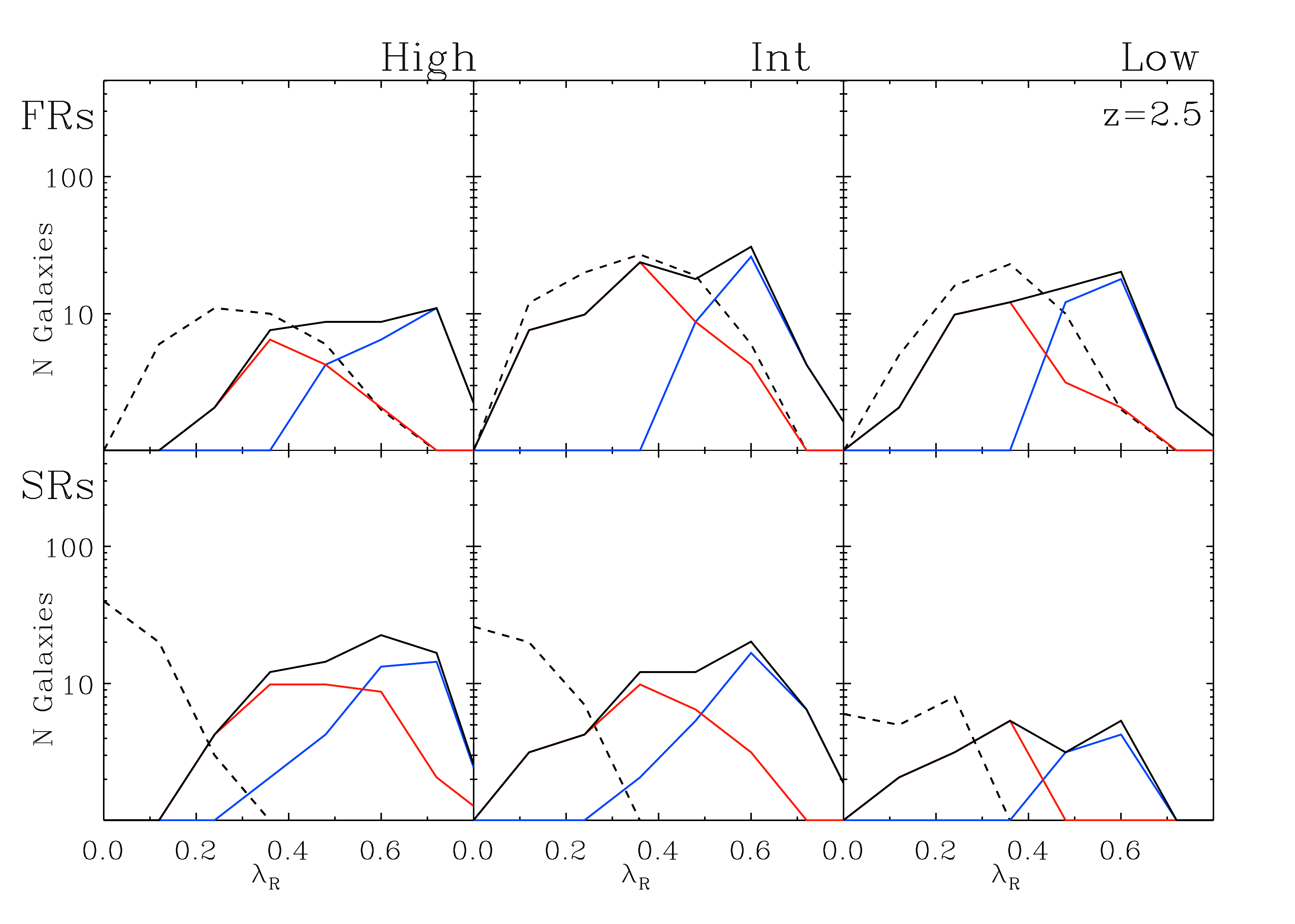}
\hspace{5mm}
\includegraphics[width=\columnwidth]{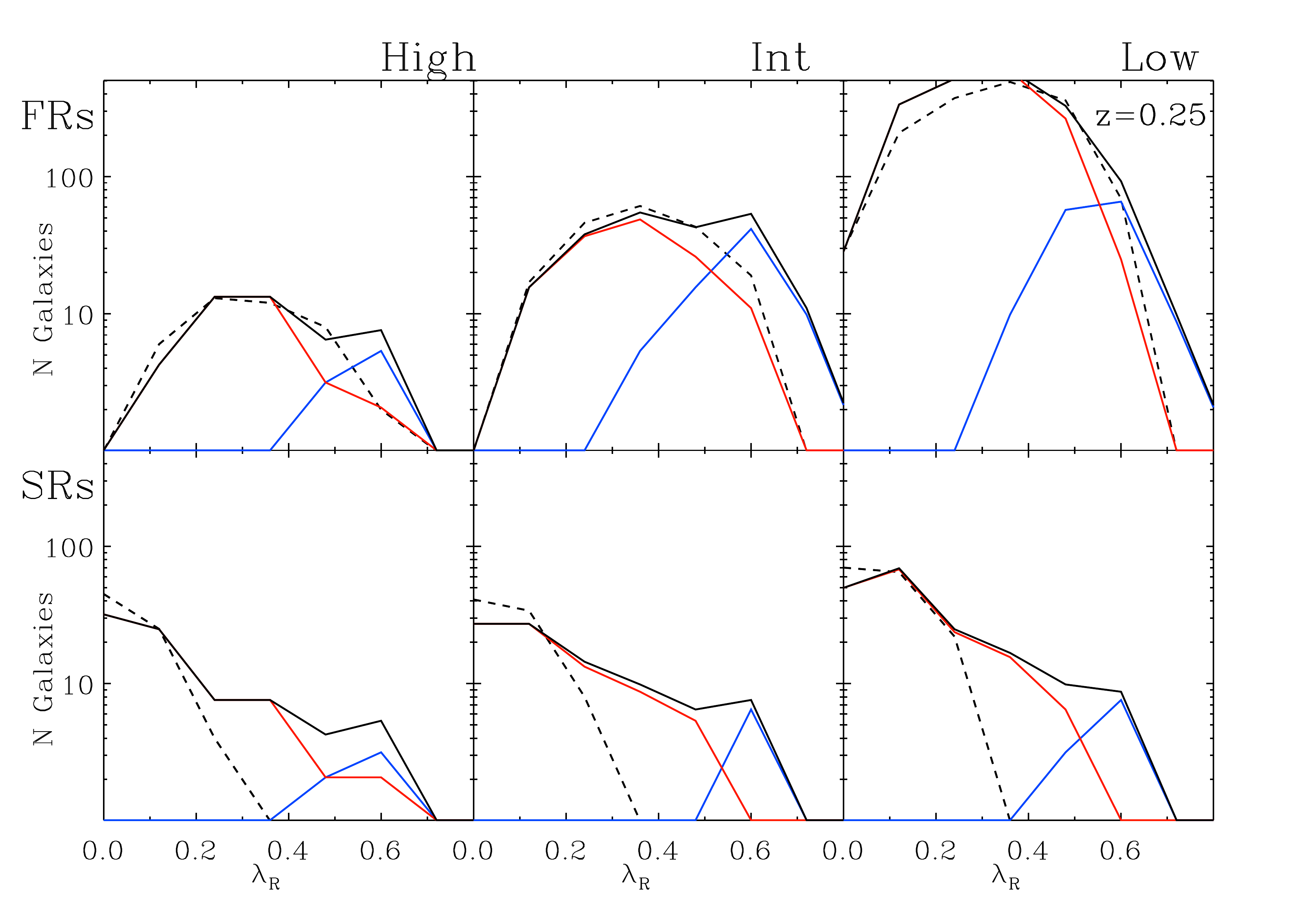}
\caption{As in Figure~\ref{fig:ProgType_z1}, we show the distribution of main
  progenitor galaxies in terms of their $\lambda_\mathrm{R}$
  parameter. Progenitors are shown at $z = 2.5$ (left-hand panel) and $z =
  0.25$ (right-hand panel). While at $z =2.5$ progenitor distributions are
  very similar, at $z = 0.25$ progenitors of present day FRs and SRs have
  clearly different distributions.}
\label{fig:ProgType_z}
\end{figure*}

\section{Merger Histories of Fast and Slow Rotators}
\label{sec:History}
\subsection{How progenitor galaxy morphology relates to present day properties}
\label{subsec:ProgType}

We begin by separating our sample of 3,207 elliptical galaxies at $z = 0$ into FRs and SRs based on our cutoff criteria (see Figure~\ref{fig:Lambda_z0}). We then separate these populations further into high, intermediate and low mass brackets. We follow the main progenitor of each of these galaxies back in time and compare how the distribution of galaxies overall, and the fraction of spiral and elliptical galaxies, vary with mass and kinematic properties.

Examining the main progenitors at $z = 1$, as a function of stellar mass
(left-hand panel) and $\lambda_\mathrm{R}$ parameter (right-hand panel) at $z
= 0$, as shown in Figure~\ref{fig:ProgType_z1}, we see that overall there is very little difference between the population of galaxies that will eventually diverge to become fast or slow rotators. The highest mass galaxies show the greatest differences, with slightly more high mass and slow spinning ellipticals among the progenitors of SRs, but overall the total distribution as well as the fractions and distributions of spirals and ellipticals, agree well between FRs and SRs.

This points to the conclusion that the rotational properties of elliptical
galaxies are mostly determined by the latter half of their evolution. The lack
of strong trends in their progenitors suggests that it is the merger histories
of the galaxies that are crucial in determining their $z = 0$ kinematics, as
this is the major, stochastic, differentiating factor between the later
evolution of any two galaxies with similar properties at $z = 1$. In the later
sections we will thus limit our analysis mostly to evolution since $z = 1$.

\begin{figure*}
\begin{center}
\includegraphics[height=0.8\columnwidth]{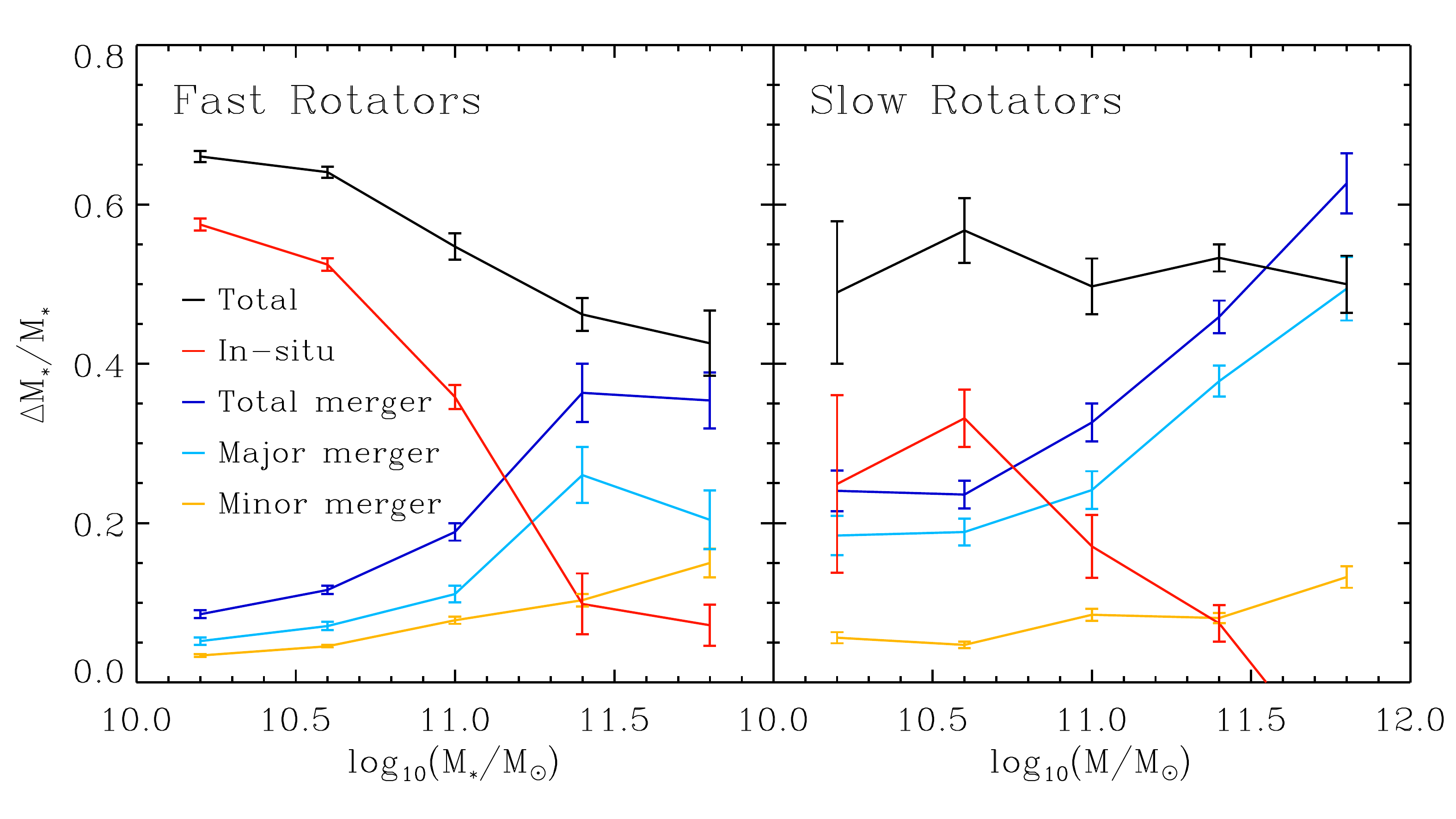}
\caption{Fraction of mass at $z = 0$ accrued by each elliptical galaxy since
  $z = 1$, as a function of total mass at $z = 0$, separated into FRs (left)
  and SRs (right). Different curves, as denoted on the legend, show mass
  fractions from minor, major and all mergers, in-situ star formation, as
  well as the total mass fractions. Points show mean values for all galaxies in each bin, and error bars show error in the mean.}
\label{fig:MassAcc}
\end{center}
\end{figure*}

\begin{figure*}
\begin{center}
\includegraphics[height=0.8\columnwidth]{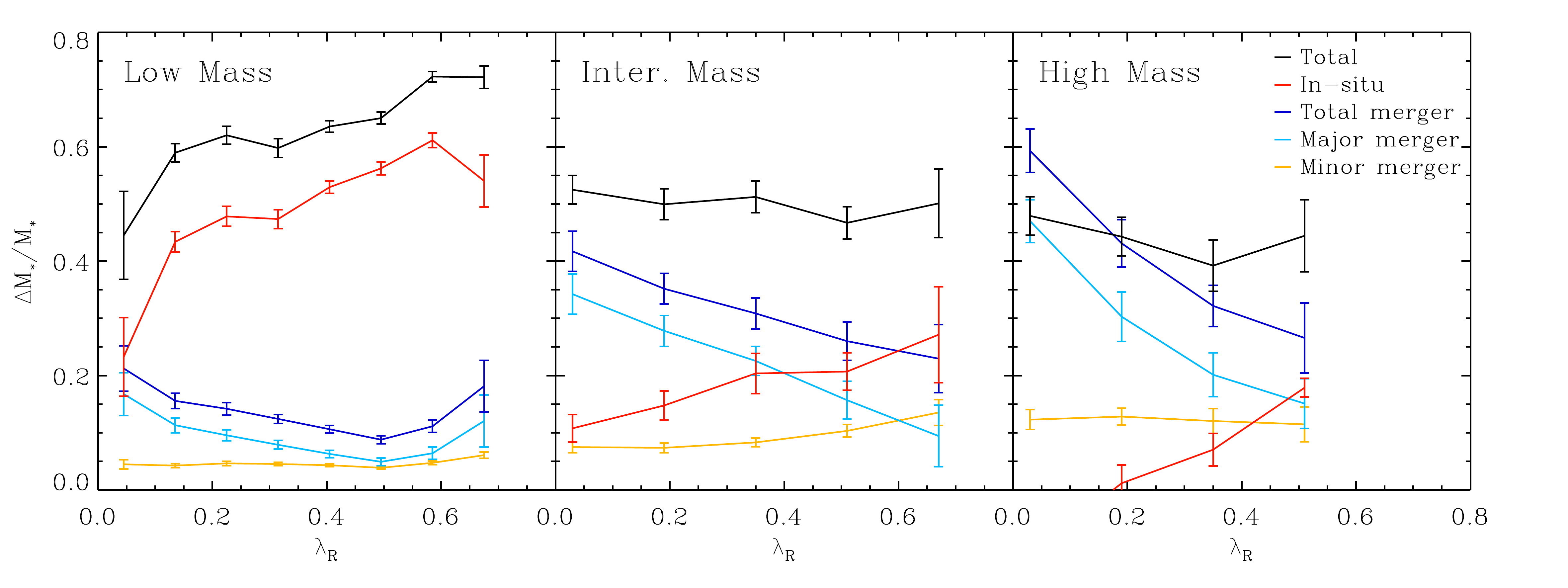}
\vspace{-0.5cm}
\caption{Similar plot to Figure~\ref{fig:MassAcc} with galaxies now separated
  into low ($M_*<10^{11} M_\odot$), intermediate ($10^{11} M_\odot < M_* < 10^{11.5} M_\odot$) and high mass ($M_* > 10^{11.5} M_\odot$) brackets and plotted as a function of their $\lambda_\mathrm{R}$
  values at $z = 0$. Note that the range of values of $\Delta M/M$ is higher for low mass galaxies and that there are not sufficient numbers of high mass galaxies at $\lambda_\mathrm{R}>0.5$ for it to be useful to plot these.}
\label{fig:LambdaAcc}
\end{center}
\end{figure*}

In Figure~\ref{fig:ProgType_z} we show the distribution of progenitors at $z =
2.5$ (left-hand panel) and $z = 0.25$ (right-hand panel) with respect to their
$\lambda_\mathrm{R}$ parameter. In accord with our previous discussion, at $z
= 2.5$ there is almost no difference 
between FR and SR populations. Despite this redshift being the epoch in which
mergers are most prevalent the lack of any divide between the two populations
suggests that either the FR and SR properties are only strongly dependent on
recent mergers, or that the conditions in which galaxies are evolving, and
mergers are occurring, changes significantly after this period. At $z = 0.25$
there is now a clear difference between SR and FR progenitors, with FRs in
general being faster spinning and having a higher fraction of spirals. High
mass galaxies have almost converged upon their $z = 0$ distribution, but the
same is not true for low and intermediate mass galaxies. If rotational
properties truly are determined by some mass cutoff it makes sense that the
highest mass galaxies, which have crossed this threshold earlier in time,
converge sooner.

An interesting side note in these plots is the re-occurring peak in the
distribution of spiral galaxies at $\lambda_\mathrm{R} \approx 0.6$ suggesting
spiral galaxies have relatively uniform rotational properties over all
redshifts and mass ranges considered. Note further that, as we discussed
previously, the resolution and feedback modelling limitations lead to an
overestimation of the fraction of elliptical galaxies at low mass (for further
details see Appendix~\ref{ap:Spirals}). 

\begin{figure*}
\begin{center}
\includegraphics[height=0.85\columnwidth]{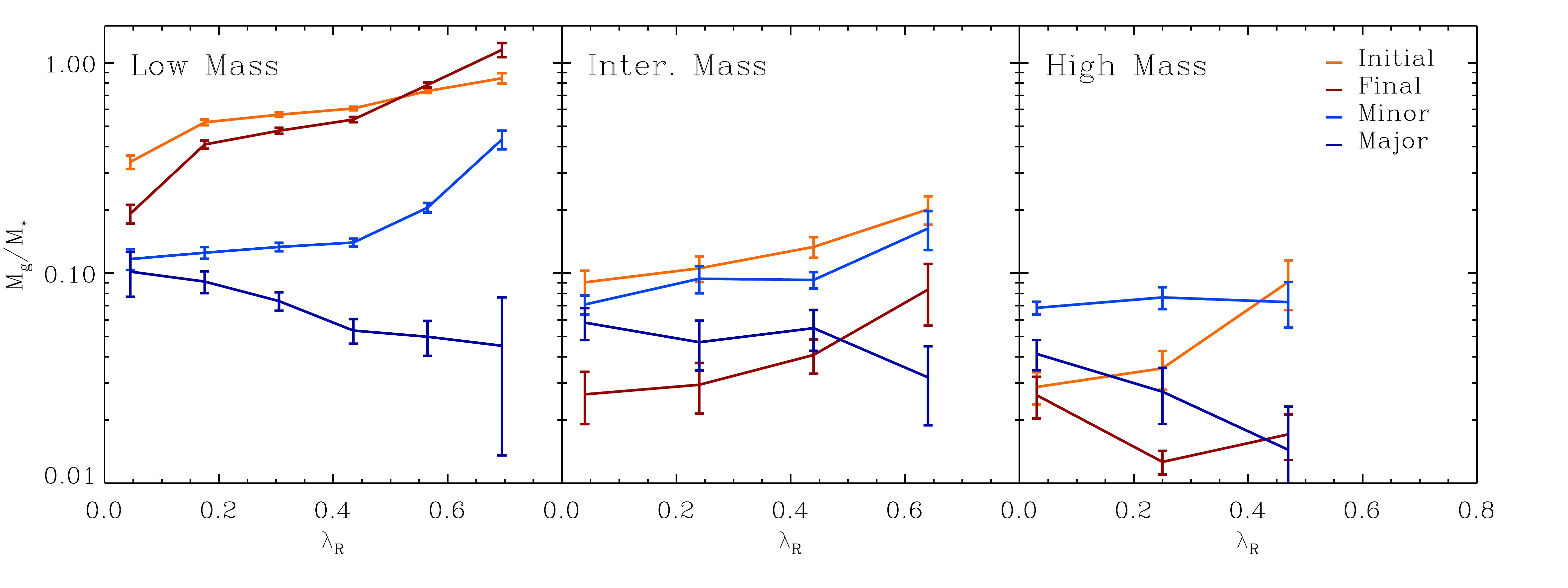}
\vspace{-0.5cm}
\caption{The fraction of gas mass to stellar mass (within two half-mass
  radii), showing relative amounts at $z=1$ and $z=0$ along with the accreted
  gas mass through major and minor mergers during that period. Galaxies are
  separated into three brackets by their stellar mass and binned by their
  rotation properties ($\lambda_\mathrm{R}$) at $z=0$. All gas masses are
  shown relative to the stellar mass at $z=0$.} 
\label{fig:GasAcc}
\end{center}
\end{figure*}

\subsection{Examining mechanisms of stellar mass change since $z = 1$}
\label{subsec:MassAccRotators}

We first examine the relative fraction of stellar mass accreted through
major and minor mergers, and the fraction formed in situ. We follow the
accretion histories of galaxies from $z = 1$ for populations seperated by
their mass at $z = 0$ (Figure~\ref{fig:MassAcc}) and rotational properties
(Figure~\ref{fig:LambdaAcc}). As we have shown in
Section~\ref{subsec:ProgType} the kinematic properties of galaxies at $z = 0$
are mostly determined by their evolution in the period post $z = 1$, so by
analysing the processes by which they acquire stellar mass in this period we
aim to explain their present day fast and slow rotating properties.

In Figure~\ref{fig:MassAcc} we look at how FRs and SRs of different masses
accrue their stellar mass in the period since $z = 1$. In both panels we see
some similar trends, i.e. more massive galaxies experience more mergers and
gain more mass that way, whilst the amount of mass formed in-situ
decreases. We also see some strong contrasts between FRs and SRs, namely that
SRs gain much more of their mass through major mergers and much less through
in-situ star formation, though the total fraction of their mass acquired is
similar for both FRs and SRs, as they roughly double their mass in the second
half of cosmic time.

We can perform a similar analysis for the three different mass brackets detailed in Section~\ref{subsec:MassBracket}, this time
looking at how accreted mass fraction varies with respect to
$\lambda_\mathrm{R}$ values at $z = 0$, as shown in Figure~\ref{fig:LambdaAcc}.

For low mass galaxies, all except the slowest spinning have their mass change
dominated by in-situ star formation. Those with more major mergers seem to be
more disrupted. However, by examining the merger histories there are some low
mass galaxies that have little ordered rotation which have not undergone recent
major mergers. This suggests that the process of accreting gas and forming
stars, or lack thereof, may alone be able to alter the spin of these
galaxies. Minor mergers seem to have very little effect, with more minor
mergers possibly very slightly spinning up intermediate mass
galaxies. Intermediate and high mass galaxies with a significant mass
contribution from major mergers seem to be spun down, whilst in-situ star
formation is associated with a spin up. We can summarize these results as
a simple set of trends: 
\renewcommand{\labelenumi}{(\roman{enumi})}
\begin{enumerate}
\item Slower rotating galaxies have markedly more major mergers in their
  accretion history, thus major mergers are linked to a net spin-down.
\item Stellar mass change from minor mergers seems to be almost independant of
  galaxy spin, thus minor mergers do not significantly affect the spin.
\item Faster rotating galaxies have more in-situ star formation, thus star
  formation is linked to a net spin up.
\end{enumerate}

The only exception to these trends are the fastest rotating low mass galaxies,
which seem to be spun-up by major mergers as well. In fact the low mass galaxies with the least mergers have $\lambda_\mathrm{R}$ values closest to the locus of points around which low mass FRs seem to cluster (Figure~\ref{fig:Lambda_z0_Mass}). This suggests a scenario in which undisturbed galaxies tend to sit at $\lambda_\mathrm{R} \approx 0.5$ unless disturbed by mergers, which in general disrupt spin, but have some small chance of leading to a spin-up and a very fast rotating elliptical galaxy.

These trends however do not give us a clear reason why we should see two
distinct populations of FRs and SRs, rather than a continuum. To further
examine this we have to look in more detail at how mergers and in-situ star
formation act on galaxies of different masses. 

\begin{figure*}
\begin{center}
\includegraphics[height=0.95\columnwidth]{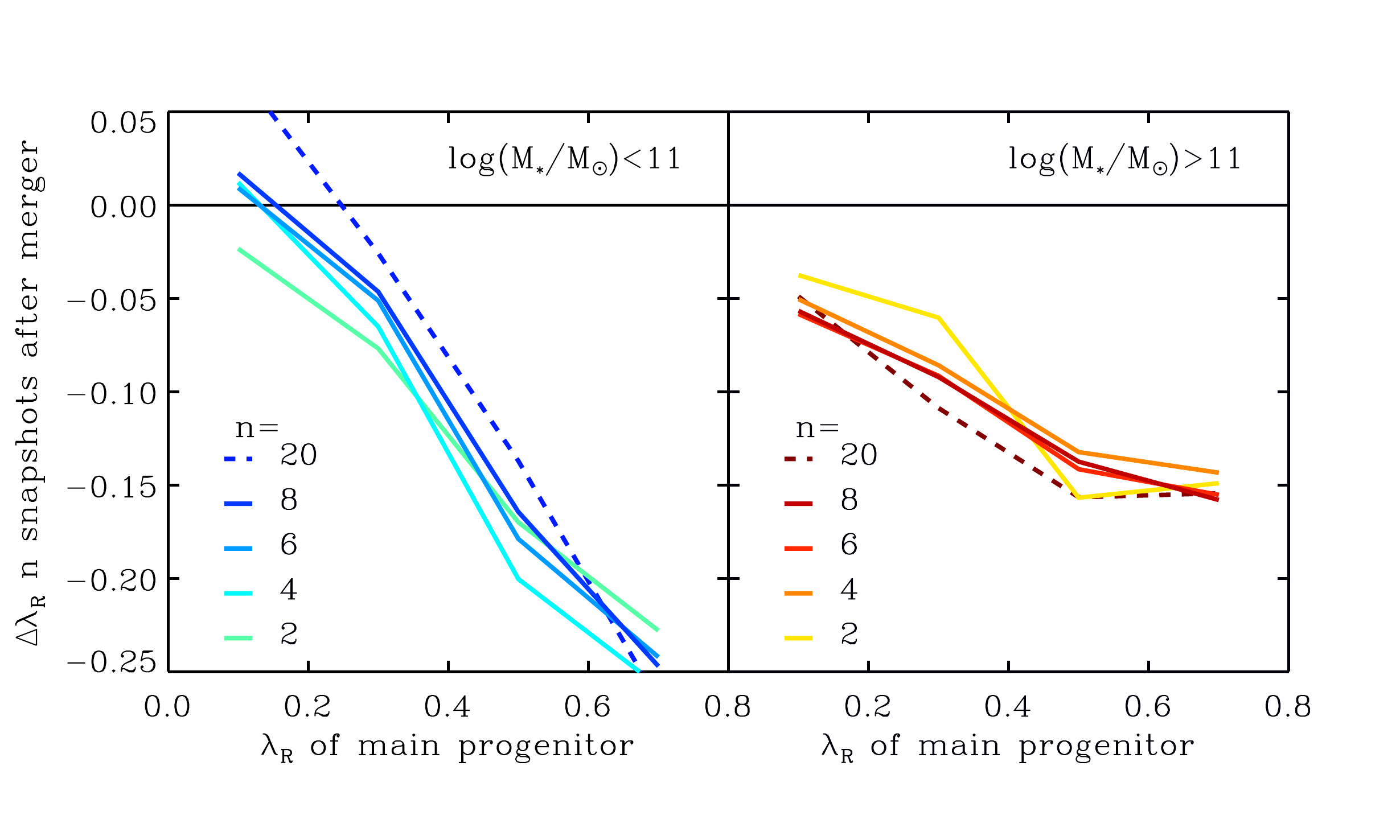}
\vspace{-0.5cm}
\caption{Change in $\lambda_\mathrm{R}$ for major mergers occurring since $z =
  1$, plotted against the initial $\lambda_\mathrm{R}$ of the more massive
  progenitor galaxy. The resulting change, relative to initial
  $\lambda_\mathrm{R}$ is shown $2$, $4$, $6$, $8$ and $20$ snapshots after
  the merger itself, and results are separated for low mass galaxies (left),
  and high mass galaxies (right). For reference snapshots are spaced in time intervals of roughly $0.1$-$0.15$ Gyr.} 
\label{fig:MajDelta}
\end{center}
\end{figure*}

\subsection{Accretion of gas since $z = 1$}

Though we trace kinematic properties of galaxies based on their stellar
component, gas also plays a large part in the evolution of a galaxy's
spin. Inflowing gas may carry a large amount of angular momentum from the
intragalactic medium (IGM)
to galaxies, and the transfer of energy and angular momentum between stars and
gas can be very important.

Figure~\ref{fig:GasAcc} shows gas fraction accreted onto galaxies (separated
into the same mass brackets) through minor and major mergers, as well as the
fraction of gas at $z = 1$ and $z = 0$, as a function of $\lambda_\mathrm{R}$
values at $z = 0$. All gas fractions are calculated with respect to the
stellar mass at $z = 0$.  

Low mass galaxies roughly retain their gas content, while the fastest spinning
low mass galaxies even slightly gain gas since $z = 1$. However, as we move
towards higher masses the fraction of gas drastically reduces, due to feedback
processes, particularly AGN feedback in the most massive galaxies, which blows
almost all the gas out from the centre of the galaxy. Note for all mass ranges
the higher the initial gas fraction the higher the final $\lambda_\mathrm{R}$. 

In contrast to the previous section there is now a clear correlation between
minor mergers and spin for low and intermediate mass galaxies. Perhaps
gas-rich minor mergers are a mechanism for replenishing a galaxy's gas supply,
or a sign of a gas-rich local environment. Gas from major mergers does not
seem to have the same effect, with slower spinning galaxies gaining more mass
from major mergers. We suggest this is because the major merger is much more
damaging to a galaxy, lowering it's spin, than any spin-up that might be
caused by the injection of gas. 

This suggests that the gas component of smaller galaxies is much more
effective at transferring angular momentum to the main galaxy than the stellar
component, and that the incoming angular momentum is well aligned with the
spin of the galaxies, leading to gas-rich minor mergers having a strong
spin-up effect. Gas from minor mergers may be a key factor in the evolution of
the fastest spinning ellipticals. 

There may be a link between the supply of gas, from accretion and minor
mergers, and the local environment. Large amounts of gas may be a sign of a
high inflow rate from the IGM, and its ability to cool and accrete faster than
it is heated and expelled by feedback processes. Incoming gas can spin up a
galaxy either by torquing the stellar component, or simply by forming new
stars with similarly high angular momentum.  

Putting together this evidence we conclude:
\renewcommand{\labelenumi}{(\roman{enumi})}
\begin{enumerate}
\item The supply of gas to a galaxy is closely linked to its spin, with more gas leading to faster spinning, more ordered rotation.
\item Gas rich minor mergers are linked to a spin up in lower mass galaxies.
\item Major mergers, regardless of gas content, are linked to a spinning down of galaxies.
\end{enumerate}

\subsection{Evolution of galaxies during mergers}
\label{subsec:Mergers}

We can roughly separate the history of a galaxy's evolution into long periods without any major merger activity and the short periods over which a major merger is rapidly changing the galaxy. 

As shown in the previous section, the rotation properties of massive
ellipticals have a strong correlation with the fraction of stellar and gas
mass accreted through major mergers. Thus, we examine, in
  Figure~\ref{fig:MajDelta} the change in $\lambda_\mathrm{R}$ following a
  major merger as a function of time. Solid curves are for times directly
    after 
  the major merger (from $2$ to $8$ snapshots afterwards, roughly corresponding
  to $0.25$~Gyrs and $1$~Gyrs, respectively), while dashed curves show the
  result after a longer period of time ($20$ snapshots, roughly corresponding
  to $2.5$~Gyrs).

For galaxies of all masses a major merger causes an immediate and drastic disruption of any ordered rotation. It should be noted that this is an average effect, and some small fraction of major mergers can still lead to a spin-up of the remnant. For all but the slowest rotating galaxies, for which the value of $\lambda_\mathrm{R}$ cannot drop any further, low mass ellipticals are more immediately disrupted by major mergers than higher mass galaxies, losing almost half their spin in a single major merger.

For lower mass galaxies, it takes around $0.5$~Gyrs ($2$ to $4$ snapshots) for
the value of $\lambda_\mathrm{R}$ to reach a minimum, after which their 
spin gradually begins to increase again, showing signs of recovering from the
effect of the merger. No such recovery of spin post-merger is seen in high
mass galaxies. Shortly after the merger they seem to settle in a slower
spinning configuration. Though each individual merger has less disruptive
effect on higher mass galaxies the fact that they do not seem to recover their
spin post merger suggests that repeated major mergers can reduce the galaxy's
spin in steps.

\begin{figure}
\begin{center}
\includegraphics[height=0.95\columnwidth]{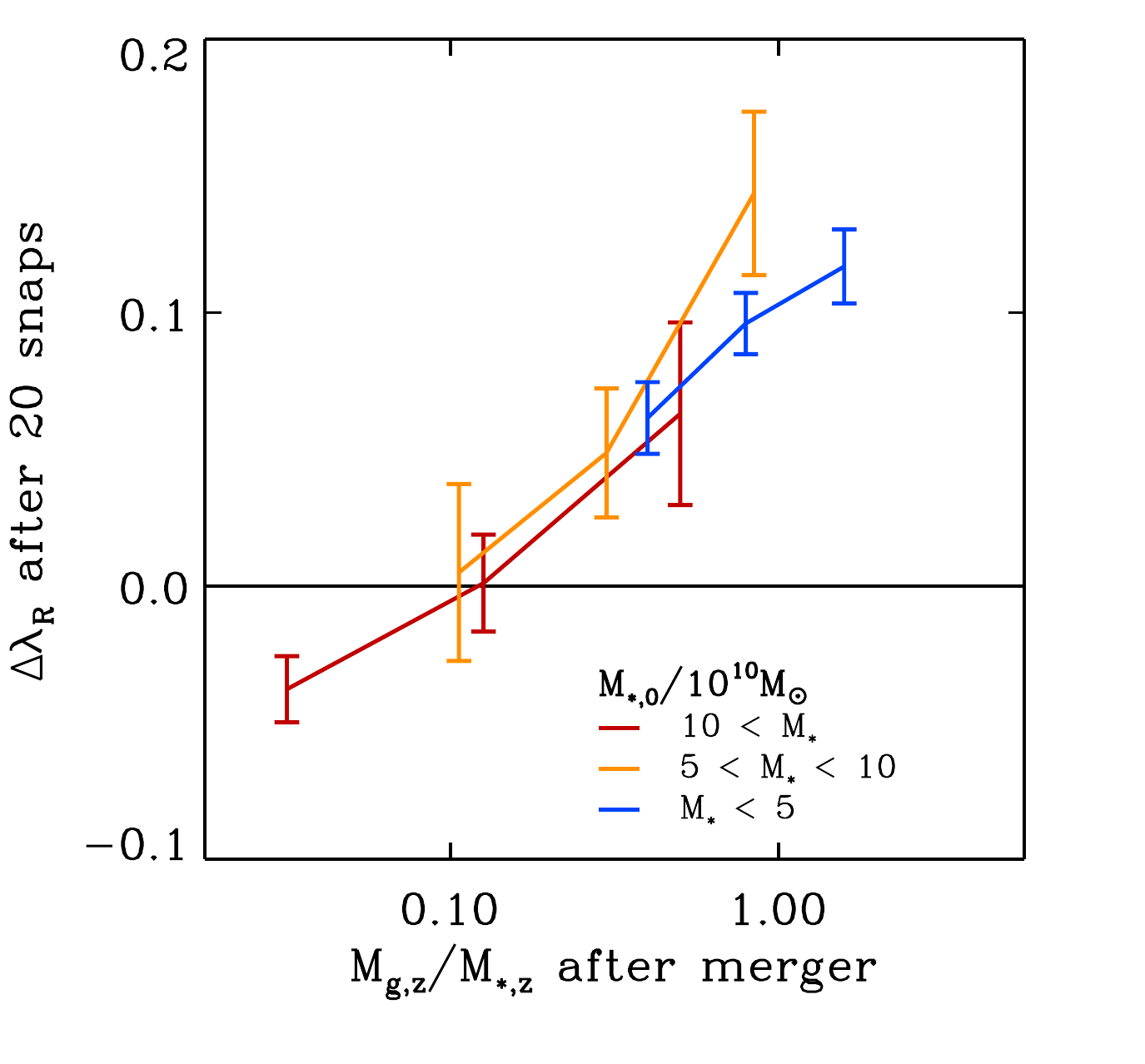}
\vspace{-0.5cm}
\caption{Change in $\lambda_\mathrm{R}$ $2$-$3$ Gyr ($20$
  snapshots) after a major merger
  occurring since $z=1$, plotted against the 
  fraction of gas mass to stellar mass (within two half-mass radii) directly following the merger (after $\sim 0.33$~Gyr or $3$ snapshots).}
\label{fig:MajDelta_Gas}
\end{center}
\end{figure}

We then investigated further the factors leading to a galaxy spinning up again after a merger. As shown in Figure~\ref{fig:MajDelta_Gas} we found a strong dependence on the gas fraction shortly after the merger, with galaxies that retained gas quickly recovering spin. This seems to be the case for all mass ranges, though as there is a strong dependence on gas fraction with galaxy mass, it would explain why we see little spin-up of merger remnants in high mass galaxies. It could also explain the dependence of spin properties on gas fraction (Figure~\ref{fig:GasAcc}) as only galaxies with sufficient gas mass would recover their spin post merger.

From this we can draw some general trends of how galaxies are affected by major mergers
\renewcommand{\labelenumi}{(\roman{enumi})}
\begin{enumerate}
\item On average major mergers strongly disrupt the spin of elliptical galaxies.
\item Galaxies which are more gas rich post-merger recover a significantly higher degree of spin.
\item Thus low mass galaxies, which are in general gas rich, begin to recover their spin post merger whilst high mass galaxies, which are much more gas poor, do not.
\end{enumerate}

We looked at the relative effect of wet and dry mergers (those for which the incoming galaxies are gas rich or gas poor) and saw very little qualitative difference. \citet{Gomez:2014aa} performed similar analysis on Illustris galaxies, specifically investigating galaxies with $\kappa$ above and below $0.5$ (corresponding by our definitions to spirals and ellipticals, respectively), and found no discernible correlation between wet/dry mergers and galaxy morphology for all but the most massive galaxies. \citet{Lagos:2017} also look at the effects of wet and dry mergers in the EAGLE simulation, finding wet (dry) mergers leading to a spin up (down), though they categorise mergers based on total gas content of the system, not just of the incoming galaxy. There seems to be much more dependence on the gas content of the main progenitor and the merger remnant, than that of the incoming galaxy.

\subsection{Evolution of galaxies without major mergers}

For most galaxies, especially in the latter half of their evolution, major
mergers are rare events and most of their lifetime is dominated by long
periods of slow evolution. As minor mergers have small effects individually,
and are frequent enough to be seen as roughly steady accretion of matter, they
will be treated as part of this gradual evolution. We have shown that a
galaxy's spin is strongly correlated with the amount of gas associated with a
galaxy, and with the amount of stellar mass formed through in-situ star
formation. Thus we want to ask the question of what happens to a galaxy that
is left to its own devices, forming stars and growing steadily in the absence
of major mergers. 

\begin{figure*}
\begin{center}
\includegraphics[height=0.95\columnwidth]{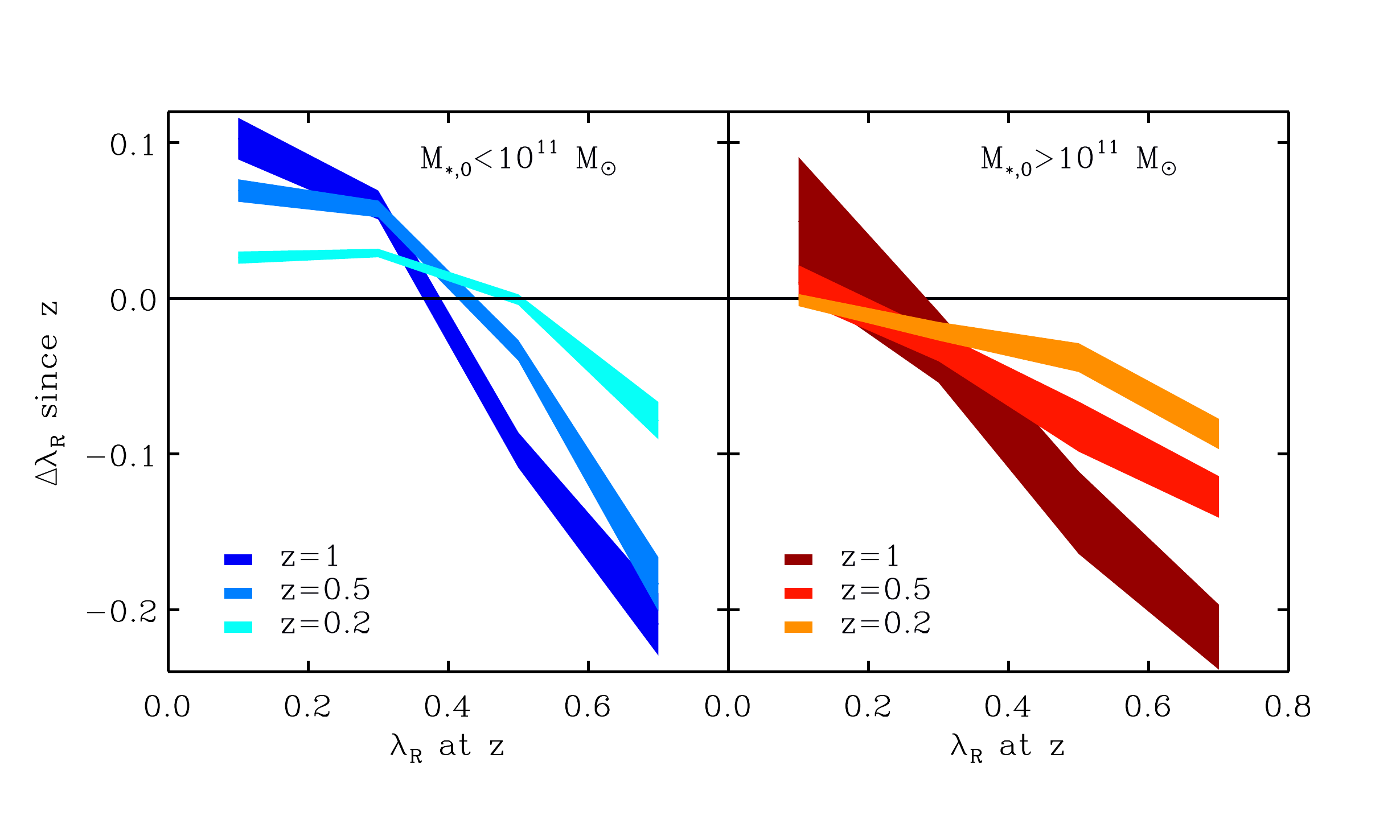}
\vspace{-0.5cm}
\caption{Change in $\lambda_\mathrm{R}$ for galaxies with negligible mass
  accretion from major mergers ($< 1\%$ of $\Delta M_*$) as a function of
  $\lambda_\mathrm{R}(z)$ over a period from a
  given redshift, $z$ (see legend) to $z = 0$.}
\label{fig:MinSituDelta}
\end{center}
\end{figure*}

Figure~\ref{fig:MinSituDelta} shows how $\lambda_\mathrm{R}$ changes over the
period leading up to $z=0$ for galaxies with little or no mass accreted
through major mergers, split between low and high mass galaxies. Roughly half
of the total sample of galaxies have no major mergers since $z=0.5$, with the
fraction being slightly smaller at higher masses. Excluding also those
galaxies with any significant mass gain via minor mergers does not change the
trends seen, but leads to increased noise. 

We see that fast spinning low mass galaxies are spun-down and slow spinning
galaxies are spun-up. This suggests low mass galaxies with no major external
influences will equilibriate to some intermediate value of
$\lambda_\mathrm{R}$. As time goes on this equilibrium value seems to
increase, and close to $z=0$ it is roughly $\lambda_\mathrm{R} = 0.5$, which
coincides well with the locus of points around which low mass galaxies are
clustered in Figure~\ref{fig:Lambda_z0_Mass}.

In contrast high mass galaxies which are initially spinning slowly are not
spun-up. Even those galaxies with the lowest $\lambda_\mathrm{R}$ show no
significant increase in their spin after $z = 1$, thus there is no similar equilibrium spin (or looked at from another perspective, the equilibrium spin value is $\lambda_\mathrm{R} \rightarrow 0$). Fast spinning massive galaxies are still naturally spun-down over time.

This behaviour could be key to the two separate populations of FRs and SRs. If
more massive galaxies tend not to be spun-up even when they are spinning
slowly, infrequent major mergers would still be effective at reducing their
spin by a large amount over time. In contrast lower mass galaxies, even if
they undergo a cataclysmic event, could then recover that spin in the long
periods between major mergers. 

\begin{figure*}
\begin{center}
\includegraphics[height=0.85\columnwidth]{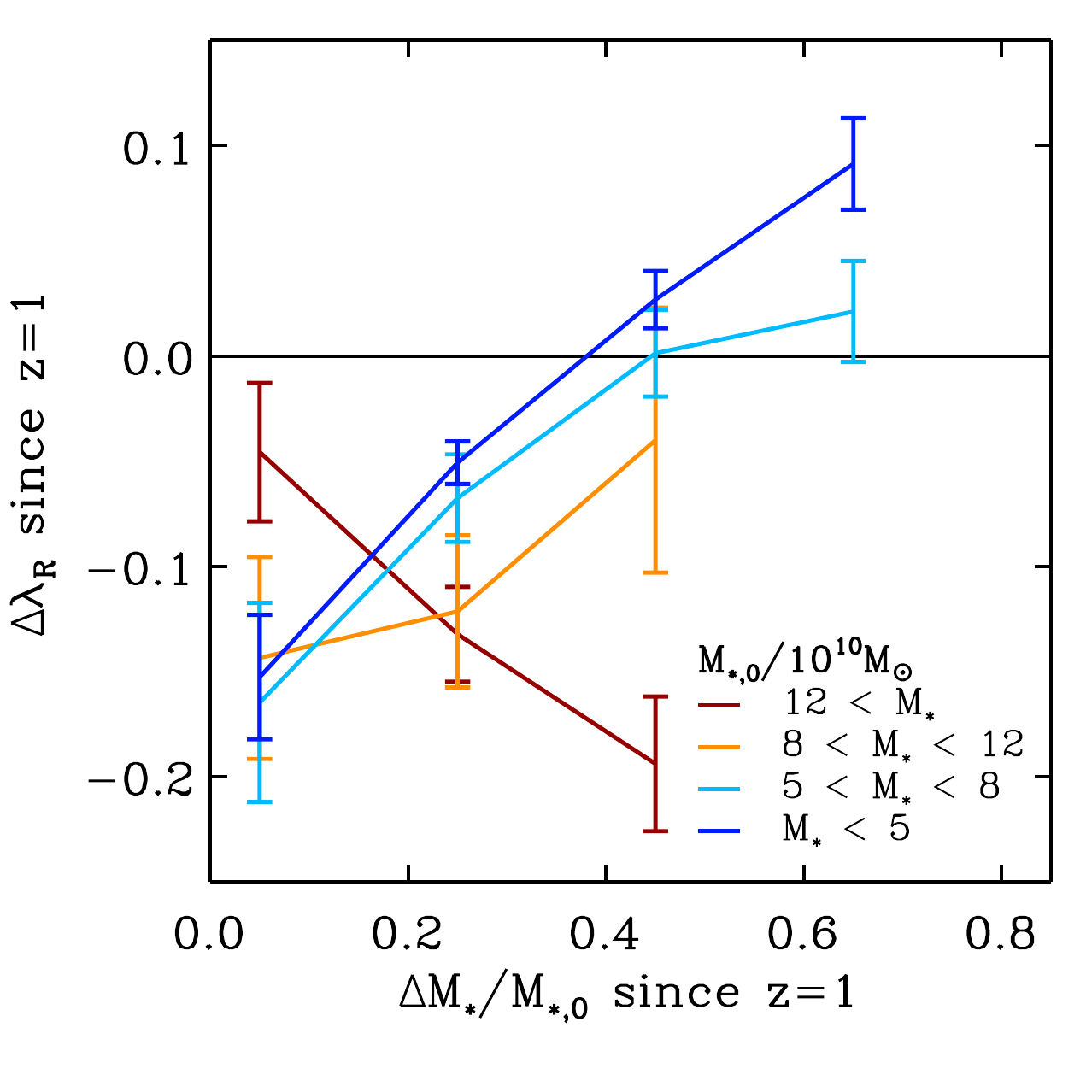}
\includegraphics[height=0.85\columnwidth]{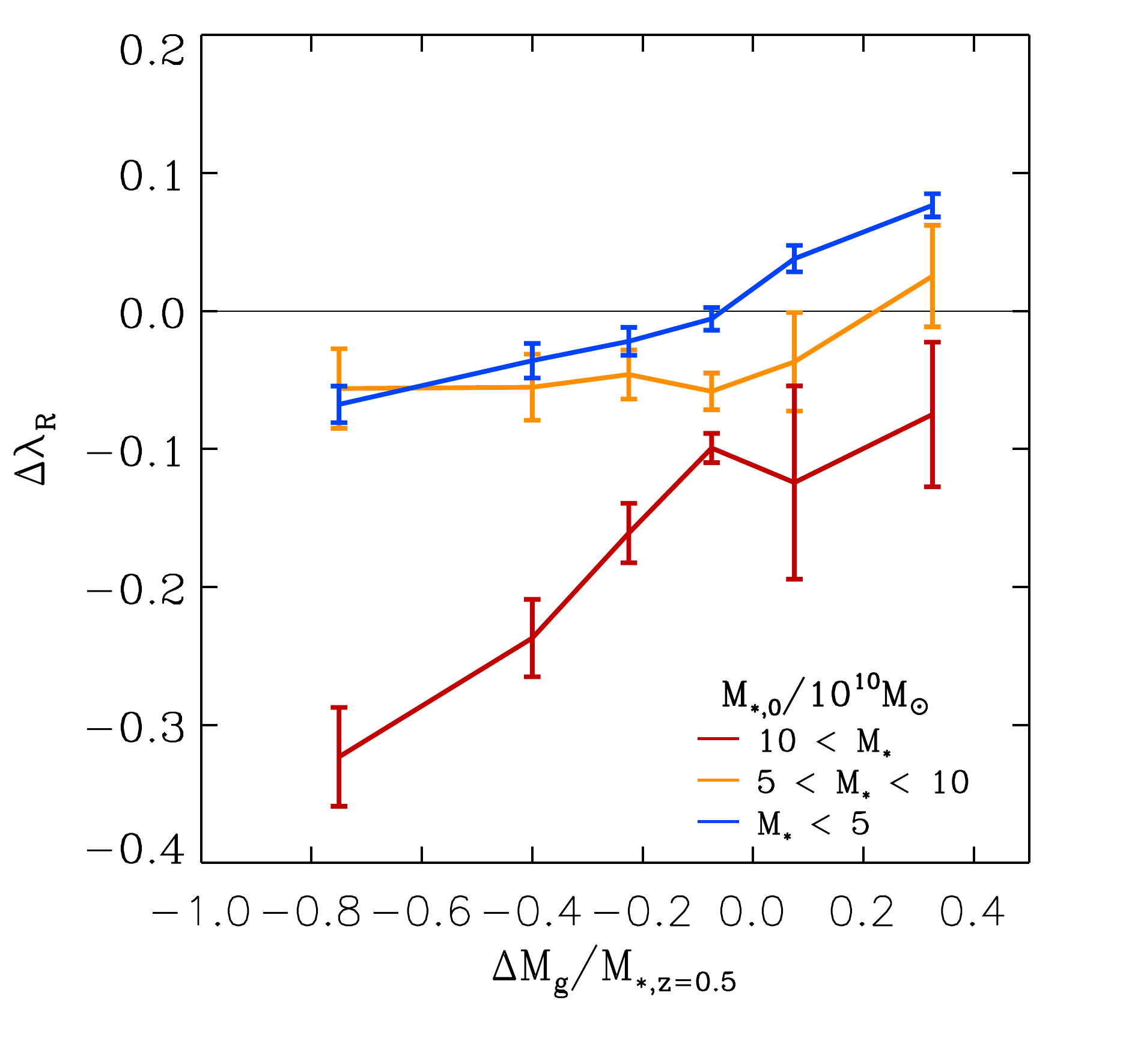}
\vspace{-0.5cm}
\caption{Left: Change in $\lambda_\mathrm{R}$ for galaxies as a function of their increase in
  stellar mass since $z = 1$, for galaxies with negligible major
  mergers. Right: Change in $\lambda_\mathrm{R}$ since $z = 0.5$ as a function
  of a galaxy's gas inflow and outflow in that period. Shown only for galaxies
  with no major mergers in this period. Change in gas mass includes gas used
  in star formation as well as galactic inflow and outflow, and is normalised
  by the stellar mass at $z=0.5$.}
\label{fig:MinSituDelta_Mass}
\end{center}
\end{figure*}

Left-hand panel of Figure~\ref{fig:MinSituDelta_Mass} shows how the spin
properties change, for galaxies in different mass bins, dependent on the
change in their stellar mass. As galaxies with major mergers are excluded, the
main avenue for these galaxies to gain mass is through minor mergers and
in-situ star formation (as before excluding galaxies with significant mass
change through minor mergers gives behaviour that is qualitatively the same,
but with much more noise in the data). 

For low and intermediate masses we see that there is a tendency for galaxies
with little or no stellar mass gain to spin-down. For high mass galaxies this
seems to be less effective, possibly because they are already slow spinning
at $z = 1$ and hence there is a minimum possible drop in
$\lambda_\mathrm{R}$. For lower mass galaxies, the more mass gained the more
the galaxy is spun-up. This is because the stellar mass gain is associated
mostly with in-situ star formation from accreting gas, which has a significant
angular momentum. 

As we move to higher mass bins the spin-up process seems to be less effective,
and for the most massive galaxies the trend is reversed: the more stellar mass
gained the more the spin is disrupted. One could interpret this by saying that
above a critical mass we transition to stellar mass growth being more
associated with minor mergers than in-situ star formation, though this would
be in contrast with the absence of any dependence on minor mergers in
Figures~\ref{fig:MassAcc} and \ref{fig:LambdaAcc}, unless many very small merging galaxies are unnoticed by the halo finder. Instead, more likely it
could be due to a change in the geometry of accretion, with incoming gas no
longer adding angular momentum coherent with the galaxies' spin, either due to
change in the local environment due to the galaxy itself (such as through
heating via AGN) or due to the galaxy's location, as higher mass galaxies tend
to sit in galaxy groups and clusters. 

Finally we compare the change in spin due to gas accretion and expulsion
(excluding mergers). Without following the flow of gas in individual cells
this is an inexact art, but we can estimate the mass of gas inflowing and
outflowing, $\Delta M_\mathrm{g}$, using $\Delta M_\mathrm{g} = M_\mathrm{g,z}
- M_\mathrm{g,0} - \Sigma \Delta M_\mathrm{g,mergers} + \Delta
M_\mathrm{*,situ}.$ This assumes that the vast majority of gas that forms
stars is external to a galaxy. Obviously this can be negative when outflow
driven by feedback blows more mass outward in this period than the galaxy is
able to accrete.

Right-hand panel of Figure~\ref{fig:MinSituDelta_Mass} shows the change in
spin of galaxies as a 
function on their ability to accrete or blow-out gas between $z = 0.5$ and $z
= 0$. Only amongst low mass galaxies inflow can dominate over outflow, and for
those the inflowing gas is linked to a spin-up. High mass galaxies tend to
expel significantly more gas than they accrete, and whilst they show similar
trends they lose much more spin in the same time period compared to their
lower mass counterparts. 

For all galaxies high level of feedback and gas
expulsion lead to a spin down as the outflowing gas takes with it angular
momentum, and tends to redistribute the angular momentum (hence disrupt the
ordered rotation) of star particles. Though net gas inflow/outflow may be
small, galaxies may still experience strong feedback and rapid accretion,
roughly cancelling each other out. Thus it is possible that high mass galaxies
are more disrupted because they are all experiencing large amounts of feedback
but replenishing some mass via accretion.

We suggest this shift in behaviour for galaxies above $\sim 10^{11} M_\odot$
to be either due to environment or feedback. These massive galaxies are less
likely to be found in cosmic filaments and more likely in clusters, leading to
a different, more isotropic, geometry of accretion. Equally, at around this
mass the dominant mode of feedback shifts from that associated with star
formation to AGN, which is a much more violent feedback mechanism. The shift
could also be due to the rising level of minor mergers, but the lack of
dependence shown earlier dissuades this idea.

Thus for a typical galaxy left to evolve gradually under accretion of gas we
see the following trends:
\renewcommand{\labelenumi}{(\roman{enumi})}
\begin{enumerate}
\item Fast spinning galaxies, of all masses, naturally spin-down over time.
\item Low mass slow spinning galaxies spin-up over time, leading to them equilibrating. High mass galaxies do not show such recovery and hence never regain their spin.
\item Gas accretion and associated star formation are linked to a spin-up of a galaxy, but in high mass galaxies, mass growth leads to a spin down.
\end{enumerate}
The qualitative results remain unchanged when examining separately those galaxies with the negligible mass gain through minor mergers, and those for which minor mergers are dominant.

\section{Conclusions and Discussion}
Using the cosmological hydrodynamic simulation Illustris, we have followed
thousands of massive elliptical galaxies back across cosmic time to understand
and explain how their rotation properties developed. Separating present day
ellipticals into fast and slow rotators (FRs and SRs), we sought to explain
how some massive galaxies maintain ordered motion and disk-like rotation,
whilst others lose order in their stellar orbits, developing complex
velocity fields dominated by dispersion. 

\subsection{Present day properties of fast and slow rotators}
We first construct a small subset of Illustris galaxies to match the
observed $\text{ATLAS}^{\text{3D}}$ sample of elliptical galaxies (E+11), finding
very similar fraction of SRs ($\sim 14\%$), which are all mostly massive
galaxies, while FRs are predominantly low mass galaxies ($M_{rm *} < 10^{11}
M_{\rm \odot}$), as found in observations as well. If we extend our analysis
to all well-resolved elliptical galaxies in Illustris (with at least 20,000 star
particles, or $M_{\rm *} \gtrsim 10^{10.5} M_{\rm \odot}$), the fraction of SRs stays
similar, indicating that we have a representative sample of SRs and FRs. While
qualitative agreement with the $\text{ATLAS}^{\text{3D}}$ survey is very
encouraging, we find a discrepancy in the ellipticities of SRs, which appear
more elongated in Illustris than in the observed sample. Similar
behaviour was seen in previous work, such as N+14 and \citet{Bois:2010aa},
which concluded that too gas poor mergers can lead to galaxies with
artificially high ellipticities.  

By separating galaxies into three mass bins, we see two clear loci around which galaxies cluster. Low-mass galaxies are a smoothly distributed population of FRs (centred on the same point as spiral galaxies would be, though with more dispersion). High-mass galaxies are mostly, but not all, part of a wholly separate distribution of SRs. Intermediate mass galaxies ($10^{11} M_\odot < M_* < 10^{11.5} M_\odot$) can fall at either locus or be smoothly distributed between them.

Looking back to earlier redshifts, the distribution of ellipticals is largely
unaltered: there are fewer and fewer high-mass galaxies and correspondingly
fewer SRs, but low- and intermediate-mass galaxies tend to lie in a similar
distribution to the one at $z = 0$, possibly with a slight shift to higher
ellipticities. The population of host galaxies and satellites also appears to
be congruent, again with the exception that there are very few high-mass
satellites and hence the vast majority of SRs are host galaxies. 

Whilst we mostly focus on a kinematic classification of fast and slow
rotators, we also examine various other characteristics that separate the two
populations. We test for central cores and cusps, boxy and disky isodensity
contours, and X-ray emission from hot gas. High-mass ellipticals obey known
observational relationships well, particularly their X-ray luminosity, whilst
lower-mass galaxies limited by resolution have much less clear trends. 

We further examine how other galaxy properties, such as stellar mass, gas
fraction, metallicity and colour, relate to kinematic properties. The fraction
of gas, compared to stellar mass, decreases almost uniformly with decreasing
spin, with star formation rate and colour closely coupled to this
trend. Stellar mass increases sharply as we transition from fast to slow
rotators, and there is also a possible gradient from low-mass spheroidal FRs
to higher-mass elongated FRs. The mass gradient is mirrored in the stellar
half-mass radii of these galaxies. SRs also have higher metallicity, both in
stars and gas. Fast-spinning spheroidal FRs are also particularly enriched,
and we suggest this may be due to starbursts during mergers that could have
also spun them up.

\subsection{Redshift evolution of fast and slow rotators}
To understand how evolutionary history impacts on the present day kinematical
properties of galaxies we have followed the first progenitor of $z = 0$
galaxies back in time, finding that there is little difference at $z = 1$
between galaxies that will become FRs and SRs and no discernible difference at
higher redshift. A divergence in the populations is apparent from $z = 0.25$,
leading us to conclude that it is evolution in the latter part of a galaxy's
life, during which the major differentiating factor is their stochastic merger
histories, that determines an elliptical galaxy's rotation properties. 

We can separate the mechanisms acting on an elliptical galaxy in this late
stage of its life and the effect they have as follows:
\renewcommand{\labelenumi}{(\alph{enumi})}
\begin{enumerate}
\item Major Mergers - we find that in general, across all mass ranges and
  regardless of gas fraction, an incoming massive galaxy tends to lead to a
  huge disruption and a slower spinning merger remnant \citep[see
    also e.g.][]{Barnes:1996aa, Cox:2006aa, Jesseit:2009aa,
    2009ApJ...705..920H}. That said, rare major mergers, potentially with
  specific geometries for the collision, can lead to an increase in spin
  \citep[see also][]{2006ApJ...636L..81N, 2009A&A...501L...9D, Bois:2010aa,
    Bois:2011aa}. The fate of a galaxy undergoing such a collision is in large
  part dependent on factors relating to encounter geometries, which we do not
  explore in this paper.  

Even after the more common highly disruptive mergers, galaxies may recover
their spin. Gas-rich galaxies, or those in a gas-rich environment, which still
have a significant gas mass directly after a merger, tend to spin back
up. This means that galaxies can undergo major mergers and remain fast
rotators. Galaxies that do not recover their spin after the merger can, by
either a single merger or repeated encounters, become completely disrupted and
end up as slow rotators. 

\item Minor Mergers - In contrast to some previous works
  \citep{2007A&A...476.1179B, 2010A&A...515A..11Q} we have found little or no
  dependence on minor mergers for determining the spin of an elliptical
  galaxy. Minor mergers account for only a small fraction of the mass growth
  of a galaxy since $z = 1$, independent of the galaxy's spin. The only strong
  trend seen is that minor mergers which bring in large amounts of gas to low-
  and intermediate-mass galaxies are associated with a spin-up and may be the
  origin of some of the fastest-spinning galaxies.

This may, however, depend on the scale of incoming body that we term a minor
merger. What we see as part of the steady accretion of material may actually
be a bombardment by bound objects so small as to not be identifiable to our
halo finders. Thus, some of the results outlined below, for steady accretion
of gas and stars, may by another nomenclature be included as an effect of
minor mergers.

 Early results from the MaNGA survey show no discernable
  differences between the $\lambda_\mathrm{R}, \epsilon$ distribution for central and satellite galaxies, Greene et al. (in prep.), and they suggest that
  this shows the amount of minor mergers (which is strongly dependent on environment) does not have a
  large impact on spin evolution.

\item Accretion of stars and gas - For low-mass galaxies, accreting gas and
  forming stars leads to a spin up, most likely from the incoming angular
  momentum of accreted material adding coherently to the spin of the
  galaxy. In higher-mass galaxies, this ceases to be the case; the rate of gas
  inflow is smaller and tends to be less than the feedback-driven outflow, and
  new stars forming and accreting seem to disrupt the galaxy's spin.

We suggest two plausible mechanisms for this dichotomy. The transition could
be environmental, with higher-mass galaxies more likely to sit at the centres
of gas-poor clusters. Thus only a small amount of gas is inflowing, and
possibly coming in at such an angle that it diminishes the net angular
momentum of the galaxy. The transition could also be dependent on the type of
feedback, as at these higher masses AGN feedback becomes dominant, and this
more violent feedback may disrupt the galaxy further or prevent incoming gas
transferring angular momentum to the galaxy effectively. Differentiating
between these two situations is beyond the scope of this work but provides an
interesting avenue for further study.
\end{enumerate}

\subsection{The origin of fast and slow rotators}
Putting together all of our results we can attempt to explain the the origin
of fast and slow rotators.\\

\textbf{Fast Rotators:} Fast-rotating elliptical galaxies are mostly lower-mass galaxies with a plentiful supply of gas that even after major mergers are able to recover their spin and tend to equilibrate to a certain degree of spin ($\lambda_\mathrm{R} \approx 0.5$). 

High levels of gas inflow can spin them up, and over periods with little
accretion they will spin down. Across the population, accretion rates do not
vary greatly, and they tend to share similar rotation properties. Major
mergers tend to disrupt their spin, though in rare fortuitous cases can lead
to faster rotation. 

In terms of spin and mass, their properties are close to those of spiral
galaxies, which sit tightly distributed around $(\lambda_\mathrm{R},\epsilon)
= (0.5,0.55)$, suggesting that FRs are part of the natural evolution from
lower mass spirals and all ellipticals may have spent part of their life as
FRs.

\textbf{Slow Rotators:} Slow-rotating elliptical galaxies are older and more
massive. They seem to have evolved from fast rotators and done so between $z =
1$ and $z = 0$. They are generally gas-poor, with most of their incoming mass
coming from mergers, and tend only to lose spin as they
grow, without recovering it.

The key divide between FRs and SRs is that the latter are not spun up by
accreting gas and stars. Even in periods without major mergers, the
accumulation of new stellar mass causes them to lose spin, and they expel gas
more quickly than it can accrete. This shift, in how the galaxy reacts to
incoming mass, may be either environmental, with changes in the local temperature, abundance and geometry of accreting gas, or it may be intrinsic, with the shift towards powerful AGN feedback preventing the transfer of angular momentum from accreting material by heating inflowing gas and disrupting or diverting its accretion. Either way, slow rotators tend to atrophy in their old age, at best retaining what spin they have, at worst slowly spinning down.

Whilst SRs can be created without mergers, these more massive galaxies also
have the highest frequency of major mergers. This repeated bombardment lowers
their degree of ordered rotation in large steps. Lower mass galaxies which
suffer uncommonly frequent or disruptive mergers can also lose spin faster
than they recover it, accounting for a small population of lower-mass SRs
(though, as stated above, these may soon regain their spin and become FRs
afresh). Major mergers may infrequently cause a spin-up, leading to high-mass
galaxies, which undergo the most mergers, having the highest spread of spin
properties. Eventually, however, as a FR grows in mass, it loses most of its
spin, whether by degree or in great bounds, and settles as a
near-completely-disordered SR.

Finally, we can compare our results with those from zoom-in simulations, such as N+14 who presented the
previous most in-depth analysis of FR and SR evolution, and with whom we find
in general excellent agreement. They present detailed histories of a small
number of galaxies and present 6 evolutionary pathways (3 each) for fast and
slow rotators. All 6 paths fit within our much broader scheme, though some are
shown to be much less frequent on a population scale (such as rare major
mergers leading to a spin-up). The only major difference between their results
and ours is that they found strong dependence on whether mergers were gas rich
or poor. We find this has little effect on a galaxy's spin, but instead that
the amount of gas the main progenitor galaxy has and retains before and after
the merger has a large effect. Where they present detailed histories of a few
specific galaxies, we show how these extend to a general cosmological
population. We can aslo compare to recent work by \citet{Choi:17} who find that galaxies spin down in the absence of mergers, in good agreement with our work, and that major and minor mergers have a major contribution to the spin down of massive galaxies, though they find a stronger cumulative effect of minor mergers than in our work.

Currently large IFU galaxy surveys such as $\textrm{ATLAS}^\textrm{3D}$,
CALIFA, MaNGA and SAMI are starting to provide comprehensive galaxy samples
with a wealth of spatially resolved, kinematical properties. By carefully
comparing large-scale hydrodynamical simulations, such as Illustris, with these
unique datasets, as we have done in this paper, we can gain new insights
into the intricate process of galaxy formation and assembly, which will allow
us to build a more complete theoretical picture of how galaxies grow and evolve.

\section*{Acknowledgements}
We would like to thank Eric Emsellem, Jenny Greene and Vicente Rodriguez-Gomez
for their excellent thoughts, comments and questions. BPM acknowledges support
from the Kavli Foundation and the German Science Foundation (DFG) for an Emmy
Noether grant. The Flatiron Institute is supported by the Simons
Foundation. DS acknowledges support by the STFC and the ERC Starting Grant
638707 ``Black holes and their host galaxies: co-evolution across cosmic
time''.





\appendix

\section{Comparison of methods for classifying ellipticals}
\label{ap:Spirals}
We use the $\kappa$ parameter (Equation~\ref{kappa}), as detailed in \citet{Sales21062012}, hereafter S+12, to separate the populations of spiral and elliptical galaxies based on their kinematics. While it performs well for high mass galaxies, well matching observations \citep{Conselice21122006} below $M_* \approx 10^{10.5} M_\odot$ it finds an unrealistically high population of bulge dominated galaxies.

Optical examination of a handful of these galaxies still shows strong
disk-like features down to $\sim 10^{10} \, {\rm M_\odot}$, but with a much
more bulge dominated structure than we would expect to see for most galaxies
at this scale. Below this mass it is hard to derive, by visual examination,
clear structure or shape. Similar results were seen in Illustris by
\citet{2015arXiv150207747S} and \citet{Gomez:2014aa}, where low mass galaxies are bulge dominated by
kinematic measures, but have SFRs and disc properties characteristic of spiral
galaxies. Dividing the population of galaxies by morphology gives a population
of spirals and ellipticals that better fits with observational data, even
at low mass ranges. Thus we also explore a range of other metrics by which to separate spirals and ellipticals.

S+12 suggests a second method to  to determine the degree of rotation of a galaxy, and therefore determine if it is a spiral or an elliptical galaxy which we test here to compare its predictions and justify our chosen methods.

The authors use the distribution of the $circularity$ parameter, $\varepsilon_i = j_{z,i}/j_{c,i}(E_i)$, for each star particle's orbit. We define $j_{c,i}(E_i)$ as the angular momentum of a star on the circular orbit which shares the same binding energy, $E_i=K_i + \Phi(r_i)$.

This is a measure of how close each star particle is to a circular orbit, as we would expect in the disk of a spiral galaxy with $\varepsilon_i$ has a maximum value of $1$. For an orbit completely out of alignment with the galaxy's total angular momentum, or with a very eccentric orbit, as we might expect in the bulge or in an elliptical galaxy $\varepsilon_i$ will tend to $0$.

Hence, by binning star particles by radius and averaging over their individual potentials, we find the circular radius $r_{c}$ such that the energy of the star particle $$E_i = K_{c,i} + \Phi(r_{c,i}) = -r_{c,i}\left.\dfrac{d\Phi}{dr}\right|_{r_{c,i}} + \Phi(r_{c,i})$$ which we then use to find $j_{c,i}=m_i\, r_{c,i}\, v_{c,i}(r_{c,i})$.

If we look at the distribution of $\varepsilon_i$ for all stars in the galaxy we see clear distinctions between bulge and disk dominated galaxies. To then classify galaxies as spirals or ellipticals we quantify the number of stars in  the bulge and the disk (this is approximately equal to the mass of stars in each as every star particle in Illustris has similar mass). We look at two different measures, $f(\varepsilon>0.5)$, the fraction of stars with $\varepsilon_i$ greater than $0.5$, i.e. the fraction of the stars in the disk, and $1-2f(\varepsilon<0)$, which quantifies the fraction of stars not in the bulge (relying on the fact that the distribution of stars in the bulge should be symmetrical around $\varepsilon=0$).

\begin{figure}
\begin{center}
\includegraphics[width=0.95\columnwidth]{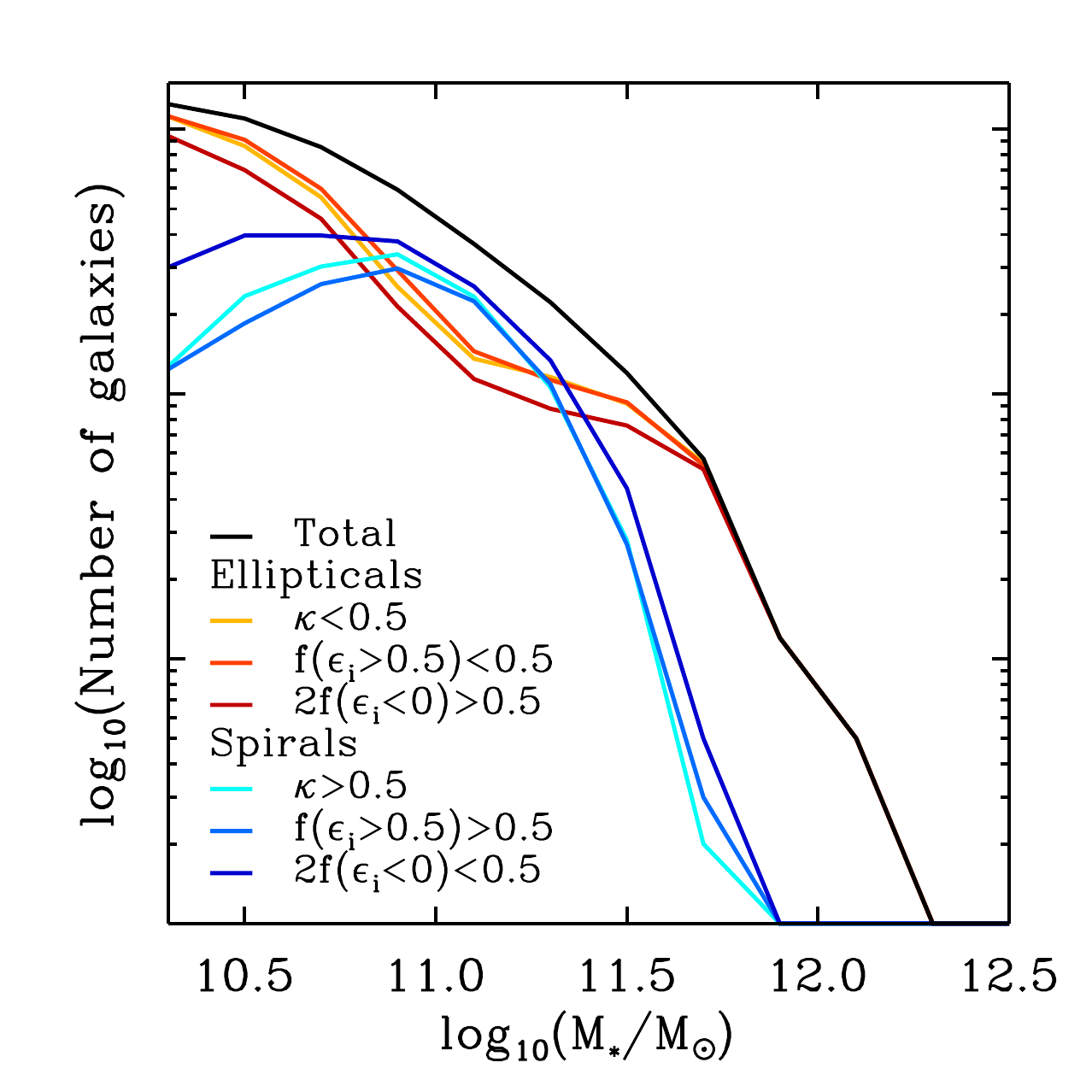}
\caption{Stellar mass function of Illustris galaxies, separated by galaxy type based on three different measures of galaxy morphology (see text). The $\kappa$ parameter is the measure we use to seperate spirals and ellipticals throughout the rest of our analysis. It can be seen it agrees well with the other methods tested.}
\label{fig:GalaxyPopulation_All}
\end{center}
\end{figure}

We take the cutoff for both of these measures to be the same as for $\kappa$. For $f(\varepsilon>0.5)$ or $1-2f(\varepsilon<0)$ greater than $0.5$ we take the galaxy to be a spiral, for values less than $0.5$ we conclude the galaxy is elliptical. In Figure~\ref{fig:GalaxyPopulation_All} is shown the stellar mass function, separated by type of galaxy as concluded by each method.

The three methods, whilst not in exact concurrence, are in good agreement. Thus to separate spirals and ellipticals in our sample we use the $\kappa$ parameter, both because it gives the intermediate values for fraction of spirals and ellipticals, and because it is more reliable and less computationally expensive when computed for a very large number of galaxies.

\begin{figure}
\begin{center}
\includegraphics[width=0.95\columnwidth]{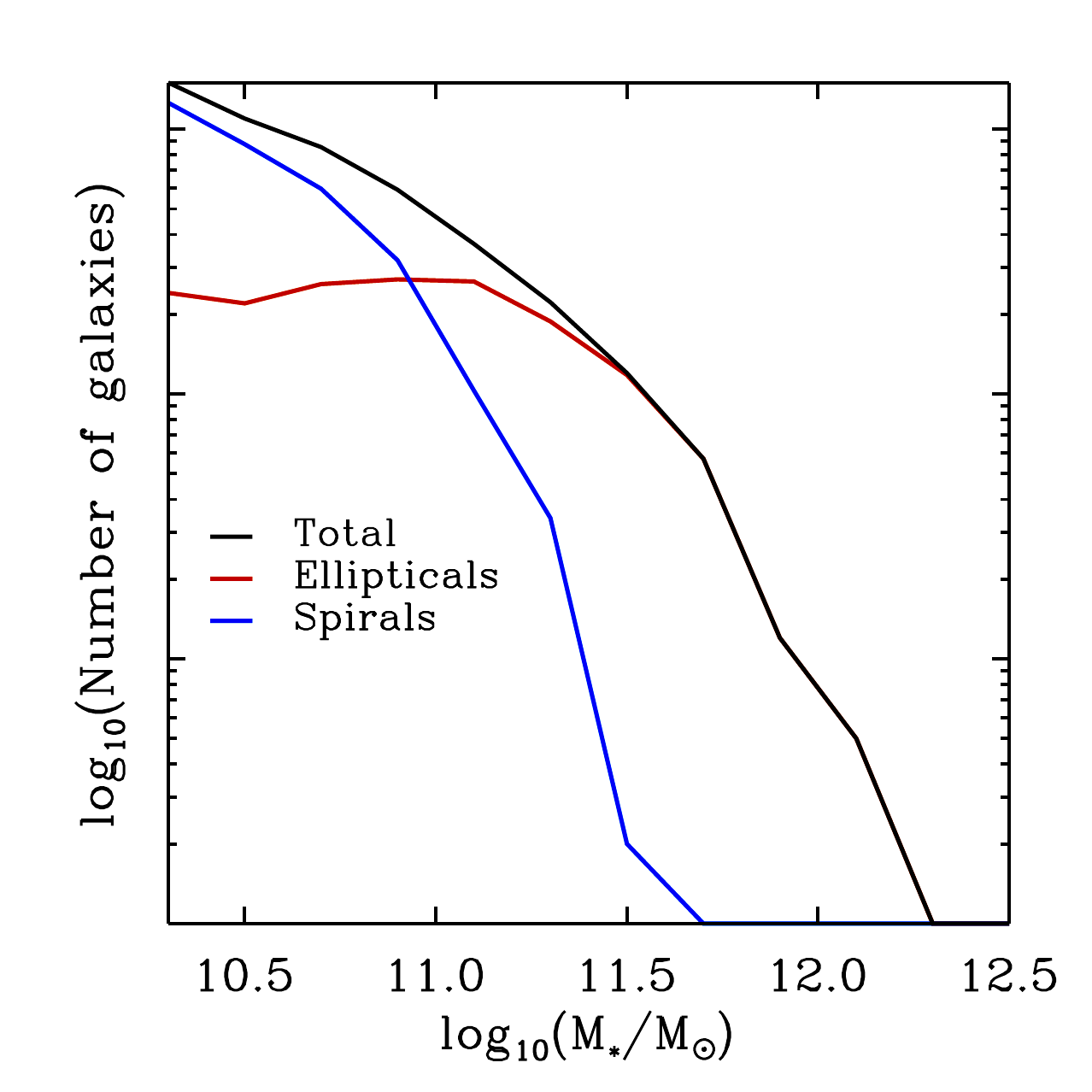}
\caption{Similar to Figure~\ref{fig:GalaxyPopulation_All} where here galaxies are split based on the cutoff in $g-r$ band luminosity from equation \ref{eq:g-r}.}
\label{fig:GalaxyPopulation_Alt}
\end{center}
\end{figure}

\begin{figure*}
\begin{center}
\includegraphics[width=0.95\textwidth]{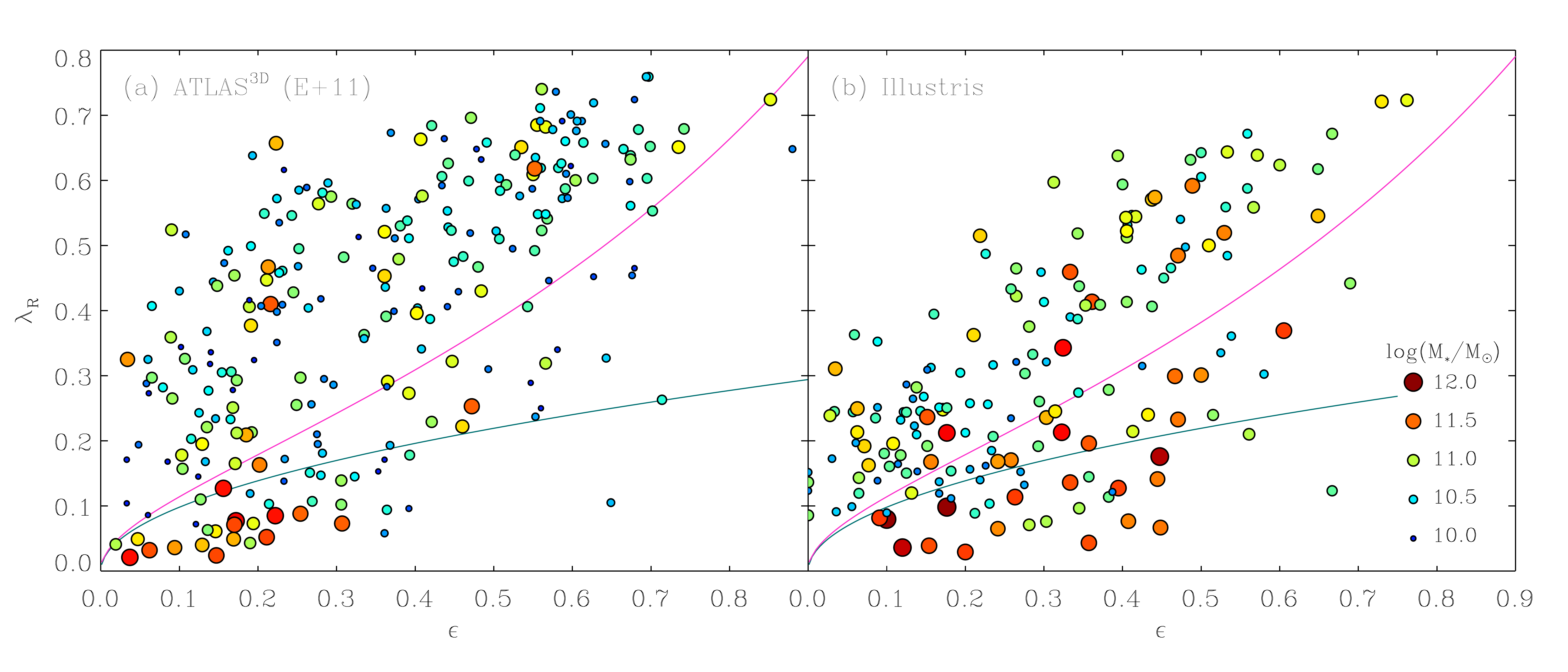}
\caption{Distribution of early-type galaxies in the $\epsilon$- $\lambda_\mathrm{R}$
  plane, as in Figure~\ref{fig:LambdaEms}, but using $g-r$ luminosity, rather than $\kappa$ as the cutoff between spirals and ellipticals. We find a smaller number of ellipticals by the same selection criteria, roughly half those in the $\text{ATLAS}^{\text{3D}}$ survey, and a similar distribution, with a slightly higher fraction of highly elongated or fast spinning galaxies compared to that found with a $\kappa$ cutoff.}
\label{fig:LambdaEms_Alt}
\end{center}
\end{figure*}

\begin{figure*}
\begin{center}
\includegraphics[width=1.0\textwidth]{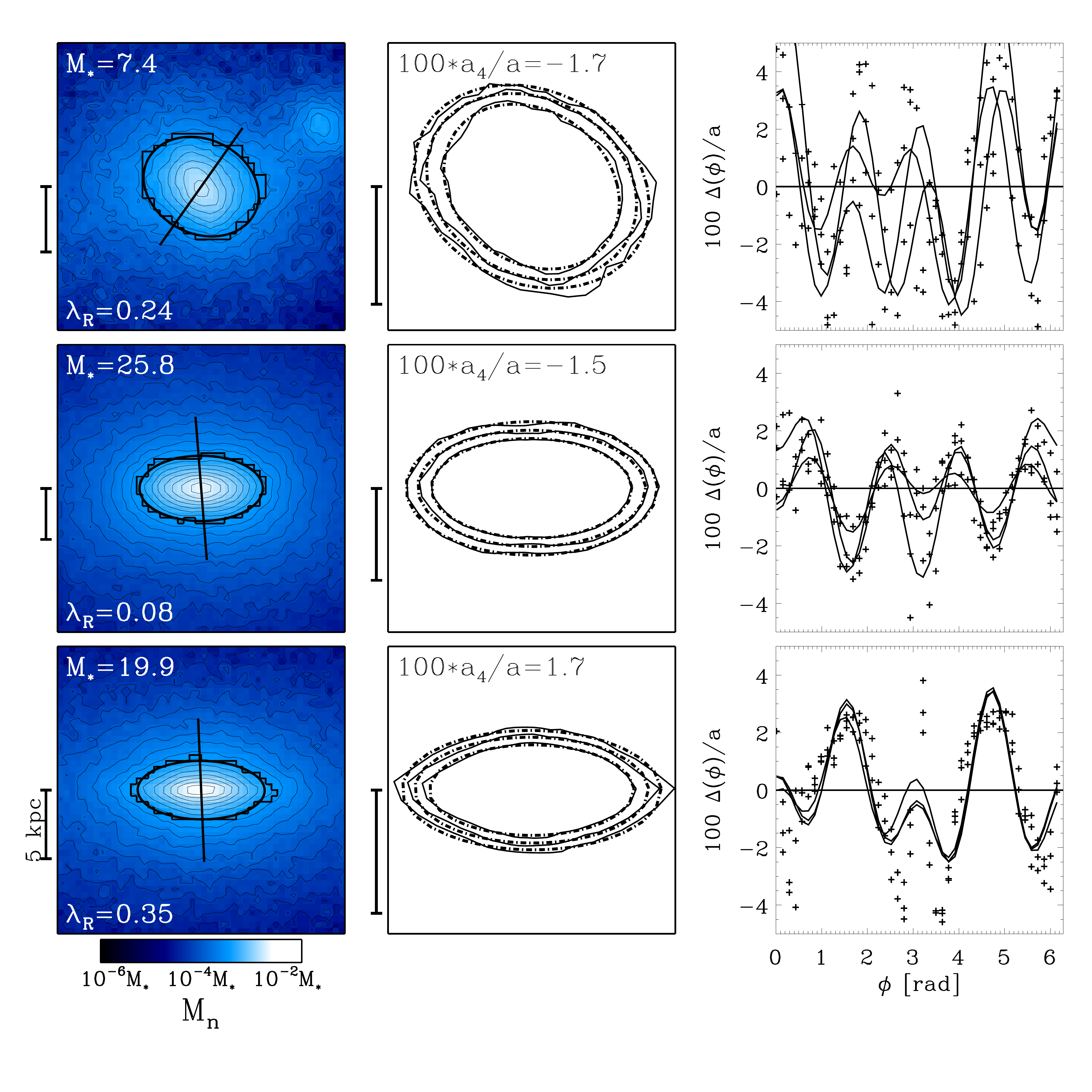}
\caption{Measuring the $a_4$ parameter for three Illustris galaxies. On the left we show projected stellar density (as in Figure~\ref{fig:GalPlots}) with a number of isodensity contours, thin black lines, added. The next panel shows a magnified image of the isodenisty contours, thin black line, and fitted ellipses, thick dashed line, for the half mass contour and the contours with 0.6 and 0.8 times the area of the half mass contour. The rightmost panel shows the residuals of the contour from the fitted ellipse, normalised by the major axis of each ellipse, as black crosses, and the fitted Fourier expansion, as a thin black line. All scale bars are 5kpc in length.}
\label{fig:GalPlot_Contour}
\end{center}
\end{figure*}

We also present numerical results from our tests using $g-r$ band luminosity as the divide between spiral and elliptical galaxies, with a cut defined by equation \ref{eq:g-r} (Figures~\ref{fig:LambdaEms_Alt} and \ref{fig:GalaxyPopulation_Alt}).

We also experimented with making a cutoff between star forming and quiescent
galaxies based on their magnitude in the $g$ and $r$ bands. Galaxies for which
$g-r$ is large are redder with little star formation and these quiescent
galaxies are often ellipticals, whilst smaller $g-r$ colours characterise
bluer, star forming galaxies, which in observational surveys correspond well
with the population of spirals \citep{Kormendy:2016aa}.

We define a cut between star forming galaxies as those below the line
\begin{equation}
    \label{eq:g-r}
    \log_{10}(g-r) = 0.1 \log_{10}(M_*) - 0.5\,,
\end{equation}
and quiescent galaxies as those above, motivated by the distribution of the
whole sample of galaxies in Illustris \citep{Vogelsberger:2014aa}. The
luminosities are found by summing the individual luminosity of
all stellar particles in the galaxy. The fraction of quiescent to star
forming galaxies, by mass, is in line with observations
\citep{2013ApJ...777...18M} over the whole mass range examined (down to halos
with stellar mass of $\sim 10^{10} \, {\rm M_\odot}$).

Figure~\ref{fig:LambdaEms_Alt} is a comparison with the galaxy population found in the $\text{ATLAS}^{\text{3D}}$ survey, as presented in Section~\ref{sec:ATLAS3D} using a cutoff defined by $\kappa$. We find less galaxies classified as ellipticals in the same volume, roughly half the previous number, with a wider spread in the distribution. There are more, faster spinning galaxies included though still few as rapidly rotating as in the $\text{ATLAS}^{\text{3D}}$ survey. The population of high mass galaxies is essentially unchanged, still with particularly elongated high mass SRs as mentioned in Section~\ref{sec:ATLAS3D}.

 Figure~\ref{fig:GalaxyPopulation_Alt} shows the stellar mass function derived using a colour cut. Whilst the results are more in line with the observed stellar mass function at low masses we believe this is due to the dependence of both on galaxy mass, rather than any strong link between luminosity and kinematic behaviour. \citet{Bottrell:2017} performed similar analysis on carefully constructed mock images of Illustris galaxies and found a persistent disagreement with kinematic and photometric measures of bulge and disk fractions at stellar masses lower that $10^{11} M_\odot$.

However the $g-r$ band is sensitive to dust and to the radius out to which luminosity is measured. The $g-r$ luminosity is strongly dependent on the mass, as is the stellar mass function, and hence we see our $g-r$ classification giving the expected fractions of spirals and ellipticals. However $\kappa$ better represents the morphology and kinematics of the individual galaxies and it is the measure we use throughout the rest of this work.

\section{Fitting isodensity contours to find boxy and disky ellipticals}
\label{ap:Contour}

We seek to expand the Fourier series of the residuals of a galaxy's isodensity contours (equation \ref{eqn:Fourier}) to find the sign and magnitude of the $a_4$ co-efficient. This tells us if the elliptical is boxy ($a_4 < 0$) or disky ($a_4 > 0$). The fitting is done by a non-linear least squares fit using a Levenburg-Marquadt algorithm, giving equal weight to each residual.

For each galaxy we perform this analysis for 3 contours with areas
equal to 0.6, 0.8 and 1.0 times the area of projected half-mass area of the
galaxy and average the results. An example of three such contours is shown in
Figure~\ref{fig:GalPlot_Contour}, from top to bottom we have a a poorly resolved galaxy classified
as a boxy elliptical, a clear boxy elliptical, and a clear disky
elliptical. The deviations from an ellipse are very slight and thus
noise introduced by substructure in the halo, binning into pixels and random
variations can lead to significant variation. The resolution needed to
accurately measure $a_4$ is significantly higher than our cutoff for
well-resolved galaxies, appearing to only be well suited for galaxies above
$10^{11}\, {\rm M_\odot}$ or higher.

\section{Characterising galaxies with a constant degree of rotational support}
\label{ap:Magenta}
Here we reproduce the derivation, presented in \citet{Cappellari:2007aa} and E+11, of an approximate relationship between $\lambda_\mathrm{R}$ and $\epsilon$ for galaxies with a constant degree of rotational support.

If the majority of stars are on circular orbits, i.e. a high degree of rotational support, we can predict, given the anisotropy of those orbits, the shape of the galaxy. Assuming some constant relation between the anisotropy ($\delta$) and $\epsilon$, $\delta = c \epsilon$, we can divide galaxies with a high degree of rotational support will have a high $c$ (tending to 1) and those with more radial less ordered orbits will have $c$ tending to 0.

Thus we can compare FRs and SRs via their rotational support, over a range of ellipticities, by finding the form of the relationship $\delta = c \epsilon$ for a given constant $c$ and converting it into a relationship between $\epsilon$ and $\lambda_\mathrm{R}$.

The global anisotropy parameter for an edge on system
\citep{2005MNRAS.363..937B} is
\begin{equation}
\label{delta}
    \delta=1-\dfrac{1+\left(V/\sigma)\right)^2}{\Omega \left(1-\alpha \left(V/\sigma)\right)^2\right)}\,,
\end{equation}
where $\alpha$ is a dimensionless constant dependent on the density profile, taken as $\sim 0.15$ from comparison to isotropic models, 
\begin{equation}
\left(\dfrac{V}{\sigma}\right)^2 = \dfrac{\langle V^2 \rangle}{\langle \sigma^2 \rangle}\,,
\end{equation}
and $\Omega$ is a measure of the ratio of potential of the system in the plane of the galaxy compared to perpendicular to the plane and is given by 
\begin{equation}
\label{omega}
    \Omega(e) = \dfrac{1}{2} \dfrac{\arcsin(e) - e\sqrt{1-e^2}}{e\sqrt{1-e^2} - (1-e^2)\arcsin(e)}\,,
\end{equation}
where $e(\epsilon)=\sqrt{1-(1-\epsilon)^2}$
\citep{Cappellari:2007aa}. Rearranging gives
\begin{equation}
\left(\dfrac{V}{\sigma}\right)^2 = \dfrac{\Omega(1-\delta)-1}{1+\alpha \Omega(1-\delta)}\,.
\end{equation}
Finally, E+11 combine this with the relationship between $(V/\sigma)$ and $\lambda_\mathrm{R}$,
\begin{equation}
\label{vsigma}
\lambda_\mathrm{R} \equiv \dfrac{\langle R\vert V\vert \rangle}{\langle R \sqrt{V^2 + \sigma^2}\rangle} \approx \dfrac{k(V/\sigma)}{\sqrt{1+k^2(V/\sigma)^2}}\,,
\end{equation}
where $k \approx 1.1$ is found to be in good agreement with observations and models.

\citet{Cappellari:2007aa} found a value of $c$, the constant of
proportionality between ellipticity and anisotropy, of roughly $0.7$
characterised FRs well. 

This relationship is shown as the magenta line in Figure~\ref{fig:LambdaEms} and a linear scaling of it is used to determine a cutoff between FRs and SRs for galaxies viewed edge on (equation~\ref{newcutoff}).


\bsp	
\label{lastpage}

\begin{thebibliography}{}
\makeatletter
\relax
\def\mn@urlcharsother{\let\do\@makeother \do\$\do\&\do\#\do\^\do\_\do\%\do\~}
\def\mn@doi{\begingroup\mn@urlcharsother \@ifnextchar [ {\mn@doi@}
  {\mn@doi@[]}}
\def\mn@doi@[#1]#2{\def\@tempa{#1}\ifx\@tempa\@empty \href
  {http://dx.doi.org/#2} {doi:#2}\else \href {http://dx.doi.org/#2} {#1}\fi
  \endgroup}
\def\mn@eprint#1#2{\mn@eprint@#1:#2::\@nil}
\def\mn@eprint@arXiv#1{\href {http://arxiv.org/abs/#1} {{\tt arXiv:#1}}}
\def\mn@eprint@dblp#1{\href {http://dblp.uni-trier.de/rec/bibtex/#1.xml}
  {dblp:#1}}
\def\mn@eprint@#1:#2:#3:#4\@nil{\def\@tempa {#1}\def\@tempb {#2}\def\@tempc
  {#3}\ifx \@tempc \@empty \let \@tempc \@tempb \let \@tempb \@tempa \fi \ifx
  \@tempb \@empty \def\@tempb {arXiv}\fi \@ifundefined
  {mn@eprint@\@tempb}{\@tempb:\@tempc}{\expandafter \expandafter \csname
  mn@eprint@\@tempb\endcsname \expandafter{\@tempc}}}

\bibitem[\protect\citeauthoryear{{Barnes}}{{Barnes}}{1988}]{Barnes:1988aa}
{Barnes} J.~E.,  1988, \mn@doi [\apj] {10.1086/166593}, \href
  {http://adsabs.harvard.edu/abs/1988ApJ...331..699B} {331, 699}

\bibitem[\protect\citeauthoryear{{Barnes}}{{Barnes}}{1998}]{Barnes:1998aa}
{Barnes} J.~E.,  1998, in {R.~C.~Kennicutt} ed., Saas-Fee Advanced Course 26:
  Galaxies: Interactions and Induced Star Formation. pp 275--+

\bibitem[\protect\citeauthoryear{{Barnes} \& {Hernquist}}{{Barnes} \&
  {Hernquist}}{1996}]{Barnes:1996aa}
{Barnes} J.~E.,  {Hernquist} L.,  1996, \mn@doi [\apj] {10.1086/177957}, \href
  {http://adsabs.harvard.edu/abs/1996ApJ...471..115B} {471, 115}

\bibitem[\protect\citeauthoryear{{Behroozi}, {Conroy}  \&
  {Wechsler}}{{Behroozi} et~al.}{2010}]{Behroozi:2010aa}
{Behroozi} P.~S.,  {Conroy} C.,   {Wechsler} R.~H.,  2010, \mn@doi [\apj]
  {10.1088/0004-637X/717/1/379}, \href
  {http://adsabs.harvard.edu/abs/2010ApJ...717..379B} {717, 379}

\bibitem[\protect\citeauthoryear{{Bekki} \& {Shioya}}{{Bekki} \&
  {Shioya}}{1997}]{Bekki:1997aa}
{Bekki} K.,  {Shioya} Y.,  1997, \mn@doi [\apjl] {10.1086/310541}, \href
  {http://adsabs.harvard.edu/abs/1997ApJ...478L..17B} {478, L17+}

\bibitem[\protect\citeauthoryear{{Bender}}{{Bender}}{1988}]{Bender:1988aa}
{Bender} R.,  1988, \aap, \href
  {http://adsabs.harvard.edu/abs/1988A\%26A...193L...7B} {193, L7}

\bibitem[\protect\citeauthoryear{{Bender} \& {Nieto}}{{Bender} \&
  {Nieto}}{1990}]{Bender:1990aa}
{Bender} R.,  {Nieto} J.,  1990, \aap, \href
  {http://adsabs.harvard.edu/abs/1990A\%26A...239...97B} {239, 97}

\bibitem[\protect\citeauthoryear{{Bender}, {Doebereiner}  \&
  {Moellenhoff}}{{Bender} et~al.}{1988}]{Bender:1988ab}
{Bender} R.,  {Doebereiner} S.,   {Moellenhoff} C.,  1988, \aaps, \href
  {http://adsabs.harvard.edu/abs/1988A\%26AS...74..385B} {74, 385}

\bibitem[\protect\citeauthoryear{{Bender}, {Surma}, {Doebereiner},
  {Moellenhoff}  \& {Madejsky}}{{Bender} et~al.}{1989}]{Bender:1989aa}
{Bender} R.,  {Surma} P.,  {Doebereiner} S.,  {Moellenhoff} C.,   {Madejsky}
  R.,  1989, \aap, \href {http://adsabs.harvard.edu/abs/1989A\%26A...217...35B}
  {217, 35}

\bibitem[\protect\citeauthoryear{{Bendo} \& {Barnes}}{{Bendo} \&
  {Barnes}}{2000}]{Bendo:2000aa}
{Bendo} G.~J.,  {Barnes} J.~E.,  2000, \mn@doi [\mnras]
  {10.1046/j.1365-8711.2000.03475.x}, \href
  {http://adsabs.harvard.edu/abs/2000MNRAS.316..315B} {316, 315}

\bibitem[\protect\citeauthoryear{{Binney}}{{Binney}}{2005}]{2005MNRAS.363..937B}
{Binney} J.,  2005, \mn@doi [MNRAS] {10.1111/j.1365-2966.2005.09495.x}, \href
  {http://adsabs.harvard.edu/abs/2005MNRAS.363..937B} {363, 937}

\bibitem[\protect\citeauthoryear{{Bois} et~al.}{{Bois}
  et~al.}{2010}]{Bois:2010aa}
{Bois} M.,  et~al., 2010, \mn@doi [\mnras] {10.1111/j.1365-2966.2010.16885.x},
  \href {http://adsabs.harvard.edu/abs/2010MNRAS.406.2405B} {406, 2405}

\bibitem[\protect\citeauthoryear{{Bois} et~al.}{{Bois}
  et~al.}{2011}]{Bois:2011aa}
{Bois} M.,  et~al., 2011, \mn@doi [\mnras] {10.1111/j.1365-2966.2011.19113.x},
  \href {http://adsabs.harvard.edu/abs/2011MNRAS.416.1654B} {416, 1654}

\bibitem[\protect\citeauthoryear{{Boselli} et~al.,}{{Boselli}
  et~al.}{2014}]{Boselli:14}
{Boselli} A.,  et~al., 2014, \mn@doi [\aap] {10.1051/0004-6361/201424419},
  \href {http://adsabs.harvard.edu/abs/2014A%26A...570A..69B} {570, A69}

\bibitem[\protect\citeauthoryear{{Bottrell}, {Torrey}, {Simard}  \&
  {Ellison}}{{Bottrell} et~al.}{2017}]{Bottrell:2017}
{Bottrell} C.,  {Torrey} P.,  {Simard} L.,   {Ellison} S.~L.,  2017, \mn@doi
  [\mnras] {10.1093/mnras/stx017}, \href
  {http://adsabs.harvard.edu/abs/2017MNRAS.tmp...30B} {}

\bibitem[\protect\citeauthoryear{{Bournaud}, {Jog}  \& {Combes}}{{Bournaud}
  et~al.}{2007}]{2007A&A...476.1179B}
{Bournaud} F.,  {Jog} C.~J.,   {Combes} F.,  2007, \mn@doi [\aap]
  {10.1051/0004-6361:20078010}, \href
  {http://adsabs.harvard.edu/abs/2007A\%26A...476.1179B} {476, 1179}

\bibitem[\protect\citeauthoryear{{Bridge} et~al.}{{Bridge}
  et~al.}{2007}]{Bridge:2007aa}
{Bridge} C.~R.,  et~al., 2007, \mn@doi [\apj] {10.1086/512029}, \href
  {http://adsabs.harvard.edu/abs/2007ApJ...659..931B} {659, 931}

\bibitem[\protect\citeauthoryear{{Bryan} et~al.}{{Bryan}
  et~al.}{2014}]{Bryan:2014aa}
{Bryan} G.~L.,  et~al., 2014, \mn@doi [\apjs] {10.1088/0067-0049/211/2/19},
  \href {http://adsabs.harvard.edu/abs/2014ApJS..211...19B} {211, 19}

\bibitem[\protect\citeauthoryear{{Cappellari} et~al.}{{Cappellari}
  et~al.}{2007}]{Cappellari:2007aa}
{Cappellari} M.,  et~al., 2007, \mn@doi [\mnras]
  {10.1111/j.1365-2966.2007.11963.x}, \href
  {http://adsabs.harvard.edu/abs/2007MNRAS.379..418C} {379, 418}

\bibitem[\protect\citeauthoryear{{Cappellari} et~al.}{{Cappellari}
  et~al.}{2011}]{Cappellari:2011aa}
{Cappellari} M.,  et~al., 2011, \mn@doi [\mnras]
  {10.1111/j.1365-2966.2010.18174.x}, \href
  {http://adsabs.harvard.edu/abs/2011MNRAS.tmp..269C} {pp 269--+}

\bibitem[\protect\citeauthoryear{{Carlberg}}{{Carlberg}}{1986}]{Carlberg:1986aa}
{Carlberg} R.~G.,  1986, \mn@doi [\apj] {10.1086/164711}, \href
  {http://adsabs.harvard.edu/abs/1986ApJ...310..593C} {310, 593}

\bibitem[\protect\citeauthoryear{{Chabrier}}{{Chabrier}}{2003}]{Chabrier:2003aa}
{Chabrier} G.,  2003, \mn@doi [\pasp] {10.1086/376392}, \href
  {http://adsabs.harvard.edu/abs/2003PASP..115..763C} {115, 763}

\bibitem[\protect\citeauthoryear{{Choi} \& {Yi}}{{Choi} \&
  {Yi}}{2017}]{Choi:17}
{Choi} H.,  {Yi} S.~K.,  2017, \mn@doi [\apj] {10.3847/1538-4357/aa5e4b}, \href
  {http://adsabs.harvard.edu/abs/2017ApJ...837...68C} {837, 68}

\bibitem[\protect\citeauthoryear{Conselice}{Conselice}{2006}]{Conselice21122006}
Conselice C.~J.,  2006, \mn@doi [Monthly Notices of the Royal Astronomical
  Society] {10.1111/j.1365-2966.2006.11114.x}, 373, 1389

\bibitem[\protect\citeauthoryear{{Cox}, {Dutta}, {Di Matteo}, {Hernquist},
  {Hopkins}, {Robertson}  \& {Springel}}{{Cox} et~al.}{2006}]{Cox:2006aa}
{Cox} T.~J.,  {Dutta} S.~N.,  {Di Matteo} T.,  {Hernquist} L.,  {Hopkins}
  P.~F.,  {Robertson} B.,   {Springel} V.,  2006, \mn@doi [\apj]
  {10.1086/507474}, \href {http://adsabs.harvard.edu/abs/2006ApJ...650..791C}
  {650, 791}

\bibitem[\protect\citeauthoryear{{Crain} et~al.}{{Crain}
  et~al.}{2015}]{Crain:2015aa}
{Crain} R.~A.,  et~al., 2015, \mn@doi [\mnras] {10.1093/mnras/stv725}, \href
  {http://adsabs.harvard.edu/abs/2015MNRAS.450.1937C} {450, 1937}

\bibitem[\protect\citeauthoryear{{Davies}, {Efstathiou}, {Fall}, {Illingworth}
  \& {Schechter}}{{Davies} et~al.}{1983}]{Davies:1983aa}
{Davies} R.~L.,  {Efstathiou} G.,  {Fall} S.~M.,  {Illingworth} G.,
  {Schechter} P.~L.,  1983, \mn@doi [\apj] {10.1086/160757}, \href
  {http://adsabs.harvard.edu/abs/1983ApJ...266...41D} {266, 41}

\bibitem[\protect\citeauthoryear{{Davis}, {Efstathiou}, {Frenk}  \&
  {White}}{{Davis} et~al.}{1985}]{Davis:1985aa}
{Davis} M.,  {Efstathiou} G.,  {Frenk} C.~S.,   {White} S.~D.~M.,  1985,
  \mn@doi [\apj] {10.1086/163168}, \href
  {http://ads.ari.uni-heidelberg.de/abs/1985ApJ...292..371D} {292, 371}

\bibitem[\protect\citeauthoryear{{De Lucia} \& {Blaizot}}{{De Lucia} \&
  {Blaizot}}{2007}]{DeLucia:2007aa}
{De Lucia} G.,  {Blaizot} J.,  2007, \mn@doi [\mnras]
  {10.1111/j.1365-2966.2006.11287.x}, \href
  {http://adsabs.harvard.edu/abs/2007MNRAS.375....2D} {375, 2}

\bibitem[\protect\citeauthoryear{{Dekel} \& {Birnboim}}{{Dekel} \&
  {Birnboim}}{2006}]{2006MNRAS.368....2D}
{Dekel} A.,  {Birnboim} Y.,  2006, \mn@doi [\mnras]
  {10.1111/j.1365-2966.2006.10145.x}, \href
  {http://adsabs.harvard.edu/abs/2006MNRAS.368....2D} {368, 2}

\bibitem[\protect\citeauthoryear{{Di Matteo}, {Jog}, {Lehnert}, {Combes}  \&
  {Semelin}}{{Di Matteo} et~al.}{2009}]{2009A&A...501L...9D}
{Di Matteo} P.,  {Jog} C.~J.,  {Lehnert} M.~D.,  {Combes} F.,   {Semelin} B.,
  2009, \mn@doi [\aap] {10.1051/0004-6361/200912354}, \href
  {http://adsabs.harvard.edu/abs/2009A\%26A...501L...9D} {501, L9}

\bibitem[\protect\citeauthoryear{{Dolag}, {Borgani}, {Murante}  \&
  {Springel}}{{Dolag} et~al.}{2009}]{2009MNRAS.399..497D}
{Dolag} K.,  {Borgani} S.,  {Murante} G.,   {Springel} V.,  2009, \mn@doi
  [\mnras] {10.1111/j.1365-2966.2009.15034.x}, \href
  {http://adsabs.harvard.edu/abs/2009MNRAS.399..497D} {399, 497}

\bibitem[\protect\citeauthoryear{{Dubois} et~al.}{{Dubois}
  et~al.}{2014}]{Dubois:2014aa}
{Dubois} Y.,  et~al., 2014, \mn@doi [\mnras] {10.1093/mnras/stu1227}, \href
  {http://adsabs.harvard.edu/abs/2014MNRAS.444.1453D} {444, 1453}

\bibitem[\protect\citeauthoryear{{Ellis} \& {O'Sullivan}}{{Ellis} \&
  {O'Sullivan}}{2006}]{2006MNRAS.367..627E}
{Ellis} S.~C.,  {O'Sullivan} E.,  2006, \mn@doi [\mnras]
  {10.1111/j.1365-2966.2005.09982.x}, \href
  {http://adsabs.harvard.edu/abs/2006MNRAS.367..627E} {367, 627}

\bibitem[\protect\citeauthoryear{{Ellison}, {Mendel}, {Patton}  \&
  {Scudder}}{{Ellison} et~al.}{2013}]{Ellison:2013aa}
{Ellison} S.~L.,  {Mendel} J.~T.,  {Patton} D.~R.,   {Scudder} J.~M.,  2013,
  \mn@doi [\mnras] {10.1093/mnras/stt1562}, \href
  {http://adsabs.harvard.edu/abs/2013MNRAS.435.3627E} {435, 3627}

\bibitem[\protect\citeauthoryear{{Emsellem} et~al.}{{Emsellem}
  et~al.}{2007a}]{Emsellem:2007aa}
{Emsellem} E.,  et~al., 2007a, \mn@doi [\mnras]
  {10.1111/j.1365-2966.2007.11752.x}, \href
  {http://adsabs.harvard.edu/abs/2007MNRAS.379..401E} {379, 401}

\bibitem[\protect\citeauthoryear{{Emsellem} et~al.}{{Emsellem}
  et~al.}{2007b}]{2007MNRAS.379..401E}
{Emsellem} E.,  et~al., 2007b, \mn@doi [MNRAS]
  {10.1111/j.1365-2966.2007.11752.x}, \href
  {http://adsabs.harvard.edu/abs/2007MNRAS.379..401E} {379, 401}

\bibitem[\protect\citeauthoryear{{Emsellem} et~al.,}{{Emsellem}
  et~al.}{2011}]{Emsellem:2011aa}
{Emsellem} E.,  et~al., 2011, \mn@doi [\mnras]
  {10.1111/j.1365-2966.2011.18496.x}, \href
  {http://adsabs.harvard.edu/abs/2011MNRAS.414..888E} {414, 888}

\bibitem[\protect\citeauthoryear{{Faber} et~al.}{{Faber}
  et~al.}{1997}]{Faber:1997aa}
{Faber} S.~M.,  et~al., 1997, \mn@doi [\aj] {10.1086/118606}, \href
  {http://adsabs.harvard.edu/abs/1997AJ....114.1771F} {114, 1771}

\bibitem[\protect\citeauthoryear{{Fall} \& {Efstathiou}}{{Fall} \&
  {Efstathiou}}{1980}]{Fall:1980aa}
{Fall} S.~M.,  {Efstathiou} G.,  1980, \mnras, \href
  {http://adsabs.harvard.edu/abs/1980MNRAS.193..189F} {193, 189}

\bibitem[\protect\citeauthoryear{{Genel} et~al.}{{Genel}
  et~al.}{2014}]{Genel:2014aa}
{Genel} S.,  et~al., 2014, \mn@doi [\mnras] {10.1093/mnras/stu1654}, \href
  {http://adsabs.harvard.edu/abs/2014MNRAS.445..175G} {445, 175}

\bibitem[\protect\citeauthoryear{{Gerhard}}{{Gerhard}}{1981}]{Gerhard:1981aa}
{Gerhard} O.~E.,  1981, \mnras, \href
  {http://adsabs.harvard.edu/abs/1981MNRAS.197..179G} {197, 179}

\bibitem[\protect\citeauthoryear{{Guo}, {White}, {Li}  \&
  {Boylan-Kolchin}}{{Guo} et~al.}{2010}]{Guo:2010aa}
{Guo} Q.,  {White} S.,  {Li} C.,   {Boylan-Kolchin} M.,  2010, \mn@doi [\mnras]
  {10.1111/j.1365-2966.2010.16341.x}, \href
  {http://adsabs.harvard.edu/abs/2010MNRAS.404.1111G} {404, 1111}

\bibitem[\protect\citeauthoryear{{Hernquist}}{{Hernquist}}{1993}]{Hernquist:1993aa}
{Hernquist} L.,  1993, \mn@doi [\apj] {10.1086/172686}, \href
  {http://adsabs.harvard.edu/abs/1993ApJ...409..548H} {409, 548}

\bibitem[\protect\citeauthoryear{{Hinshaw} et~al.}{{Hinshaw}
  et~al.}{2013}]{Hinshaw:2013aa}
{Hinshaw} G.,  et~al., 2013, \mn@doi [\apjs] {10.1088/0067-0049/208/2/19},
  \href {http://adsabs.harvard.edu/abs/2013ApJS..208...19H} {208, 19}

\bibitem[\protect\citeauthoryear{{Hirschmann}, {Dolag}, {Saro}, {Bachmann},
  {Borgani}  \& {Burkert}}{{Hirschmann} et~al.}{2014}]{Hirschmann:2014aa}
{Hirschmann} M.,  {Dolag} K.,  {Saro} A.,  {Bachmann} L.,  {Borgani} S.,
  {Burkert} A.,  2014, \mn@doi [\mnras] {10.1093/mnras/stu1023}, \href
  {http://adsabs.harvard.edu/abs/2014MNRAS.442.2304H} {442, 2304}

\bibitem[\protect\citeauthoryear{{Hoffman}, {Cox}, {Dutta}  \&
  {Hernquist}}{{Hoffman} et~al.}{2009}]{2009ApJ...705..920H}
{Hoffman} L.,  {Cox} T.~J.,  {Dutta} S.,   {Hernquist} L.,  2009, \mn@doi
  [\apj] {10.1088/0004-637X/705/1/920}, \href
  {http://adsabs.harvard.edu/abs/2009ApJ...705..920H} {705, 920}

\bibitem[\protect\citeauthoryear{{Hopkins}}{{Hopkins}}{2015}]{Hopkins:2015aa}
{Hopkins} P.~F.,  2015, \mn@doi [\mnras] {10.1093/mnras/stv195}, \href
  {http://adsabs.harvard.edu/abs/2015MNRAS.450...53H} {450, 53}

\bibitem[\protect\citeauthoryear{{Hopkins}, {Hernquist}, {Cox}, {Dutta}  \&
  {Rothberg}}{{Hopkins} et~al.}{2008}]{Hopkins:2008aa}
{Hopkins} P.~F.,  {Hernquist} L.,  {Cox} T.~J.,  {Dutta} S.~N.,   {Rothberg}
  B.,  2008, \mn@doi [\apj] {10.1086/587544}, \href
  {http://adsabs.harvard.edu/abs/2008ApJ...679..156H} {679, 156}

\bibitem[\protect\citeauthoryear{{Hu}, {Naab}, {Walch}, {Moster}  \&
  {Oser}}{{Hu} et~al.}{2014}]{Hu:2014aa}
{Hu} C.-Y.,  {Naab} T.,  {Walch} S.,  {Moster} B.~P.,   {Oser} L.,  2014,
  \mn@doi [\mnras] {10.1093/mnras/stu1187}, \href
  {http://adsabs.harvard.edu/abs/2014MNRAS.443.1173H} {443, 1173}

\bibitem[\protect\citeauthoryear{{Jesseit}, {Naab}, {Peletier}  \&
  {Burkert}}{{Jesseit} et~al.}{2007}]{Jesseit:2007aa}
{Jesseit} R.,  {Naab} T.,  {Peletier} R.~F.,   {Burkert} A.,  2007, \mn@doi
  [\mnras] {10.1111/j.1365-2966.2007.11524.x}, \href
  {http://adsabs.harvard.edu/abs/2007MNRAS.376..997J} {376, 997}

\bibitem[\protect\citeauthoryear{{Jesseit}, {Cappellari}, {Naab}, {Emsellem}
  \& {Burkert}}{{Jesseit} et~al.}{2009}]{Jesseit:2009aa}
{Jesseit} R.,  {Cappellari} M.,  {Naab} T.,  {Emsellem} E.,   {Burkert} A.,
  2009, \mn@doi [\mnras] {10.1111/j.1365-2966.2009.14984.x}, \href
  {http://adsabs.harvard.edu/abs/2009MNRAS.397.1202J} {397, 1202}

\bibitem[\protect\citeauthoryear{{Joachimi}, {Semboloni}, {Bett}, {Hartlap},
  {Hilbert}, {Hoekstra}, {Schneider}  \& {Schrabback}}{{Joachimi}
  et~al.}{2013}]{2013MNRAS.431..477J}
{Joachimi} B.,  {Semboloni} E.,  {Bett} P.~E.,  {Hartlap} J.,  {Hilbert} S.,
  {Hoekstra} H.,  {Schneider} P.,   {Schrabback} T.,  2013, \mn@doi [MNRAS]
  {10.1093/mnras/stt172}, \href
  {http://adsabs.harvard.edu/abs/2013MNRAS.431..477J} {431, 477}

\bibitem[\protect\citeauthoryear{{Khandai}, {Di Matteo}, {Croft}, {Wilkins},
  {Feng}, {Tucker}, {DeGraf}  \& {Liu}}{{Khandai} et~al.}{2015}]{Khandai2015aa}
{Khandai} N.,  {Di Matteo} T.,  {Croft} R.,  {Wilkins} S.,  {Feng} Y.,
  {Tucker} E.,  {DeGraf} C.,   {Liu} M.-S.,  2015, \mn@doi [\mnras]
  {10.1093/mnras/stv627}, \href
  {http://adsabs.harvard.edu/abs/2015MNRAS.450.1349K} {450, 1349}

\bibitem[\protect\citeauthoryear{{Kormendy}}{{Kormendy}}{2016}]{Kormendy:2016aa}
{Kormendy} J.,  2016, \mn@doi [Galactic Bulges] {10.1007/978-3-319-19378-6_16},
  \href {http://adsabs.harvard.edu/abs/2016ASSL..418..431K} {418, 431}

\bibitem[\protect\citeauthoryear{{Kormendy} \& {Bender}}{{Kormendy} \&
  {Bender}}{1996}]{Kormendy:1996aa}
{Kormendy} J.,  {Bender} R.,  1996, \mn@doi [\apjl] {10.1086/310095}, \href
  {http://adsabs.harvard.edu/abs/1996ApJ...464L.119K} {464, L119+}

\bibitem[\protect\citeauthoryear{{Kormendy} \& {Bender}}{{Kormendy} \&
  {Bender}}{2012}]{Kormendy:2012aa}
{Kormendy} J.,  {Bender} R.,  2012, \mn@doi [\apjs]
  {10.1088/0067-0049/198/1/2}, \href
  {http://adsabs.harvard.edu/abs/2012ApJS..198....2K} {198, 2}

\bibitem[\protect\citeauthoryear{{Kormendy}, {Fisher}, {Cornell}  \&
  {Bender}}{{Kormendy} et~al.}{2009}]{Kormendy:2009aa}
{Kormendy} J.,  {Fisher} D.~B.,  {Cornell} M.~E.,   {Bender} R.,  2009, \mn@doi
  [\apjs] {10.1088/0067-0049/182/1/216}, \href
  {http://adsabs.harvard.edu/abs/2009ApJS..182..216K} {182, 216}

\bibitem[\protect\citeauthoryear{{Krajnovic} et~al.}{{Krajnovic}
  et~al.}{2011}]{Krajnovic:2011aa}
{Krajnovic} D.,  et~al., 2011, preprint, \href
  {http://adsabs.harvard.edu/abs/2011arXiv1102.3801K} {} (\mn@eprint {arXiv}
  {1102.3801})

\bibitem[\protect\citeauthoryear{{Lagos}, {Theuns}, {Stevens}, {Cortese},
  {Padilla}, {Davis}, {Contreras}  \& {Croton}}{{Lagos}
  et~al.}{2017}]{Lagos:2017}
{Lagos} C.~d.~P.,  {Theuns} T.,  {Stevens} A.~R.~H.,  {Cortese} L.,  {Padilla}
  N.~D.,  {Davis} T.~A.,  {Contreras} S.,   {Croton} D.,  2017, \mn@doi
  [\mnras] {10.1093/mnras/stw2610}, \href
  {http://adsabs.harvard.edu/abs/2017MNRAS.464.3850L} {464, 3850}

\bibitem[\protect\citeauthoryear{{Lake}}{{Lake}}{1989}]{Lake:1989aa}
{Lake} G.,  1989, \mn@doi [\aj] {10.1086/115074}, \href
  {http://adsabs.harvard.edu/abs/1989AJ.....97.1312L} {97, 1312}

\bibitem[\protect\citeauthoryear{{Larson}}{{Larson}}{1969}]{Larson:1969aa}
{Larson} R.~B.,  1969, \mn@doi [\mnras] {10.1093/mnras/145.4.405}, \href
  {http://adsabs.harvard.edu/abs/1969MNRAS.145..405L} {145, 405}

\bibitem[\protect\citeauthoryear{{Lauer} et~al.}{{Lauer}
  et~al.}{1995}]{Lauer:1995aa}
{Lauer} T.~R.,  et~al., 1995, \mn@doi [\aj] {10.1086/117719}, \href
  {http://adsabs.harvard.edu/abs/1995AJ....110.2622L} {110, 2622}

\bibitem[\protect\citeauthoryear{{Merritt}}{{Merritt}}{2006}]{2006ApJ...648..976M}
{Merritt} D.,  2006, \mn@doi [\apj] {10.1086/506139}, \href
  {http://adsabs.harvard.edu/abs/2006ApJ...648..976M} {648, 976}

\bibitem[\protect\citeauthoryear{{Mo}, {van den Bosch}  \& {White}}{{Mo}
  et~al.}{2010}]{2010gfe..book.....M}
{Mo} H.,  {van den Bosch} F.~C.,   {White} S.,  2010, {Galaxy Formation and
  Evolution, Cambridge University Press}.
Cambridge University Press, 2010

\bibitem[\protect\citeauthoryear{{Moster}, {Somerville}, {Maulbetsch}, {van den
  Bosch}, {Macci{\`o}}, {Naab}  \& {Oser}}{{Moster}
  et~al.}{2010}]{Moster:2010aa}
{Moster} B.~P.,  {Somerville} R.~S.,  {Maulbetsch} C.,  {van den Bosch} F.~C.,
  {Macci{\`o}} A.~V.,  {Naab} T.,   {Oser} L.,  2010, \mn@doi [\apj]
  {10.1088/0004-637X/710/2/903}, \href
  {http://adsabs.harvard.edu/abs/2010ApJ...710..903M} {710, 903}

\bibitem[\protect\citeauthoryear{{Moster}, {Macci{\`o}}, {Somerville}, {Naab}
  \& {Cox}}{{Moster} et~al.}{2011}]{Moster:2011aa}
{Moster} B.~P.,  {Macci{\`o}} A.~V.,  {Somerville} R.~S.,  {Naab} T.,   {Cox}
  T.~J.,  2011, \mn@doi [\mnras] {10.1111/j.1365-2966.2011.18984.x}, \href
  {http://adsabs.harvard.edu/abs/2011MNRAS.415.3750M} {415, 3750}

\bibitem[\protect\citeauthoryear{{Moster}, {Naab}  \& {White}}{{Moster}
  et~al.}{2013}]{Moster:2013aa}
{Moster} B.~P.,  {Naab} T.,   {White} S.~D.~M.,  2013, \mn@doi [\mnras]
  {10.1093/mnras/sts261}, \href
  {http://adsabs.harvard.edu/abs/2013MNRAS.428.3121M} {428, 3121}

\bibitem[\protect\citeauthoryear{{Moster}, {Macci{\`o}}  \&
  {Somerville}}{{Moster} et~al.}{2014}]{Moster:2014aa}
{Moster} B.~P.,  {Macci{\`o}} A.~V.,   {Somerville} R.~S.,  2014, \mn@doi
  [\mnras] {10.1093/mnras/stt1702}, \href
  {http://adsabs.harvard.edu/abs/2014MNRAS.437.1027M} {437, 1027}

\bibitem[\protect\citeauthoryear{{Muzzin} et~al.}{{Muzzin}
  et~al.}{2013}]{2013ApJ...777...18M}
{Muzzin} A.,  et~al., 2013, \mn@doi [The Astrophysical Journal]
  {10.1088/0004-637X/777/1/18}, \href
  {http://adsabs.harvard.edu/abs/2013ApJ...777...18M} {777, 18}

\bibitem[\protect\citeauthoryear{{Naab} \& {Burkert}}{{Naab} \&
  {Burkert}}{2003}]{Naab:2003aa}
{Naab} T.,  {Burkert} A.,  2003, \mn@doi [\apj] {10.1086/378581}, \href
  {http://adsabs.harvard.edu/abs/2003ApJ...597..893N} {597, 893}

\bibitem[\protect\citeauthoryear{{Naab}, {Burkert}  \& {Hernquist}}{{Naab}
  et~al.}{1999}]{Naab:1999aa}
{Naab} T.,  {Burkert} A.,   {Hernquist} L.,  1999, \mn@doi [\apjl]
  {10.1086/312275}, \href {http://adsabs.harvard.edu/abs/1999ApJ...523L.133N}
  {523, L133}

\bibitem[\protect\citeauthoryear{{Naab}, {Jesseit}  \& {Burkert}}{{Naab}
  et~al.}{2006a}]{Naab:2006aa}
{Naab} T.,  {Jesseit} R.,   {Burkert} A.,  2006a, \mn@doi [\mnras]
  {10.1111/j.1365-2966.2006.10902.x}, \href
  {http://adsabs.harvard.edu/abs/2006MNRAS.372..839N} {372, 839}

\bibitem[\protect\citeauthoryear{{Naab}, {Khochfar}  \& {Burkert}}{{Naab}
  et~al.}{2006b}]{2006ApJ...636L..81N}
{Naab} T.,  {Khochfar} S.,   {Burkert} A.,  2006b, \mn@doi [\apjl]
  {10.1086/500205}, \href {http://adsabs.harvard.edu/abs/2006ApJ...636L..81N}
  {636, L81}

\bibitem[\protect\citeauthoryear{{Naab}, {Johansson}, {Ostriker}  \&
  {Efstathiou}}{{Naab} et~al.}{2007}]{2007ApJ...658..710N}
{Naab} T.,  {Johansson} P.~H.,  {Ostriker} J.~P.,   {Efstathiou} G.,  2007,
  \mn@doi [\apj] {10.1086/510841}, \href
  {http://adsabs.harvard.edu/abs/2007ApJ...658..710N} {658, 710}

\bibitem[\protect\citeauthoryear{{Naab} et~al.}{{Naab}
  et~al.}{2014}]{Naab:2014aa}
{Naab} T.,  et~al., 2014, \mn@doi [\mnras] {10.1093/mnras/stt1919}, \href
  {http://adsabs.harvard.edu/abs/2014MNRAS.444.3357N} {444, 3357}

\bibitem[\protect\citeauthoryear{{Negroponte} \& {White}}{{Negroponte} \&
  {White}}{1983}]{Negroponte:1983aa}
{Negroponte} J.,  {White} S.~D.~M.,  1983, \mnras, \href
  {http://adsabs.harvard.edu/abs/1983MNRAS.205.1009N} {205, 1009}

\bibitem[\protect\citeauthoryear{{Nelson} et~al.,}{{Nelson}
  et~al.}{2015}]{2015A&C....13...12N}
{Nelson} D.,  et~al., 2015, \mn@doi [Astronomy and Computing]
  {10.1016/j.ascom.2015.09.003}, \href
  {http://adsabs.harvard.edu/abs/2015A\%26C....13...12N} {13, 12}

\bibitem[\protect\citeauthoryear{{Partridge} \& {Peebles}}{{Partridge} \&
  {Peebles}}{1967}]{Partridge:1967aa}
{Partridge} R.~B.,  {Peebles} P.~J.~E.,  1967, \mn@doi [\apj] {10.1086/149079},
  \href {http://adsabs.harvard.edu/abs/1967ApJ...147..868P} {147, 868}

\bibitem[\protect\citeauthoryear{{Pellegrini}}{{Pellegrini}}{2005}]{2005MNRAS.364..169P}
{Pellegrini} S.,  2005, \mn@doi [\mnras] {10.1111/j.1365-2966.2005.09549.x},
  \href {http://adsabs.harvard.edu/abs/2005MNRAS.364..169P} {364, 169}

\bibitem[\protect\citeauthoryear{{Qu}, {Di Matteo}, {Lehnert}, {van Driel}  \&
  {Jog}}{{Qu} et~al.}{2010}]{2010A&A...515A..11Q}
{Qu} Y.,  {Di Matteo} P.,  {Lehnert} M.,  {van Driel} W.,   {Jog} C.~J.,  2010,
  \mn@doi [\aap] {10.1051/0004-6361/200913559}, \href
  {http://adsabs.harvard.edu/abs/2010A\%26A...515A..11Q} {515, A11}

\bibitem[\protect\citeauthoryear{{Remus}, {Dolag}, {Naab}, {Burkert},
  {Hirschmann}, {Hoffmann}  \& {Johansson}}{{Remus}
  et~al.}{2017}]{Remus:2017aa}
{Remus} R.-S.,  {Dolag} K.,  {Naab} T.,  {Burkert} A.,  {Hirschmann} M.,
  {Hoffmann} T.~L.,   {Johansson} P.~H.,  2017, \mn@doi [\mnras]
  {10.1093/mnras/stw2594}, \href
  {http://adsabs.harvard.edu/abs/2017MNRAS.464.3742R} {464, 3742}

\bibitem[\protect\citeauthoryear{{Robertson}, {Bullock}, {Cox}, {Di Matteo},
  {Hernquist}, {Springel}  \& {Yoshida}}{{Robertson}
  et~al.}{2006}]{Robertson:2006aa}
{Robertson} B.,  {Bullock} J.~S.,  {Cox} T.~J.,  {Di Matteo} T.,  {Hernquist}
  L.,  {Springel} V.,   {Yoshida} N.,  2006, \mn@doi [\apj] {10.1086/504412},
  \href {http://adsabs.harvard.edu/abs/2006ApJ...645..986R} {645, 986}

\bibitem[\protect\citeauthoryear{{Rodriguez-Gomez} et~al.,}{{Rodriguez-Gomez}
  et~al.}{2015}]{2015MNRAS.449...49R}
{Rodriguez-Gomez} V.,  et~al., 2015, \mn@doi [\mnras] {10.1093/mnras/stv264},
  \href {http://adsabs.harvard.edu/abs/2015MNRAS.449...49R} {449, 49}

\bibitem[\protect\citeauthoryear{{Rodriguez-Gomez} et~al.,}{{Rodriguez-Gomez}
  et~al.}{2016a}]{Gomez:2014aa}
{Rodriguez-Gomez} V.,  et~al., 2016a, preprint, \href
  {http://adsabs.harvard.edu/abs/2016arXiv160909498R} {} (\mn@eprint {arXiv}
  {1609.09498})

\bibitem[\protect\citeauthoryear{{Rodriguez-Gomez} et~al.,}{{Rodriguez-Gomez}
  et~al.}{2016b}]{Rodriguez-Gomez:2016aa}
{Rodriguez-Gomez} V.,  et~al., 2016b, \mn@doi [\mnras] {10.1093/mnras/stw456},
  \href {http://adsabs.harvard.edu/abs/2016MNRAS.458.2371R} {458, 2371}

\bibitem[\protect\citeauthoryear{Sales, Navarro, Theuns, Schaye, White, Frenk,
  Crain  \& Dalla~Vecchia}{Sales et~al.}{2012}]{Sales21062012}
Sales L.~V.,  Navarro J.~F.,  Theuns T.,  Schaye J.,  White S. D.~M.,  Frenk
  C.~S.,  Crain R.~A.,   Dalla~Vecchia C.,  2012, \mn@doi [MNRAS]
  {10.1111/j.1365-2966.2012.20975.x}, 423, 1544

\bibitem[\protect\citeauthoryear{{Schaye} et~al.}{{Schaye}
  et~al.}{2015}]{Schaye:2015aa}
{Schaye} J.,  et~al., 2015, \mn@doi [\mnras] {10.1093/mnras/stu2058}, \href
  {http://adsabs.harvard.edu/abs/2015MNRAS.446..521S} {446, 521}

\bibitem[\protect\citeauthoryear{{Schweizer}}{{Schweizer}}{1982}]{Schweizer:1982aa}
{Schweizer} F.,  1982, \mn@doi [\apj] {10.1086/159573}, \href
  {http://adsabs.harvard.edu/abs/1982ApJ...252..455S} {252, 455}

\bibitem[\protect\citeauthoryear{{Sijacki}, {Vogelsberger}, {Kere{\v s}},
  {Springel}  \& {Hernquist}}{{Sijacki} et~al.}{2012}]{Sijacki:2012}
{Sijacki} D.,  {Vogelsberger} M.,  {Kere{\v s}} D.,  {Springel} V.,
  {Hernquist} L.,  2012, \mn@doi [MNRAS] {10.1111/j.1365-2966.2012.21466.x},
  \href {http://adsabs.harvard.edu/abs/2012MNRAS.424.2999S} {424, 2999}

\bibitem[\protect\citeauthoryear{{Sijacki}, {Vogelsberger}, {Genel},
  {Springel}, {Torrey}, {Snyder}, {Nelson}  \& {Hernquist}}{{Sijacki}
  et~al.}{2015}]{Sijacki:2015aa}
{Sijacki} D.,  {Vogelsberger} M.,  {Genel} S.,  {Springel} V.,  {Torrey} P.,
  {Snyder} G.~F.,  {Nelson} D.,   {Hernquist} L.,  2015, \mn@doi [\mnras]
  {10.1093/mnras/stv1340}, \href
  {http://adsabs.harvard.edu/abs/2015MNRAS.452..575S} {452, 575}

\bibitem[\protect\citeauthoryear{{Snyder} et~al.}{{Snyder}
  et~al.}{2015}]{2015arXiv150207747S}
{Snyder} G.~F.,  et~al., 2015, arXiv:1502.07747, \href
  {http://adsabs.harvard.edu/abs/2015arXiv150207747S} {}

\bibitem[\protect\citeauthoryear{{Springel}}{{Springel}}{2010}]{Springel:2010aa}
{Springel} V.,  2010, \mn@doi [\mnras] {10.1111/j.1365-2966.2009.15715.x},
  \href {http://adsabs.harvard.edu/abs/2010MNRAS.401..791S} {401, 791}

\bibitem[\protect\citeauthoryear{{Springel} \& {Hernquist}}{{Springel} \&
  {Hernquist}}{2003}]{Springel2003}
{Springel} V.,  {Hernquist} L.,  2003, \mn@doi [\mnras]
  {10.1046/j.1365-8711.2003.06206.x}, \href
  {http://adsabs.harvard.edu/abs/2003MNRAS.339..289S} {339, 289}

\bibitem[\protect\citeauthoryear{{Springel}, {Yoshida}  \& {White}}{{Springel}
  et~al.}{2001}]{2001NewA....6...79S}
{Springel} V.,  {Yoshida} N.,   {White} S.~D.~M.,  2001, \mn@doi [\na]
  {10.1016/S1384-1076(01)00042-2}, \href
  {http://adsabs.harvard.edu/abs/2001NewA....6...79S} {6, 79}

\bibitem[\protect\citeauthoryear{{Tasca} et~al.}{{Tasca}
  et~al.}{2014}]{Tasca:2014aa}
{Tasca} L.~A.~M.,  et~al., 2014, \mn@doi [\aap] {10.1051/0004-6361/201321507},
  \href {http://adsabs.harvard.edu/abs/2014A\%26A...565A..10T} {565, A10}

\bibitem[\protect\citeauthoryear{{Tenneti}, {Mandelbaum}, {Di Matteo}, {Feng}
  \& {Khandai}}{{Tenneti} et~al.}{2014}]{Tenneti:2014aa}
{Tenneti} A.,  {Mandelbaum} R.,  {Di Matteo} T.,  {Feng} Y.,   {Khandai} N.,
  2014, \mn@doi [\mnras] {10.1093/mnras/stu586}, \href
  {http://adsabs.harvard.edu/abs/2014MNRAS.441..470T} {441, 470}

\bibitem[\protect\citeauthoryear{{Teyssier}}{{Teyssier}}{2002}]{Teyssier:2002aa}
{Teyssier} R.,  2002, \mn@doi [\aap] {10.1051/0004-6361:20011817}, \href
  {http://adsabs.harvard.edu/abs/2002A\%26A...385..337T} {385, 337}

\bibitem[\protect\citeauthoryear{{Toomre}}{{Toomre}}{1977}]{Toomre:1977aa}
{Toomre} A.,  1977, in {B.~M.~Tinsley \& R.~B.~Larson} ed., Evolution of
  Galaxies and Stellar Populations. pp 401--+

\bibitem[\protect\citeauthoryear{{Toomre} \& {Toomre}}{{Toomre} \&
  {Toomre}}{1972}]{Toomre:1972aa}
{Toomre} A.,  {Toomre} J.,  1972, \mn@doi [\apj] {10.1086/151823}, \href
  {http://adsabs.harvard.edu/abs/1972ApJ...178..623T} {178, 623}

\bibitem[\protect\citeauthoryear{{Vogelsberger}, {Genel}, {Sijacki}, {Torrey},
  {Springel}  \& {Hernquist}}{{Vogelsberger}
  et~al.}{2013}]{2013MNRAS.436.3031V}
{Vogelsberger} M.,  {Genel} S.,  {Sijacki} D.,  {Torrey} P.,  {Springel} V.,
  {Hernquist} L.,  2013, \mn@doi [\mnras] {10.1093/mnras/stt1789}, \href
  {http://adsabs.harvard.edu/abs/2013MNRAS.436.3031V} {436, 3031}

\bibitem[\protect\citeauthoryear{{Vogelsberger} et~al.}{{Vogelsberger}
  et~al.}{2014}]{Vogelsberger:2014aa}
{Vogelsberger} M.,  et~al., 2014, \mn@doi [\mnras] {10.1093/mnras/stu1536},
  \href {http://adsabs.harvard.edu/abs/2014MNRAS.444.1518V} {444, 1518}

\bibitem[\protect\citeauthoryear{{Welker}, {Devriendt}, {Dubois}, {Pichon}  \&
  {Peirani}}{{Welker} et~al.}{2014}]{Welker:2014aa}
{Welker} C.,  {Devriendt} J.,  {Dubois} Y.,  {Pichon} C.,   {Peirani} S.,
  2014, \mn@doi [\mnras] {10.1093/mnrasl/slu106}, \href
  {http://adsabs.harvard.edu/abs/2014MNRAS.445L..46W} {445, L46}

\bibitem[\protect\citeauthoryear{{White}}{{White}}{1979}]{White:1979aa}
{White} S.~D.~M.,  1979, \mn@doi [\apjl] {10.1086/182920}, \href
  {http://adsabs.harvard.edu/abs/1979ApJ...229L...9W} {229, L9}

\makeatother
\end{thebibliography}
\end{document}